\documentclass{ar-1col}

\usepackage{url}
\usepackage{natbib}
\usepackage{amssymb}
\usepackage{verbatim}
\usepackage{subcaption}
\usepackage{amsmath,siunitx}
\usepackage{breqn}
\usepackage{appendix}
\usepackage{upgreek}
\usepackage{soul}
\usepackage[normalem]{ulem}
\usepackage{xr}
\usepackage{pdfpages}

\jname{Annu. Rev. Astron. Astrophys.} 
\jyear{2026}
\doi{10.1146/annurev-astro-052722-104242}

\newcommand{\cii}{[C\textsc{ii}]}
\newcommand{\ciilam}{[C\textsc{ii}]~$158\,\mu{\rm m}$}
\newcommand{\oiii}{[O\textsc{iii}]}
\newcommand{\oiiilam}{[O\textsc{iii}]~$88\,\mu{\rm m}$}
\newcommand{\niii}{[N\textsc{iii}]}
\newcommand{\oi}{[O\textsc{i}]}
\newcommand{\nii}{[N\textsc{ii}]}

\newcommand{\lcii}{$L_{[{\rm C}\,\textsc{ii}]}$}
\newcommand{\loiii}{$L_{[{\rm O}\,\textsc{iii}]}$}
\newcommand{\lir}{$L_{\rm IR}$}
\newcommand{\lsun}{${\rm L}_{\odot}$}
\newcommand{\betad}{$\beta_{\rm d}$}
\newcommand{\betauv}{$\beta_{\rm UV}$}
\newcommand{\Td}{$T_{\rm d}$}
\newcommand{\muv}{$M_{\rm UV}$}
\newcommand{\lmstar}{${\rm log}_{10}(M_{\star}/{\rm M}_{\odot})$}
\newcommand{\mdsm}{$M_{\rm d}/{M}_{\star}$}
\newcommand{\lsfr}{${\rm log}_{10}(SFR_{\star}/{\rm M}_{\odot}/{\rm yr})$}
\newcommand{\llir}{${\rm log}_{10}(L_{\rm IR}/{\rm L}_{\odot})$}
\newcommand{\irxb}{${\rm IRX}$--$\beta$}
\newcommand{\irxm}{${\rm IRX}$--$M_{\star}$}
\newcommand{\fobs}{$f_{\rm obs}$}
\newcommand{\sfrunits}{${\rm M}_{\odot}/{\rm yr}$}
\newcommand{\numsources}{154}
\newcommand{\sfrtot}{$\rm SFR_{\rm UV + IR}$}
\newcommand{\sfruv}{$\rm SFR_{\rm UV}$}
\newcommand{\sfrir}{$\rm SFR_{\rm IR}$}
\newcommand{\athree}{A$^3$COSMOS}

\begin{document}

\markboth{Smit \& Bowler}{High-redshift Universe with ALMA }

\title{The ALMA View of High-Redshift Galaxy Formation}

\author{Renske Smit$^1$ and Rebecca A.A. Bowler$^2$ 
\affil{$^1$Astrophysics Research Institute, Liverpool John Moores University, 146 Brownlow Hill, Liverpool L3 5RF, UK; email: r.smit@ljmu.ac.uk}
\affil{$^2$ Jodrell Bank Centre for Astrophysics, Department of Physics and Astronomy, The University of Manchester, Manchester, M13 9PL, UK; email: rebecca.bowler@manchester.ac.uk}
}

\vspace{-0.1cm}
\begin{abstract}
\vspace{-0.1cm}
The advent of routine operations with the Atacama Large Millimeter/submillimeter Array (ALMA) in the last decade has led to a revolution in the direct study of the inter-stellar medium (ISM) in `normal' high-redshift galaxies in the rest-frame far-infrared (FIR). 
This review summarizes the observational literature on $z > 6.5$ sources observed with ALMA and NOEMA (NOrthern Extended Millimeter Array).
The main findings are as follows:

\vspace{0.15cm}

\hangindent=.3cm$\bullet$
Cool gas consistently scales with star-formation in a wide range of galaxy environments, similar to local relations, suggesting fundamental processes of star-formation stay relatively constant over 13 billion years of cosmic time. 

\hangindent=.3cm$\bullet$
Significant metal enrichment is present in galaxies just $\sim300-400$ million years after the Big Bang.

\hangindent=.3cm$\bullet$
There is a trend to higher obscured fraction of star formation with stellar mass and SFR already in place at $z \simeq 7$, with measurements of the cosmic SFR density showing that $> 10$ percent of the SFR density is obscured at this epoch.  

\hangindent=.3cm$\bullet$ 
The estimated dust masses compared to the stellar mass suggest that rapid dust enrichment occurs, likely from supernova with little dust destruction and/or rapid growth in the ISM, which is in agreement with maximal predictions from models.

\hangindent=.3cm$\bullet$ 
In individual sources, cold disks are already in existence, with their gas disks being more extended and smooth than their observed stellar counterparts. 

\end{abstract}

\maketitle

\tableofcontents

\section{INTRODUCTION}

\subsection{Background}
\label{sec:background}
The earliest stages of galaxy formation and evolution in the first billion years
of cosmic time was characterised by the formation of the first stars - generating the first heavy elements and dust grains - the first black holes and the first galaxies.
This in combination with the Universe-wide phase change called the Epoch of Reionization (EoR), when the inter-galactic medium (IGM) transitioned from being predominantly neutral to ionized (see reviews by~\citealp{Stark16, Dayal18, Robertson22}).

\begin{marginnote}
\entry{EoR}{epoch of reionisation}
\entry{IGM}{intergalactic medium}
\entry{$\Lambda$CDM}{dark energy ($\Lambda$) and cold dark matter}
\entry{DM}{dark matter}
\end{marginnote}

In $\Lambda$CDM cosmology, many differences between high-redshift galaxy formation and local galaxy evolution are directly linked to the evolution of the underlying dark matter haloes across cosmic time. 
First, the \textit{specific} accretion rate ($\dot{M}/M$) of dark matter (DM) and gas from the cosmic web onto haloes increases steadily towards higher redshift \citep{Somerville15}. 
As a result the (minor) merger rate, as well as the total gas fractions within galaxies are expected to increase with redshift. 
Furthermore, for a given depletion rate of gas into stars, the expected specific star-formation rate (sSFR=SFR/$M_\ast$) should roughly follow a similar evolution, such that galaxies at $z=7$ have $\sim1.5-2$ dex higher SFR compared to $z=0$ galaxies at fixed stellar mass  \citep{Stark16}. 
The increased accretion rates, as well as the higher SFR and associated stellar feedback per unit mass is expected to increase the turbulence in the ISM~\citep{F"orsterSchreiber20}.
Second, as haloes collapse at higher redshift against the denser background of the expanding Universe, it is predicted that the gas that settles into galaxies will be of higher density.
Assuming conservation of angular momentum of the gas in the halo as it cools onto a disk, galaxy sizes will scale with the virial radii of the halo, such that $z=7$ star-forming galaxies are expected to be $\sim5-8\times$ more compact than their $z=0$ counterparts at fixed stellar mass \citep{Dayal18, Crain23}, further contributing to higher gas and SFR surface density in the early Universe. 
The combined effect of higher gas fractions and higher turbulence in more compact, denser galaxies increases the ISM pressure and the conditions under which stars form. 
This in turn leads to more massive star clusters and more massive stars forming, which can have a galaxy wide impact through their stellar feedback processes.  

In addition to galaxy evolution effects linked to the cosmic web within the expanding Universe, galaxies in the first billion years are affected by the timescales needed to build up significant metal  and dust reservoirs after the first stars form. 
In particular, Asymptotic Giant Branch (AGB) stars are thought to produce a significant fraction of the dust in our local Universe, however, their timescales spent on the stellar main sequence ranges from a hundred million to billions of years, allowing for only a minor contribution at very early times \citep{Schneider24}. 
As a result there remain large gaps in our understanding of the rate of dust build-up and the dust grain composition - and resulting dust extinction law - in high-redshift galaxies. 
Finally, the increasing temperature of the CMB  heats all but the most self-shielded dust regions, impacting gas cooling during the star-formation process \citep{Bate25}. 

Given these dramatic changes from the physical processes at lower redshift, a great number of open questions on high-redshift galaxy formation remain: How early do the first galaxies build up their metal and dust reservoirs? 
How does the ISM (and circumgalactic medium; CGM) of these early systems allow for escape of ionising photons from massive stars into the IGM? 
Are the local scaling relations that are fundamental to star formation already in place in the first billion years of cosmic time? 
How is dust obscuration affecting our understanding of the physical properties of early galaxies? 
Can settled galaxy disks already form or are the majority of early systems dominated by turbulence?

\begin{marginnote}
\entry{ISM}{interstellar medium}
\entry{CMB}{cosmic microwave background}
\entry{ALMA}{Atacama Large Millimeter/submillimeter Array}
\end{marginnote}

The \emph{James Webb Space Telescope} (JWST) is providing new constraints on the stellar and hot gas properties of galaxies in the EoR. 
In particular, findings of bursty star-formation histories (SFHs) and an over-abundance of luminous galaxies as early as $z=14$ have been discovered at the time of writing \citep[e.g.][]{Carniani24}. 
However, in order to test the current theoretical framework out to the highest redshifts, it is simultaneously  critical to constrain the presence and properties of  dust and gas - fundamental building blocks of galaxy growth that submillimeter telescopes in the Atacama Large Millimeter/submillimeter Array (ALMA) era are uniquely capable of detecting. 

\begin{figure}

    \includegraphics[width=1.0\textwidth, trim = 1cm 2cm 10cm 5cm,clip]{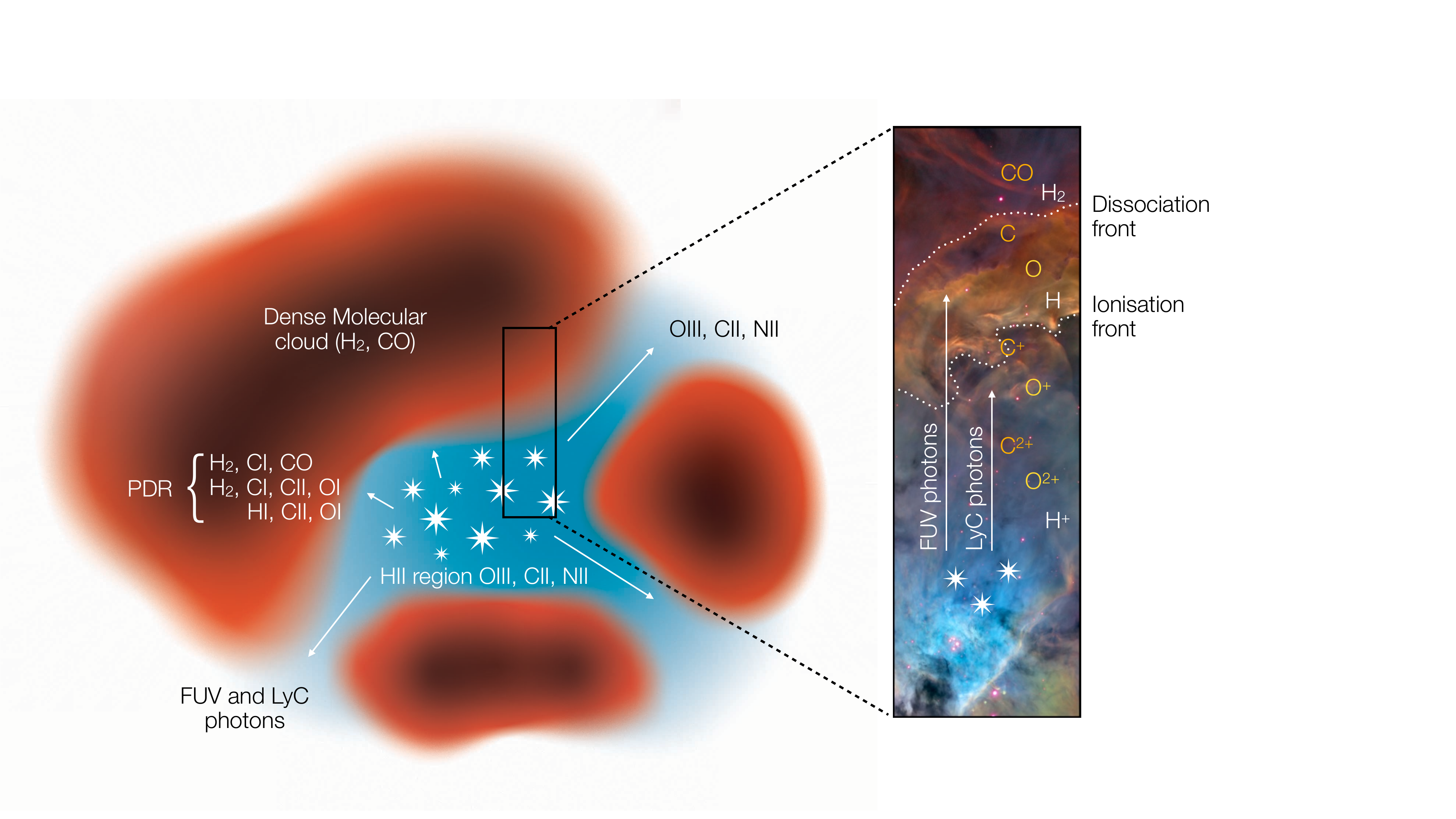}
    \caption{
   An illustration of the multi-phase gas and dust across a galaxy within the EoR.
    Young star clusters are shown as the white stars, which ionize HII regions around them.
    Channels of ionized gas where Lyman continuum photons escape are also shown.
    Dense molecular regions are shown in red, which are the sites of a significant proportion of the dust grains.
    The edges of the molecular regions are PDRs, where \cii\ is predominantly produced.
    The complex geometry of the stars, ionized and neutral gas, and dust, impacts the multi-wavelength properties of galaxies within the EoR.
    }
    \label{fig:cartoon}
\end{figure}

Despite accounting for only 1\% of the ISM mass, dust grains not only scatter and absorb starlight and re-radiate the absorbed
energy as mid-infrared (MIR) aromatic features and thermal FIR continuum, but within the life cycle of the ISM, dust grains facilitate numerous chemical reactions, including the formation of the H2 molecule, and are responsible for the heating of the gas in photo-dissociation regions, energetically coupling the stellar radiation field to the gas outside of ionised regions \citep{Galliano18}.
Cosmic gas is mostly comprised of Hydrogen in its neutral (HI) and molecular (H2) phases, the latter is the one most directly associated with star-formation, whereas the neutral phase is considered the long-term gas reservoir. 
The gas and dust properties of galaxies at $z \simeq 7$ are intricately linked with the cosmic reionization process, as the escape of ionizing photons from massive stars can be regulated by the properties and geometry of gas and dust surrounding star-forming regions \citep[][]{Dayal18,Robertson22}(see {\bf Figure~\ref{fig:cartoon}}).

\begin{marginnote}
\entry{NOEMA}{NOrthern Extended Millimeter Array}
\end{marginnote}

\subsection{Motivation}
\label{sec:motivation}
The past decade, since the commissioning of ALMA in 2014, has seen a revolution in the observations of cold gas and dust within galaxies at very high redshifts.
Here we aim to summarise the current state of the field by creating a compilation of the published ALMA and NOrthern Extended Millimeter Array (NOEMA) observations targeting the $z > 6.5$ population to-date\footnote{This redshift chosen to enable a focus on the EoR, but also to include the rich datasets that have targeted the interval $z = 6.5$--$7.0$ (e.g. Reionization Era Bright Emission Line Survey or REBELS).}, allowing us to summarise the current state-of-the-art and identify the key open questions to be answered in the next decade of ALMA.
Our philosophy for this review is to connect the historically relatively separate communities of high-redshift galaxy observations, based on the redshifted rest-frame Ultra-Violet (UV) light, and detections of dusty sources in the millimeter-submillimeter wavelengths.
The key to this is to introduce both the basics of ALMA observing and give an overview of the rest-frame UV selection techniques that that are required to select the target sample for follow-up in the sub-mm. 
The detailed properties of the target high-redshift galaxies and quasars from a mainly rest-frame UV/optical stand-point have been presented in~\citet{Stark16},~\citet{Ouchi20} and~\citet{Fan23}. 
And we refer the reader to~\citet{Robertson22} for a review of galaxies contributing to reionization (pre--\emph{JWST}).
We do not discuss in detail lower redshift dusty galaxies or the theoretical background of dust emission, recent reviews on these topics are~\citet{Hodge20, Salim20} and~\citet{Schneider24}.
The review of cold gas emission by~\citet{Carilli13} summarised the state-of-the-art FIR line emission measurements available prior to ALMA, which included a single source (a quasar host galaxy) at $z > 6.5$.
Due to observational limitations, there were no detections of line or continuum emission from more `normal' galaxies with existing facilities (mainly the IRAM Plateau de Bure interferometer; PdBI).
In the context of this review, we define `normal' galaxies as those with SFRs of $\lesssim 100\,{\rm M_{\odot}}{\rm yr}^{-1}$ and without active black holes. 
Far-infrared properties of Quasars are discussed in \citet{Fan23}. 
A complementary review on FIR fine-structure lines can be found in~\citet{Decarli25}.

In Section \ref{sec:methods} we discuss various analysis methods and concepts important for interpreting FIR line and continuum observations obtained through interferometry. In Section \ref{sec:sample} we present the compilation of $z>6.5$ galaxies used in this review, including their  selection methods and the largest high-redshift observational ALMA programs that have contributed to the sample. We present an overview of  emission lines accessible to ALMA at $z>6.5$ in Section \ref{sec:lines}, including their physical interpretation and the observational progress to date. We discuss dust continuum observations and dust build-up in Section \ref{sect:dust}, while we summarize the progress on resolved observations and kinematics in Section \ref{sec:resolved}. Finally, we discuss how ALMA has addressed the questions posed in the previous Section and what the future outlook is for ALMA observations in the next decade in Section \ref{sec:discussion}.

\begin{figure}
    \centering
    \includegraphics[width=0.95\textwidth]{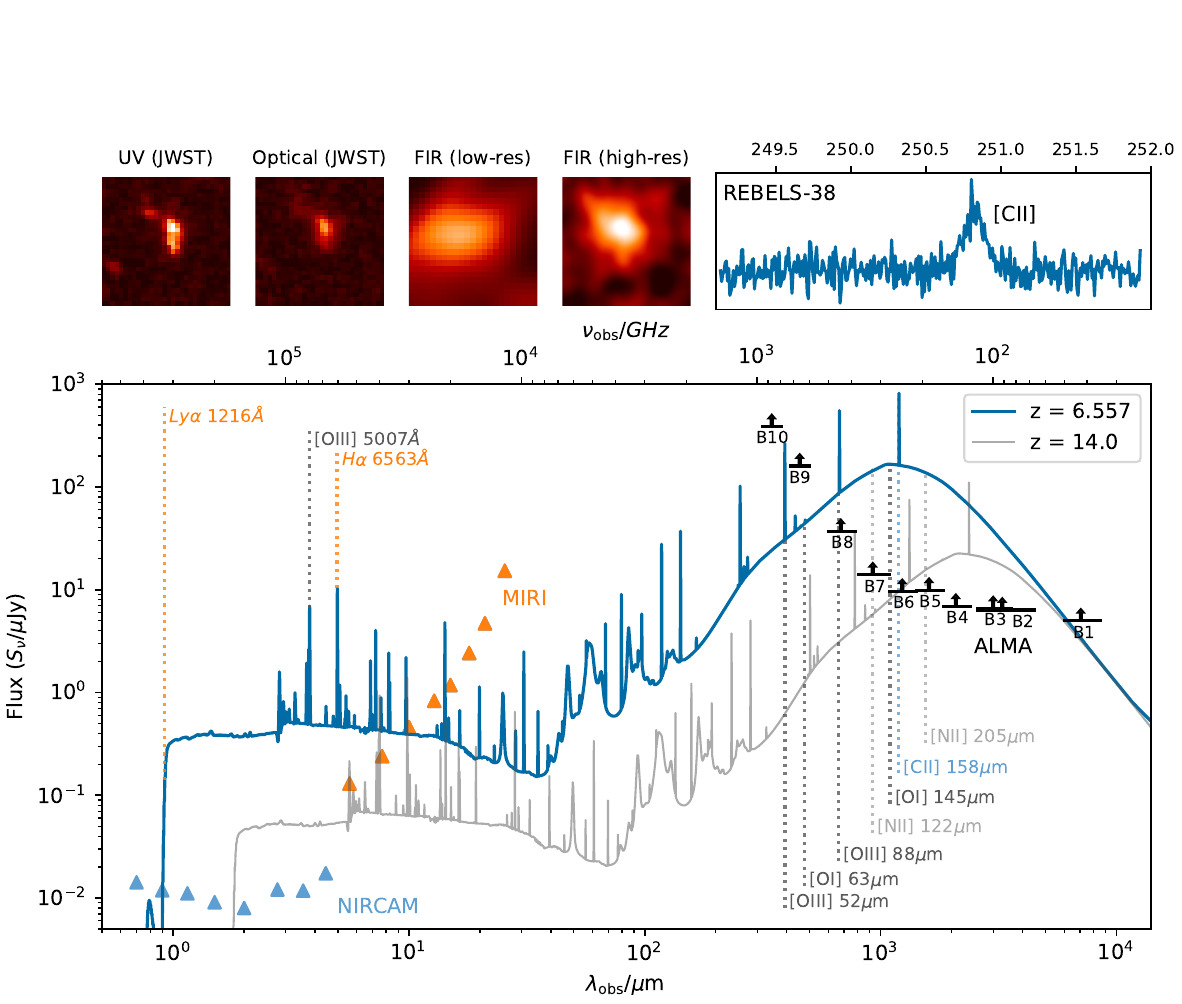}
    \caption{An illustration of the regions of a $z = 7$ galaxy SED observed by different facilities.
    The observed~\emph{JWST} and ALMA data from REBELS-38 are shown in the upper images, with the \cii~data for this galaxy on the right.
    In the main plot we show the best-fit SED model for REBELS-38.  SED model taken from~\citet{Fisher25}.
    The wavelength of the ALMA observing bands are shown as the black arrows labeled from Band 1 to 10, and common FIR emission lines are highlighted with the vertical lines. 
    The arrows show the sensitivity achieved by~\emph{JWST} (blue and orange) and ALMA (black labeled) in 10000 seconds of observation.}
    \label{fig:sed}
\end{figure}

\begin{table}
\tabcolsep7.5pt
\caption{Parameters for line transitions accessible to ALMA at $z>6.5$}
\begin{center}
\begin{tabular}{@{}l|c|c|c|c|c|c@{}}
\hline
 &&&& ionization   & excitation & $n_{\rm crit}$\\
Species  & transition & $\lambda$ {(}$\mu$m) & $\nu$ {(}GHz) & potential {(}eV) & potential {(}K) &  {(}cm$^{-3}$)$^{\rm a}$\\
\hline
\multicolumn{6}{c}{PDR/neutral gas}\\
\hline
\cii   & $^2P_{3/2}\rightarrow {^2P_{1/2}}$ & 157.74 & 1900.54  & 11.3-24.4 & 91 & 2.8$\times 10^3$ \\
\hline

\oi    & $^3P_1\rightarrow {^3P_2}$ & 63.18 & 4744.77 & 0-13.6 & 228 & 4.7$\times 10^5$ \\
      & $^3P_0\rightarrow {^3P_1}$ & 145.53 & 2060.07  &  & 329 & 9.4$\times 10^4$ \\
\hline
 [C{\sc i}]   &  $^3P_2\rightarrow {^3P_1}$ & 370.42 & 809.34 & 0-11.3 & 63 & 1.2$\times 10^3$  \\
      & $^3P_1\rightarrow {^3P_0}$ & 609.14 & 492.16 &  & 24 & 4.7$\times 10^2$ \\
\hline
CO  & $J = 7-6$  & 371.65 & 806.65 &   & 155 & 3.9$\times 10^5$  \\
    &  $J = 6-5$ & 433.56 & 691.47  &  & 116 & 2.6$\times 10^5$ \\
    &  $J = 5-4$ & 520.2 & 576.27  &  & 83 & 1.7$\times 10^5$ \\
    &  $J = 4-3$ & 650.3  & 461.04  &  & 55 & 8.7$\times 10^4$ \\
    &  $J = 3-2$ & 867.0 & 345.80  &  & 33 & 3.6$\times 10^4$ \\
\hline
\multicolumn{6}{c}{Ionised gas}\\
\hline
\cii   & $^2P_{3/2}\rightarrow {^2P_{1/2}}$ & 157.74 & 1900.54 & 11.3-24.4 & 91 & 50 \\
\hline
 \oiii & $^3P_2\rightarrow {^3P_1}$ & 51.82 & 5785.88 & 35.1-54.9 & 440 & 3.6$\times 10^3$  \\
       & $^3P_1\rightarrow {^3P_0}$ & 88.36 & 3393.01  &  & 163 & 510 \\
        \hline
        
 \nii & $^3P_2\rightarrow {^3P_1}$ & 121.90 & 2459.38 & 14.5-29.6 & 188 & 310  \\
 & $^3P_1\rightarrow {^3P_0}$ & 205.18 & 1461.13 &  & 70 & 48  \\
\hline
 [N{\sc iii}] & $^2P_{3/2}\rightarrow {^2P_{1/2}}$ & 57.32 & 5230.43 & 29.6-47.4 & 251 & 2.1$\times 10^3$  \\
\hline
\end{tabular}
\end{center}
\begin{tabnote}
$^{\rm a}$For PDR/neutral gas the collision partners are H{\sc i} and $H_2$ (assuming $T_{\rm gas}=100\,K$), while for ionised gas the collision partner is electrons.
\end{tabnote}
\label{tab:lines}
\end{table}

\section{OBSERVATIONAL METHODS}\label{sec:methods}
In this section we present a summary of the key observational methods used in the detection and study of FIR continuum and line emission in galaxies and quasars at high redshifts.
An overview of the multiwavelength continuum and line emission expected from a high-redshift star-forming galaxy is shown in \textbf{Figure~\ref{fig:sed}}.

\subsection{(Sub)Millimeter Line Observations}
\begin{figure}
    \centering
    \includegraphics[width=1.0\textwidth]{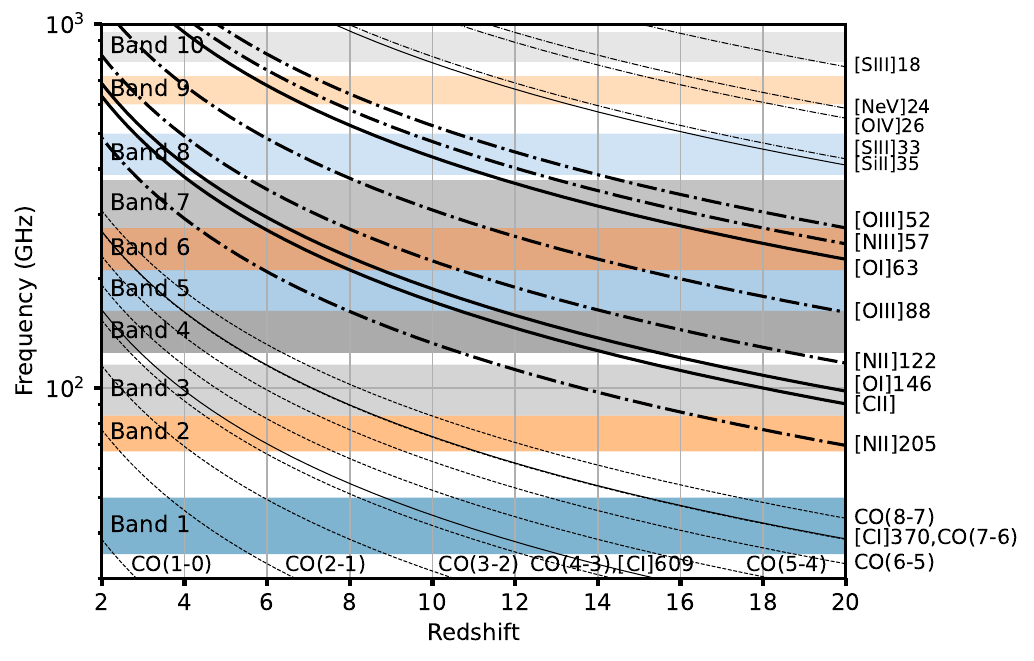}
    \caption{Overview of lines that are accessible with ALMA at $z=2-20$.}
    \label{fig:ALMAbands}
\end{figure}

Working with line luminosities at (sub)millimeter wavelengths comes with definitions and units that are not typically used in optical/near-infrared (NIR) studies. These differences have emerged due to historic reasons of the distinctly different observational techniques \citep{Carilli13,Solomon05}, but also differences between the phases of the ISM that are probed, with optical/NIR line studies focusing on the ionised gas (typically H{\sc ii} regions around young star clusters), while low-redshift submillimeter line studies have focused on molecular gas. 
At high-redshift, ALMA can now target a variety of phases in galaxies at $z>6$; from ionised gas (through \oiii, \nii), to neutral gas (\cii, \oi) and even molecular gas (\cii, C{\sc i}, CO). An overview of lines is given in \textbf{Table \ref{tab:lines}} and \textbf{Figure \ref{fig:ALMAbands}}. An elaborate discussion of the excitation mechanisms of the different lines, the use of critical density in the submillimeter and optical/near-infrared literature and the impact of the ionisation field on the different gas phases is given in the {\bf Supplemental Text} \citep[see also][]{Decarli25}.

The submillimeter line luminosity, $L_{\rm line}$ (in $L_\odot$), can be derived from the observed flux density as 

\begin{equation}
    L_{\rm line} = 1.04 \times 10^{-3} \, S_{\rm line} \Delta v \, D_L^2 \, \nu_{0} \, (1+z)^{-1} 
\end{equation}

Here, $S_{\rm line} \Delta v$ is the \textit{velocity} integrated line flux in units of $\rm Jy\, km\, s^{-1}$, $\nu_{0}$ is the rest-frame frequency ($\nu_{\rm obs} = \nu_{0} \, (1+z)^{-1}$) in GHz and $D_L$ is the luminosity distance in Mpc. 

The velocity integrated line flux, $S_{\rm line} \Delta v$, is almost exclusively used in submillimeter/radio studies and can be converted to cgs units of $\rm erg\, s^{-1}\, cm^{-2}$ by multiplying with $10^{-14}\cdot \nu_{\rm obs}/c$, where $\nu_{\rm obs}$ is the observed frequency in GHz and  $c$ is in $\rm km/s$. Note that as a result, the luminosity ratio of two lines, $L_{\rm line 1}$/$L_{\rm line 2}$, scales as ($S_{\rm line 1} \Delta v$)/($S_{\rm line 2} \Delta v$) $\cdot$ $\nu_{\rm line 1}/\nu_{\rm line 2}$, rather than the direct ratio of the submillimeter line fluxes.

\subsection{Dust Continuum Emission}\label{sect:contmod}

Through heating of the dust grains, the energy from young stars is re-radiated in the FIR and observed as a modified blackbody (MBB) form (see {\bf Supplemental Figure 1}).
Through assuming a functional form of the spectral-energy distribution (SED) in the FIR region, and integrating, the infrared luminosity (\lir) can be determined.
As the majority of $z > 6.5$ sources with ALMA observations have only one continuum band detection or non-detection, in practice a simplified form is used when estimating the \lir.
For a detailed overview of dust composition and different dust SEDs we refer the reader to~\citet{Draine03} and~\citet{Casey14}.
In the general form, we can model the flux observed at rest-frame frequency $\nu$ for a given dust temperature $T_{\rm d}$ as:

\begin{equation}
    S(\nu, T_{\rm d}) = \frac{(1-e^{-\tau(\nu)})\nu^3}{e^{h\nu/kT_{\rm d}}-1}.
\end{equation}

\begin{marginnote}
\entry{Infrared luminosity}{$L_{\rm IR}: 8$--$1000\,\mu{\rm m}$} 
\entry{Far-infrared luminosity}{$L_{\rm FIR}: 42.5$--$122.5\,\mu{\rm m}$}
\end{marginnote}

Here the optical depth is given by: $\tau(\nu) = (\nu/\nu_0)^{\beta_{\rm d}} = (\lambda_0/\lambda)^{\beta_{\rm d}}$, where \betad\ is the dust spectral emissivity index that depends on the dust composition and $\nu_{0}$ ($\lambda_{0}$) is the frequency (wavelength) where the optical depth equals unity~\citep{Draine03}.
Measured values of this wavelength at lower redshift show $\lambda_0 = 50$--$200$ microns~\citep{Simpson17}. 
It is typically assumed that the observed fluxes at high redshift with ALMA are at $\lambda> \lambda_0$ such that the dust can be assumed to be optically thin.
In the optically thin limit the function becomes:
\begin{equation}
    S(\nu, T_{\rm d}) = \frac{\nu^{(\beta_{\rm d} + 3)}}{e^{h\nu/kT_{\rm d}}-1}.
\end{equation}

An additional power law component has been advocated at lower redshifts, as well as multiple temperature components~\citep{Casey12}.
However, for observations at $z > 6.5$ where in the majority of cases a single band detection or upper limit is obtained, a simple optically thin MBB is typically assumed.

In the limit of a Planck function (with \betad $ = 0$), the dust temperature is related to the observed peak in the SED via Wien's displacement law with $\lambda_{\rm peak} \propto 1/T_{\rm d}$.
In the case of MBB forms however, the temperature defined by the observed peak in the SED ($T_{\rm peak}$) differs from the dust temperature input into the SED formalism ($T_{\rm SED}$) as shown in {\rm Supplemental Figure 1}.
In this work we quote $T_{\rm d}$ as the second of these parameters, the SED temperature, however caution must be taken in comparing temperatures derived in different formalisms.
In general $T_{\rm d} > T_{\rm peak}$ as discussed by~\citet{Casey12}. 

Due to the increasing temperature of the CMB at very high redshifts, the effect of a) an increased dust temperature due to heating by CMB photons and b) the reduced contrast of FIR observations when viewed against an increased CMB background, must be taken into account.
We refer the reader to~\citet{daCunha13} and~\citet{Ota14} for a detailed overview of these effects and how to correct for them.
In addition, some studies (particularly at $z = 3$--$5$) use empirically derived SEDs in the FIR, or fit low-redshift dusty galaxy templates to the observations (e.g.~\citealp{AlvarezMarquez19}).
The dust mass can be constrained from the Rayleigh-Jeans tail of the infrared SED, at high wavelength beyond the peak and where the emission is expected to be optically thin.
Here the observed flux depends on the mass of dust, the dust mass absorption coefficient and the luminosity distance ($d_{\rm L}$) as:
\begin{equation}
    S(\nu, T) = \frac{\kappa_{\nu}\,M_{\rm d} (B_{\nu}(T_{\rm d}) - B_{\nu}(T_{\rm CMB}))}{d_{\rm L}^2}
\end{equation}
The dust mass is very sensitive to the assumed dust temperature with $M_{\rm d} \propto L_{\rm IR}\, T_{\rm d}^{-(4+\beta_{\rm d})}$.
The dust mass absorption coefficient can vary in the range $5 < \kappa_{0} < 30 {\rm cm}^{2}{g}^{-1}$~\citep{Hirashita14}, with a frequency dependence given by $\kappa_{\nu} = \kappa_{0} (\nu/\nu_0)^{\beta_{\rm d}} $ where the reference frequency corresponds to $\lambda_0 = 158 \mu {\rm m}$.
We take $\kappa_{0} = 8.94 \,{\rm cm}^{2}{\rm g}^{-1} $ as appropriate for dust ejected by supernova (e.g. following~\citealt{Schouws22}).
There has been considerable discussion about the dust temperature and more recently the \betad\ emissivity parameter recent years, thanks to the combination of multiple bands with ALMA constraining the long wavelength decline and peak of the SED.
We discuss the observed parameters in Section~\ref{sect:dust}.

\begin{figure}
    \centering
    \includegraphics[width=\linewidth]{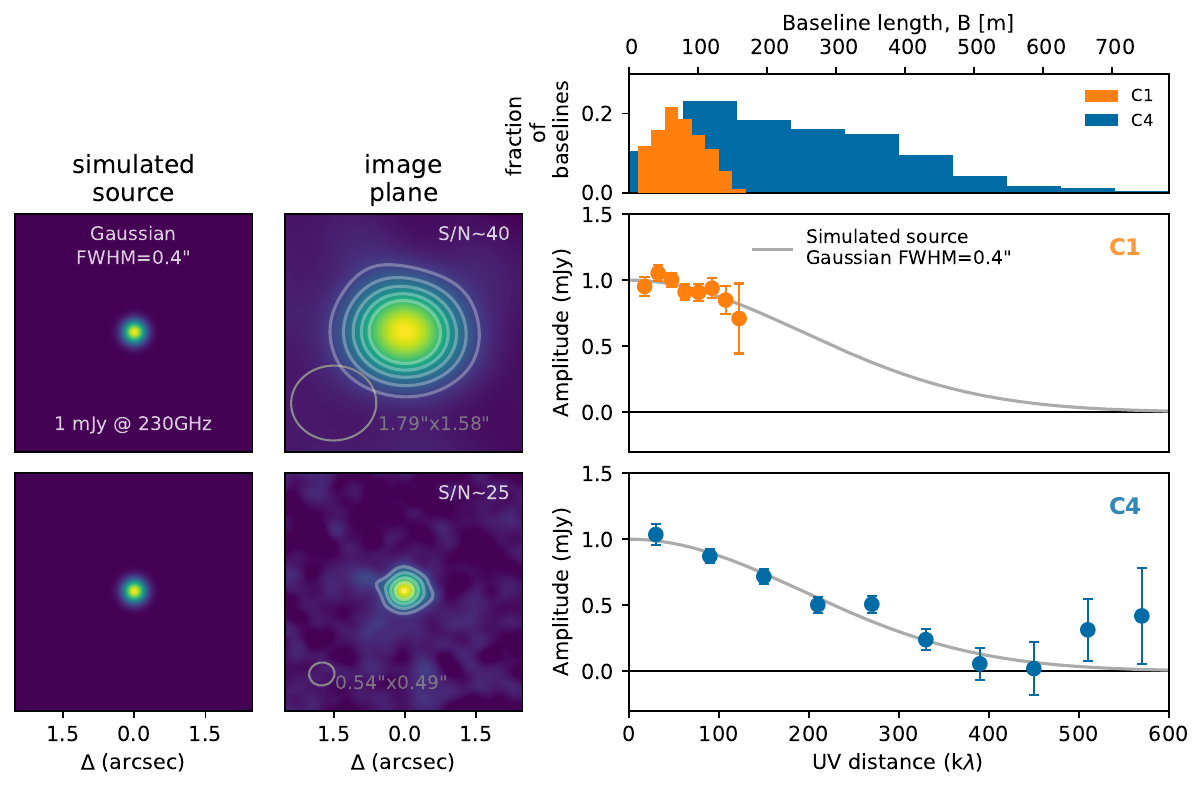}
    \caption{Simulated Gaussian source with $R_e\sim 1\rm kpc$ at $z\sim7$ (FWHM=0.4") at 230GHz (left columns). We use two configurations (C1 and C4; Synthesized Beams of $\sim\ang{;;1.7}$ and $\sim\ang{;;0.5}$) to simulate the observations in the image plane, at fixed rms/Beam (middle panels). The peak S/N reaches 40 for C1 and goes down by $\sim$40\% for C4. White contours are drawn at 5, 10, 15, 20 and 25$\sigma$ and the beam size is indicated in the bottom left corner of each panel. The changing S/N can most easily be understood by comparing the baseline distance distribution of each configuration with the size of the source in the {\sc uv} plane (right column). Grey lines show the intrinsic signal and coloured points the binned mock observations, while the top panel shows the fraction of baselines as a function of baseline length.
    }
    \label{fig:interferometry}
\end{figure}

\subsection{Interferometry}
In interferometric observations the signals received by different antennas are correlated and recorded in the so-called `{\sc uv}-plane' or Fourier space; with the complex visibilities consisting of an amplitude (providing flux information) and a phase (providing positional information). Each individual baseline between antennas has a resolution related to the baseline length, $B$: $\theta\sim\lambda/B$ (while larger structures are resolved out).

The ALMA large array consisting of 12 meter diameter antennas (the `12m array') typically carries out observations with 40-50 antennas, dividing the effective collecting area over $\sim$1000 baselines. 
ALMA is designed with different configurations that are able to perform unresolved detection experiments of high-redshift galaxies as well as resolved observations. 
In \textbf{Figure~\ref{fig:interferometry}} we simulate a high-redshift source in the image plane and {\sc uv}-plane to illustrate these different observations. 
The signal in the {\sc uv}-plane is not affected by a point spread function (PSF) (see right most panels of \textbf{Figure~\ref{fig:interferometry}}), however, with the Fourier transform to the image plane, the PSF or `Beam' emerges from the finite sampling of the {\sc uv} plane by the baseline distribution (see middle panels of \textbf{Figure~\ref{fig:interferometry}}). The detection experiment (top panels of \textbf{Figure~\ref{fig:interferometry}}) with $\theta_{\rm Beam}>2\times\theta_{\rm source}$ uses all baselines near the maximum amplitude of the source in the {\sc uv} plane, while the experiment with modest structural information (bottom panels of \textbf{Figure~\ref{fig:interferometry}}), with $\theta_{\rm Beam}\sim\theta_{\rm source}$, divides its baselines over the full source structure in the {\sc uv} plane, at the cost of signal-to-noise (S/N). The {\bf Supplemental Text} provides additional information on {\sc uv}-plane modeling, ALMA configurations, the imaging process and observing strategies for high-redshift sources with ALMA.

\section{SAMPLE OF ALMA-OBSERVED $z > 6.5$ SOURCES}\label{sec:sample}
In this section we present a compilation of the ALMA follow-up of $z > 6.5$ galaxies and quasars to-date.
These \numsources\ sources are then taken as the basis for the rest of the review, where we provide a consistent computation of the IR luminosity and obscured SFRs.

\begin{marginnote}
\entry{LBGs}{Lyman-break galaxies}
\entry{LAEs}{Lyman-$\alpha$ emitters}
\end{marginnote}

\subsection{Sample selection}\label{sect:selection}
The majority of known $z > 6.5$ galaxies and quasars have been selected based on their rest-frame UV emission, which probes the unobscured light from young stars.
For reionization-era sources the rest-UV is redshifted into the near-infrared (NIR) filters of telescopes like~\emph{Hubble Space Telescope (HST)} and now~\emph{JWST} in space, as well as NIR capable ground-based facilities like VISTA and UKIRT.
Once detected in the NIR, high-redshift galaxy and quasar candidates are separated from the more numerous interloper populations (consisting of mainly red $z \simeq 1$--$3$ galaxies and ultra-cool brown dwarfs) through the Lyman-break technique.
Here the key spectral features are the blue rest-frame UV slope dominated by young stellar populations (parameterised as $\beta_{\rm UV}$, with $F_{\lambda} \propto \lambda^{\beta_{\rm UV}}$), and the strong Lyman-$\alpha$ break at 1216\AA~resulting from absorption by the predominantly neutral intergalactic medium at these redshifts. 
Lyman-break galaxies (LBGs) that are typically selected from deep field programs show SFRs of the order of 1 to 100 \sfrunits, blue colours ($\beta_{\rm UV} \simeq -2$), and stellar masses of the order of \lmstar$=8$--$10$ (see review by~\citealp{Stark16}).
An additional class of sources is given by Lyman-$\alpha$ emitters (LAEs; see~\citealp{Ouchi20} for a review), of which large samples have been selected using narrow-band photometric filters.
Here a photometric excess in a narrow ($\lambda_{\rm obs} <100$ \AA) band in comparison to an encompassing broad-band signals a strong emission line, where a source with rest-frame equivalent width of $EW_{0} > 20\,$\AA\ is typically defined as an LAE.
LAES are shown to have lower dust attenuation and stellar masses, but higher sSFR, than LBGs at the same redshift and rest-frame UV luminosity (see review by~\citealp{Ouchi20}).
High-redshift~\emph{candidate} quasars, LBGs and LAEs once selected are typically followed-up with spectroscopic instruments (e.g. in the rest-UV or optical with~\emph{JWST} or ground-based facilities, or FIR with ALMA) to confirm the redshift through emission lines.
Small samples of sources have been selected based on their far-infrared emission, but these have exclusively been serendipitous sources found within the ALMA observations of LBGs or quasars (e.g.~\citealp{Fudamoto21, Wang24a}).

\subsection{Deep field programs}
ALMA has targeted deep extragalactic fields to provide mm/sub-mm constraints in combination with a wealth of auxiliary data from UV to radio telescopes.
The first deep fields were performed in the~\emph{Hubble} Ultra Deep Field as part of the pilot ALMA SPECtroscopic Survey in the Hubble UDF (ASPECS;~\citealp{Walter16}) and ``ALMA UDF''~\citep{Dunlop17} programs, followed in Cycle 4 by the full ASPECS survey~\citep{ Decarli20}.
These programs aimed to utilize the previous observations in this field to determine the line and continuum emission from galaxies over a broad range of redshifts.
These programs identified no individual $z > 6$ detections, likely due to a combination of the strong dependence of obscured SFR on stellar mass (e.g.~\citealp{Dunlop17, Bouwens16}), the flattening of the mm-counts to low luminosity~\citep{GonzalezLopez20} and the SFR-\lcii~dependence (see Section~\ref{sec:CII}).
The highest redshift object with a spectroscopic redshift detected in the dust continuum in ASPECS is $z = 3.711$~\citep{Bouwens20} and in the HUDF $z = 5.00$ in~\citet{Dunlop17}.
In addition to blank field mosaics, foreground clusters have been targeted that allow strong gravitational lensing to be exploited in the search for faint high-redshift galaxies.
The ALMA Lensing Cluster Survey (ALCS; e.g.~\citealp{Fujimoto23}) targeted 33 lensing clusters.
The highest redshift source found in the ALCS with an ALMA detection was the multiply imaged $z = 6.07$ galaxy~\citep{Fujimoto21}, Cosmic Grapes.
In addition, the Deep UNCOVER-ALMA Legacy High-Z (DUALZ;~\citealp{Fujimoto25}) targeted the Abell 2744 cluster with Cycle 9 data, detected (with the prior on the redshift) a source at $z = 6.32$.
Other programs such as MORA and HFF-ALMA are too shallow to provide strong constraints on the line or continuum properties of individual known $z > 6.5 $ galaxies in the field.
For these deep field programs it is only possible to provide stacked constraints on the dust continuum properties, and upper limits on the prevalence of line emission.
We show these constraints where relevant. 

\begin{figure}
\begin{subfigure}{.5\textwidth}
\includegraphics[height=5.5cm]{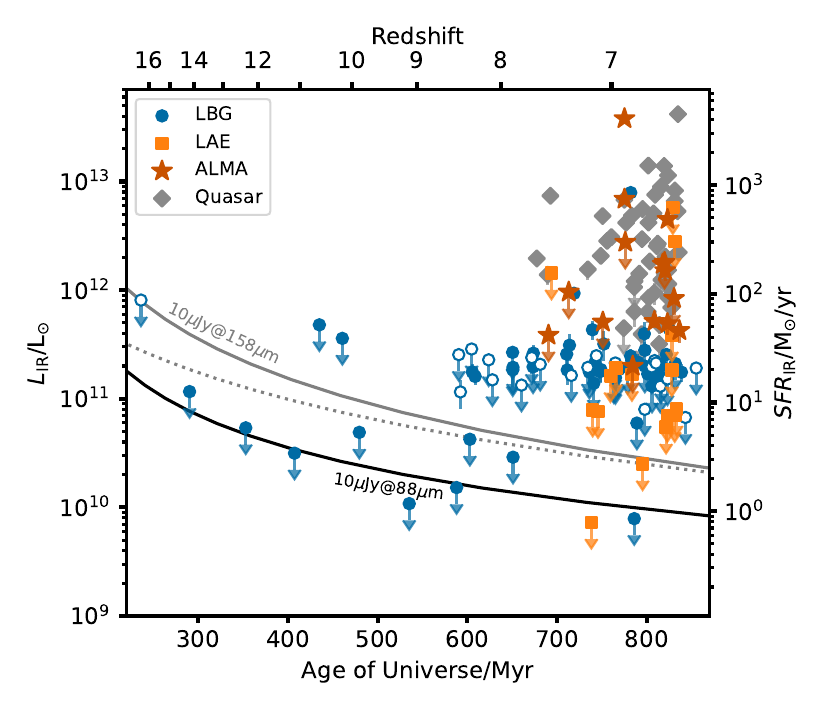}\\
  \includegraphics[height=5.5cm]{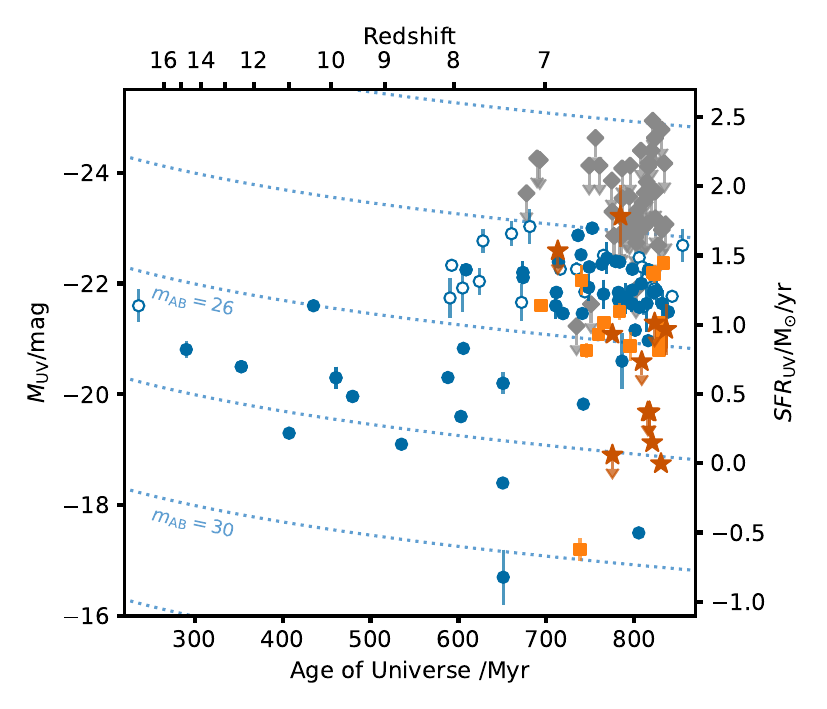}
\end{subfigure}%
\begin{subfigure}{.5\textwidth}
\includegraphics[height=5.5cm]{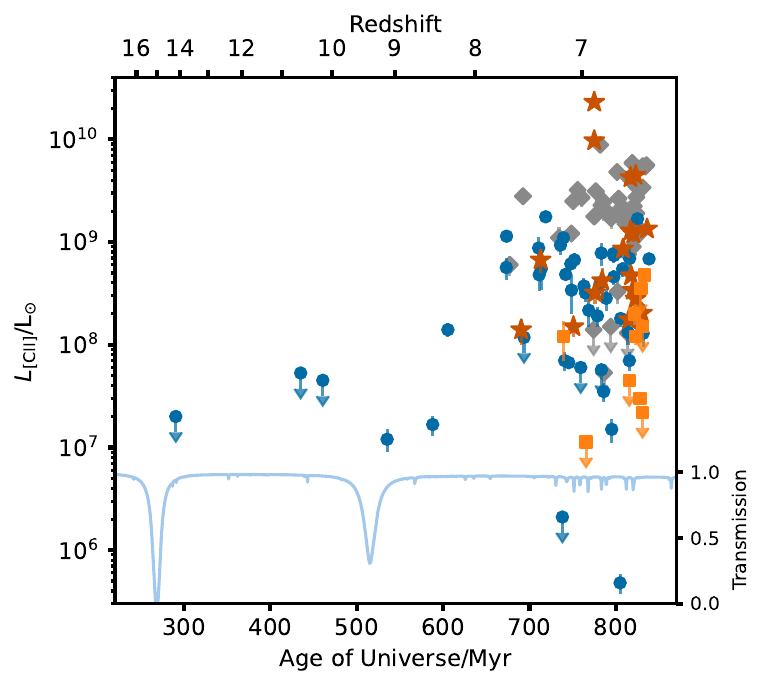}\\
  \includegraphics[height=5.5cm]{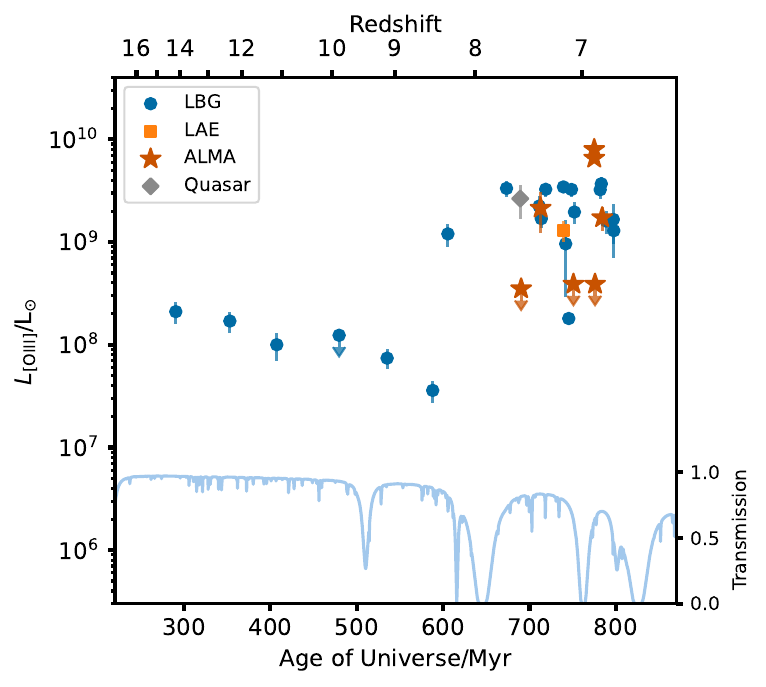}
\end{subfigure}%
\vspace{0.3cm}
\caption{An overview of the luminosities of the sample of $z > 6.5$ sources observed with ALMA and NOEMA as a function of Universe age.  From the upper left (clockwise), we show the derived infrared luminosity, the measured \lcii\ and \loiii\ respectively (with the atmosphere transmission on the right-hand axis) and the absolute UV magnitude.  For the \muv\ and \lir\ plots we also show a conversion to SFR on the right-hand axis.
For the quasars we estimate the galaxy \muv~as 10 percent of the observed quasar \muv, and show this as an upper limit.  In both plots the open symbols illustrate photometric redshifts, while filled are spectroscopic (either from FIR or UV/optical emission lines).  Upper limits are given as $3\sigma$.}
\label{fig:lir}
\end{figure}

\subsection{Targeted programs}
Numerous ALMA programs have targeted single or small ($< 10$) samples of high-redshift sources for follow-up.
The early observational programs tended to target first the \cii~line, with the dust continuum constraints coming essentially ``for free'' (e.g.~\citealp{Ota14, Schaerer15}).
The first programs were unsuccessful in finding \cii\ or dust continuum, likely as they were targeting LAEs which appear to have a lower dust content and potential \cii\ deficit.
Follow-up of luminous LBGs were more successful in detecting the dust continuum~\citep{Bowler18, Bowler22, Schouws22}, with several sources further detected in multiple bands~\citep{Knudsen17, Hashimoto19}.
These programs paved the way for several important ALMA large programs that have substantially increased the understanding of the FIR properties of $z > 6.5$ sources.
The ALMA-REBELS large program (Cycle 7;~\citealp{Bouwens22}) performed follow-up of a sample of 40 luminous $z > 6.5$ LBGs.
It complemented the deep field programs by targeting the rare, higher mass galaxy population.
Another significant ALMA large program was ASPIRE~\citep{Wang24}.
ASPIRE targeted 25 $ z \sim 7$ quasars to form a view of high-redshift galaxy properties, though measurements of the quasar host galaxy \cii\ and dust emission.
We also mention here ALPINE~\citep{LeFevre20}, a program that focused on $z = 4.5$--$6$ galaxies selected from the spectroscopically confirmed sample over the COSMOS field.
ALPINE provides a key reference for higher redshift samples, however it is important to note that the selection function is different to many $z > 6.5$ samples as a spectroscopic redshift was required prior to observing with ALMA.
Other relevant programs at $z < 6.5$ are SERENADE, which targeted extremely UV-bright galaxies found within wide fields~\citep{Mitsuhashi24}, and CRISTAL, which was a large program following up the $\sim 20$ most massive and star-forming \cii\ detected sources from ALPINE for high redshift \cii\ and dust continuum measurements~\citep{HerreraCamus25}.

\subsection{Compilation of a catalogue from the literature}\label{sect:uniform}
In this review we aim to compile the published observations of $z > 6.5$ galaxies and quasars taken by ALMA and NOEMA since these telescopes first started taking data just over a decade ago.
As introduced in Section~\ref{sect:selection}, the known samples of high-redshift galaxies fall into several distinct categories, namely sources studied through deep mosaics (e.g. ALMA-UDF, ASPECS) that covered established extragalactic fields in addition to targeted follow-up of particular galaxy populations (LAEs, LBGs, quasars, lensed LBGs etc.).
At the time of publication of this review, the number of sources in our compilation is \numsources.\footnote{For the most up-to-date compilation see https://github.com/raabowler/hzalmacat/tree/main}
We assume a $\Lambda$CDM cosmology with $\Omega_M= 0.3$, $\Omega_\Lambda = 0.7$ and $H_{0} = 70 {\rm km}{\rm s}^{-1} {\rm Mpc}^{-1}$ and a~\citet{Chabrier03} initial mass function.
We provide more detailed notes on the catalogue creation in the {\bf Supplemental Text}.

In this review we recompute the \lir~for the full sample of galaxies using an assumed FIR SED with a gently evolving \Td\ according to the~\citet{Sommovigo22} relation, normalised at $ z= 6.4$ to $T_{\rm d} = 41 {\rm K}$ (to match that derived in~\citealp{Mitsuhashi24}).
We fix the emissivity to $\beta_{\rm d} = 1.8$, following the results of~\citet{Witstok23}.
In Section~\ref{sect:dustt} we show a summary of the best-fit dust temperatures from a compilation of works at high redshift to justify this choice.
FIR line luminosities were taken directly from the relevant publication, as these are not strongly model dependent.
The other properties (e.g. rest-frame UV luminosity, stellar mass) are also included directly from the literature.
The \muv\ is uncorrected for dust attenuation.
The rest-frame UV slope \betauv\ is also taken from the individual studies, however we note that biases can arise from the measurement method (SED vs. colour derived for example;~\citealp{Faisst20}).

An overview of the sample is shown in \textbf{Figure~\ref{fig:lir}}, where we show the \lir, \muv, \lcii\ and \loiii\ as a function of redshift (see also {\bf Supplemental Table 1}).
In the last few years, the redshift record for the detection of \oiii~has been pushed to the frontier, with several sources now confirmed by ALMA at $z =11$--$14.179$~\citep{Zavala24, Schouws25}.
The majority of published observations of galaxies with measurements of the FIR continuum are upper limits.
The two highest redshift detections of the continuum in a spectroscopically confirmed galaxy are MACS0416\_Y1 $z = 8.31$ followed by REBELS-18 at $z = 7.675$~\citep{Inami22}, while there are several other higher redshift detections in galaxies with photometric redshifts (REBELS-04, REBELS-37).
J0313--1806 is the highest redshift quasar with a continuum detection~\citep{Wang24}.

\section{LINE EMISSION}\label{sec:lines}
A wide range of lines emerging from molecular, neutral and ionised gas are in principle accessible to ALMA at $z>6.5$ (see \textbf{Figure \ref{fig:ALMAbands}}). 
Some of the basic parameters of these molecular, atomic and ionic species and their line transitions are listed in \textbf{Table \ref{tab:lines}}. 
The primary challenge of detecting $z>6.5$ galaxies with ALMA is the observed faintness of these distant systems.

\subsection{Line scanning and the spectroscopic redshift frontier}\label{sec:scanning}
Due to the combination of their inherent brightness and visibility in the high transmission ALMA bands at $z>6.5$ (see {\bf Supplemental Text}), \ciilam\ and \oiiilam\ (particularly well visible above $z\gtrsim8.3$; \citealt{Bouwens22}) have been the most targeted emission lines in EoR galaxies. 
The earliest attempts at detecting these lines were sources with confident spectroscopic redshift from bright Lyman-$\alpha$ emission - the only bright spectral line accessible from the ground with optical and near-infrared detectors. 
While some of these observations were successful out to $z=7.21$ \citep{Inoue16,Knudsen16,Pentericci16,Bradavc17}, the number of Ly$\alpha$ emitters significantly declines beyond $z\sim6.5$ \citep{Ouchi20}, due to the scattering of Ly$\alpha$ photons when encountering neutral hydrogen in the IGM. 
Indeed, one of the hopes for the high-redshift community when ALMA first came online was that of being a ``redshift machine": scanning for bright lines by designing multiple ALMA set-ups with contiguous frequency coverage as a way of obtaining spectroscopic confirmation of galaxies. However, the initial faintness  of \cii\ in Ly$\alpha$ emitting galaxies \citep{Ouchi13,Maiolino15,Knudsen16} and a few unsuccessful line scans for \cii\ \citep{GonzalezLopez14, Schaerer15, Knudsen17} - 
 some with the significantly less sensitive PdBI (predecessor of NOEMA) - discouraged many further ALMA line-scanning experiments. 

\citet{Smit18} obtained the first ALMA spectroscopic confirmation of Lyman Break galaxies in the EoR without Ly$\alpha$ redshifts by targeting \cii\ in galaxies that had additional photometric redshift constraints from the Spitzer Space Telescope. Two more pilot studies with NOEMA and ALMA confirmed the utility of \cii\ as an alternative to Ly$\alpha$ \citep{Molyneux22, Schouws23}, with in particular UV-bright galaxies showing \cii-luminous line emission. A few studies have suggested that both UV luminosity and the absence of Ly$\alpha$ - which is more common in low-mass, metal-poor galaxies with low dust content - are important empirical indicators for \cii\ visibility \citep[e.g.][]{Matthee19,Harikane20}. Following on from these initial pilots, the first ALMA large program was approved to significantly increase the number of spectroscopically confirmed UV-bright galaxies in the EoR \citep[REBELS;][]{Bouwens22}. 

Beyond $z>8$, two high-profile spectroscopic confirmations targeting \oiii\ followed the initial \cii\ line-scanning results. The \oiii\ detection in a lensed galaxy at $z=9.1$ by \citet{Hashimoto18} held the record for the most distant spectroscopic line confirmation until the commissioning of JWST. Not long thereafter \citet{Tamura19} spectroscopically confirmed another brightly lensed galaxy at $z=8.3$ through \oiii\ line scanning. Until recently, the lack of bright (lensed) galaxy candidates beyond $z>8$ has limited the progress of line scanning at the redshift frontier \citep{VanLeeuwen25}.

With the ground-breaking data-sets of JWST, new opportunities for ALMA as a ``redshift machine" have emerged. While NIRSpec can observe rest-frame optical lines such as \oiii~$\lambda$5007{\AA} out to $z\sim9.4$, few bright lines are accessible to NIRSpec for galaxy candidates (selected with NIRCam) above $z>10$, with spectroscopic observations relying instead on the continuum break at the wavelength of Ly$\alpha$ for confirmation \citep[e.g.][]{CurtisLake23}. While some of the first attempts at line scanning on targets $z>10$ have been unsuccessful \citep{Harikane22,Bakx23,Fujimoto23,Kaasinen23,Popping23,Yoon23}, recent results suggest these observations did not have the required depth to detect a line \citep{Zavala24}. Indeed the first successful \oiii\ line scan above $z>10$  revealed a precision redshift of $z=14.179$ for a galaxy previously estimated to be at $z\sim14.3$ based on NIRSpec continuum break information \citep{Carniani24,Schouws25}. This major ALMA success suggests the future redshift frontier will require the synergy of JWST and ALMA spectroscopy, such that NIRSpec can provide a tight prior on the redshift from continuum observations, while ALMA provides precision redshift information from deep integrations over a small redshift range (requiring only a few frequency set-ups for a spectral line scan). 

\begin{marginnote}
    \entry{PDRs}{photodissociation regions}
\end{marginnote}

\begin{figure}
    \centering
    \includegraphics[width=0.8\textwidth]{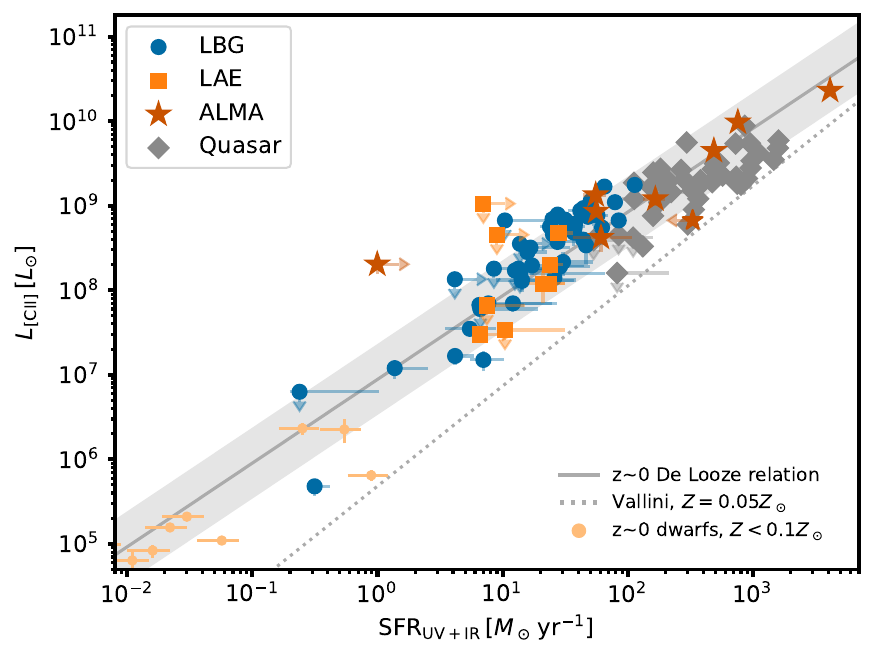}
    \caption{The relation between SFR and $L_{\rm [CII]}$ for the $z>6.5$ literature compilation, compared to the local relation and scatter measured by \citet{DeLooze14} (grey line and shaded region). Very metal-poor galaxies ($Z<0.1Z_\odot$) at $z\sim0$ are shown in yellow, while the simulated \citet{Vallini15} relation for $Z=0.05Z_\odot$ galaxies is shown as a dotted line. Upper errorbars on the SFR include 1$\sigma$ limits on dust continuum non-detections.
    }
    \label{fig:LCII_SFR}
\end{figure}

\subsection{[CII]$\lambda 158\mu$m Emission}
\label{sec:CII}
Carbon has a first ionization potential (11.3 eV) below that of Hydrogen and therefore FUV photons (6-13.6 eV) that permeate the galaxy ionize Carbon in all but the most dense, self-shielded clouds. As a result \ciilam\ traces ionized, neutral and even molecular gas, including warm, dense photodissociation regions (PDRs), the cold neutral medium and diffuse ionised medium \citep[e.g.][]{Pineda13}. As a multi-medium tracer the usefulness of \cii\ as a diagnostic tool for ISM conditions has been debated \citep[e.g.][]{Casey14}, however, its use as a kinematic tracer of the ISM has been widely recognized (see Section~\ref{sect:kinematics}).

Dedicated observing programs  have found evidence for neutral/molecular gas as the dominant emission for \ciilam\ in $z>6.5$ galaxies (see Section \ref{sec:NIIlines}). This is consistent with emission line analyses from JWST which find high ionization parameters (high \oiii/[O{\sc ii}]) are common in $z>6.5$ galaxies, suggesting the ionization structure within H{\sc ii} regions is weighted towards highly ionized species and carbon could be doubly or even triply ionized within these regions \citep[e.g.][]{Cameron23}.

These findings potentially justify the use of \cii\ as a tracer of the gas mass in distant galaxies. As traditional gas mass tracers such as CO are too faint to detect for all but the most extreme starburst galaxies at $z>6.5$, \cii\ is the only cool/cold gas tracer that is observationally accessible. We will review the empirical relations and proposed calibrations for \cii\ with respect to the SFR and to molecular and neutral gas mass.

\subsubsection{$L_{\rm [CII]}$-SFR relation}
In the local Universe, the relation between $L_{\rm [CII]}$ and SFR has been investigated in galaxies with a wide range of metallicities, star-formation rates, Active Galactic Nucleus (AGN) contribution and ISM conditions. In particular, \citet{DeLooze14} derive the following relation, based on a large compilation of $z\sim0$ dwarf galaxies \citep[in particular from the Dwarf Galaxy Survey, DGS;][]{Madden13}, AGN, and FIR-bright galaxies out to $z>0.5$\footnote{Including starbursts (SBs; massive galaxies with star-formation-powered FIR radiation), Ultra-Luminous Infrared Galaxies (ULIRGs; $L_{\rm IR}>10^{12}L_\odot$) and Dusty Star-forming Galaxies (DSFGs; rest-frame MIR or FIR selected high-redshift galaxies)}: 
\begin{equation}
      \log{{\rm SFR}/M_\odot\, \rm yr^{-1}} = (-6.99 \pm 0.14) + (1.01 \pm 0.02) \times \log{L_{\rm \cii} / L_\odot},   
\end{equation}
with 0.42 dex scatter. Either the origin for this relation could be the importance of FUV radiation from young stars in heating PDRs (see {\bf Supplemental Text}), whose cooling is dominated by \cii\ emission, or else \ciilam\ could be an important tracer of molecular gas (see Section~\ref{sec:LCII-H2}) and the $L_{\rm [CII]}$-SFR relation is a manifestation of the integrated Kennicutt-Schmidt relation \citep{Kennicutt12,DeLooze14}.
In \textbf{Figure \ref{fig:LCII_SFR}} we show the $z>6.5$ compilation with respect to this relation and find good agreement within the 1$\sigma$ scatter over $\sim$3 dex in SFR, consistent with findings from ALPINE at $z\sim4-6$ \citep{Schaerer20}, possibly suggesting the Kennicutt-Schmidt relation could hold out to the EoR. 

At low star-formation rates galaxies might have a mild bias towards low $L_{\rm [CII]}/\rm SFR$ ratios. This is consistent  with local metal-poor dwarf galaxies from the DGS, where lower SFR galaxies are typically lower metallicity ($\rm 12+[O/H]<8.0$, $Z<0.2 Z_\odot$) and have lower obscuration fractions ($\rm SFR_{\rm UV}>SFR_{\rm IR}$). This is furthermore consistent with simulations by \citet{Vallini15} that predict a systematic offset of the $L_{\rm [CII]}-\rm SFR$ relation for metal-poor galaxies (see \textbf{Figure \ref{fig:LCII_SFR}} for $Z=0.05Z_\odot$).
Other empirical indicators of low $L_{\rm [CII]}/\rm SFR$ in the DGS are a high ionisation parameter or burstiness (\citealp{Algera25}; see also Section \ref{sec:OIII/CII}), a high dust temperature \citep[see also][]{Bakx20}, a large volume filling factor of ionized gas and hot, dense neutral/PDR gas \citep[][]{DeLooze14}. 

At low- and intermediate-redshifts ULIRGs and DSFGs also have a well-known \cii\ deficit \citep{Casey14,Hodge20, Decarli25}. 
However, this effect is not clearly present in the compilation of $z>6.5$ galaxies at the high-SFR end of the $L_{\rm [CII]}$-SFR relation.

\begin{table}[h]
\tabcolsep7.5pt
\caption{\cii-to-$H_2$ conversion factor}
\label{tab1}
\begin{center}
\begin{tabular}{@{}l|c|c|c|c@{}}
\hline
$\alpha_{\rm [CII]}^{\rm a}$ &&&& \\
{(}$M_\odot/L_\odot$) & Reference(s)$^{\rm b}$ & Redshift & Target selection$^{\rm c}$ & Notes\\
\hline
31 & (1) & $z\sim2$ & MS galaxies & $M_{\rm H_2}$ from $M_{\rm dust}$\\
 & (2) & $z\sim4-6$ & LBGs & $M_{\rm H_2}$ from $M_{\rm dyn}$\\
 & (3),(4) & $z\sim6-8$ & LBGs & $M_{\rm H_2}$ from $M_{\rm dyn}$\\
34.9 & (5) & $z\sim0^{\rm b}$ & MS gal., ETGs, LIRGs,  & $M_{\rm H_2}$ from CO ($\alpha_{\rm CO}=4.3$) \\
 & & & ULIRGs, LBAs, DSFGs &  \\
 81 & (6) & $z\sim0$ & metal-poor dwarfs & modeling CO-dark gas\\
  48 & (7) & $z\sim0$ & metal-poor dwarfs & modeling CO-dark gas\\
10 &  (8) & $z\sim2-4$ & ULIRGs, QSOs & $M_{\rm H_2}$ from CO ($\alpha_{\rm CO}=1$)\\
 & (9)  & $z\sim4.5$ & SMGs & $M_{\rm H_2}$ from $M_{\rm dust}$\\
7 & (10) & $z\sim4-5$ & DSFG & $M_{\rm H_2}$ from $M_{\rm dyn}$\\
5 & (11) & $z\sim6.5-7$ & QSOs & $M_{\rm H_2}$ from CO ($\alpha_{\rm CO}=0.8$)\\
\hline
\end{tabular}
\end{center}
\begin{tabnote}
$^{\rm a}$ $M_{\rm mol}/M_\odot = \alpha_{\rm [CII]} \times L_{\rm [CII]}/L_\odot $. \\
$^{\rm b}$ The $z\sim0$ sample is complemented by a small sample of high redshift DSFGs out to $z\sim6$. \\
References: (1) \citet{Zanella18}, (2) \citet{DessaugesZavadsky20}, (3) \citet{Aravena24}, (4) \citet{Rowland24}, (5) \citet{Zhao24}, (6) \citet{Madden20}, (7) \citet{Ramambason24}, (8) \citet{Swinbank12}, (9) \citet{Gullberg18}, (10) \citet{Rizzo21}, (11) \citet{Kaasinen24}\\
Abbreviations:  MS, main sequence; LBA, Lyman-break analog; LIRG, luminous infrared galaxy ($L_{\rm IR}>10^{11}L_\odot$); ETG, early-type galaxy
\end{tabnote}
\end{table}

\subsubsection{$L_{\rm [CII]}$-$M_{\rm H_2}$ relation}
\label{sec:LCII-H2}
Molecular gas is the fuel for star formation, and given the faintness of low-$J$ CO lines \cii\ is the only feasible tracer of cold gas in larger samples at high redshift. In recent years, an increasing body of literature aims to understand the correlation between $L_{\rm [CII]}$ and $M_{\rm H_2}$. 
\citet{Zanella18} provided the most commonly used calibration of \ciilam\ as molecular gas tracer from a sample of main-sequence galaxies at $z\sim2$ and an additional literature sample from $z\sim0$ to $z\sim5$. They define the \cii-to-$H_2$ conversion factor as
\begin{equation}
    \alpha_{\rm [CII]} (M_\odot/L_\odot)=  M_{\rm H_2} / L_{\rm [CII]},
\end{equation}
an analogy to the widely used CO-to-$H_2$ conversion factor, $\alpha_{\rm CO}$. Good agreement with this conversion was found by \citet{DessaugesZavadsky20} and \citet{Aravena24} using low angular resolution dynamical constraints
from LBGs in ALPINE ($z\sim4-6$) and REBELS ($z=6-8$) respectively. \citet{Rowland24} further confirm the latter with high-resolution dynamical measurement in REBELS-25.

Furthermore, various studies have found strong correlations between \ciilam\ and CO(1-0) at low to intermediate redshifts \citep[e.g.][]{Gullberg15,Zhao24}, supporting the use of \ciilam\ as a molecular gas tracer, however, the uncertainty in the $\alpha_{\rm CO}$ factor remains one of the main systemic uncertainties behind the derived variation in $\alpha_{\rm [CII]}$. \citet{Zhao24} derive $\alpha_{\rm [CII]} =  8.12 \times \alpha_{\rm CO}$ and $\alpha_{\rm [CII]}$ = 34.9 $M_\odot$/$L_\odot$, assuming $\alpha_{\rm CO}=4.3\, M_\odot (\rm K\, km\, s^{-1}\, pc^2)^{-1}$, in good agreement with \citet{Zanella18}. 
Studies of ULIRGs and SMGs have typically found much lower values for $\alpha_{\rm [CII]}$; e.g. \citet{Swinbank12} and \citet{Kaasinen24} find 10 and 5 $M_\odot$/$L_\odot$, assuming an $\alpha_{\rm CO}$ of 1.0 and $0.8\, M_\odot (\rm K\, km\, s^{-1}\, pc^2)^{-1}$ respectively. These lower  values for $L_{\rm IR}$-bright systems are confirmed with dynamical studies by \citet{Gullberg18} and \citet{Rizzo21}, suggesting systemic changes in $\alpha_{\rm [CII]}$ with stellar mass, dust fraction and/or metallicity.
A compilation of conversion factors from the literature is presented in \textbf{Table~\ref{tab1}}.

\begin{marginnote}
    \entry{SMGs}{submillimeter galaxies}
    \entry{QSOs}{quasi-stellar objects}
\end{marginnote}

Furthermore, in local metal-poor galaxies, CO(1-0) has been suggested to be a poor tracer of the molecular reservoir due to a great fraction of `CO-dark' molecular gas, where reduced dust shielding at lower metallicity dissociates CO in all but the densest molecular cores. \citet{Madden20} show that CO-dark molecular regions are well traced by \ciilam\ and derive
$M_{\rm H_2} = 10^{2.12} \times [L_{\rm [CII]}]^{0.97}$, corresponding to  $\alpha_{\rm [CII]}=76-87\,M_\odot/L_\odot$. \citet{Ramambason24} revisit the CO-dark gas modeling using a more complex geometry of the PDR and find $\alpha_{\rm [CII]}=43-52\,M_\odot/L_\odot$.

In summary, a new picture is emerging where \ciilam\ is a good tracer of gas mass, although with modest variations 
for different galaxy samples. 
\citet{Aravena24} study the molecular gas in the REBELS sample ($z=6-8$) with the \citet{Zanella18} conversion and find 
the median molecular gas-to-stellar mass ratios and molecular gas-depletion timescales are roughly constant with redshift over the range $z=2-7$, while the cosmic density of molecular gas at $z\sim7$ continues a steep decline from $z\sim2$.

\subsubsection{$L_{\rm [CII]}$-$M_{\rm HI}$ relation}
In the optically thin limit and assuming only collisional excitation the \ciilam\ luminosity and the gas mass can theoretically be directly related for a given gas density and temperature \citep[e.g.][]{HerreraCamus21,Veilleux20}:
\begin{equation}
    \frac{M_{\rm H}}{M_\odot} = f_{\rm [CII],neutral} \cdot 0.77 \left(\frac{1.4\times 10^{-4}}{X(C^+)}\right) \left[ \frac{1+2 e^{-91.25/T} +n_{\rm crit}/n}{2 e^{-91.25/T}} \right] \frac{L_{\rm [CII]}}{L_\odot},
\end{equation}
where $X(C^+)$ is the abundance of singly ionized carbon and $C^+/H = 1.4\times 10^{-4}$ obtained from absorption spectroscopy in the Milky Way \citep{Savage96} is assumed. A common assumption is the maximal excitation case, where $n>>n_{\rm crit}$ and $T>>91 K$ such that $M_{\rm H}/M_\odot \sim f_{\rm [CII],neutral} \cdot L_{\rm [CII]}/L_\odot$ becomes a fairly conservative lower limit for the neutral gas mass traced by \ciilam. For a wider range in ISM conditions, \citet{HerreraCamus21} tabulate values of $\kappa_{\rm HI}$, where $M_{\rm H}/M_\odot = \kappa_{\rm [CII]} \cdot L_{\rm [CII]}/L_\odot$.

\citet{Heintz21} use damped Lyman-$\alpha$ (DLA) systems
at $z\gtrsim 2$ to measure both the Hydrogen and [C{\sc ii}$^*$]~1335.7{\AA} column density and (assuming $M_{\rm HI}/L_{\rm [CII]} = N_{\rm HI}/N_{\rm [CII^*]} \times 1.165 \times 10^{-4} M_\odot/L_\odot$) derive a metallicity dependent relation between \ciilam\ luminosity and the neutral gas mass  
\begin{equation}
\log{M_{\rm HI}/M_\odot} = 
\left(-0.87\pm 0.09\right) \times \log{Z/Z_\odot}
+ \left(1.48\pm 0.12\right) + \log{L_{\rm [CII]}/L_\odot }.
\end{equation}
At solar metallicity, this corresponds to $\kappa_{\rm HI}=4.1$, somewhat above the maximal excitation assumption, but broadly consistent with values reported by \citet{HerreraCamus21} for moderate gas density and temperatures ($T\sim10^2-10^3 K$, $n\sim10^3 \rm cm^{-3}$). 

Using this calibration \citet{Heintz22} study the gas fractions of REBELS galaxies at $z\sim7$ and find high gas fractions; $M_{\rm HI}/M_\ast\sim 10$ ($f_{\rm gas}\sim90\%$). They furthermore argue that at these redshifts only $\sim10\%$ of the cosmic {H\sc i} gas content is confined in galaxies and associated with the star-forming ISM.

\subsection{[OIII]$\lambda 88\mu$m Emission}
Two times ionised Oxygen requires a radiation field above 35.1 eV, typically generated by the hard stellar spectra of metal poor stars in gas with high ionisation parameter (high radiation intensity for a given particle density). Even before the launch of JWST rest-frame optical \oiii\ lines were inferred to be quite luminous in a significant fraction of the LBG population \citep[e.g.][]{Smit14,Smit15} and \oiiilam\ observations with ALMA have been successful in nearly all targeted LBGs and LAEs where a spectroscopic redshift was known and the line falls in a good transmission window of the atmosphere.   

\subsubsection{$L_{\rm [OIII]}$-SFR relation}
\begin{figure}
\begin{subfigure}{.5\textwidth}
    \includegraphics[width = 1.0\textwidth]{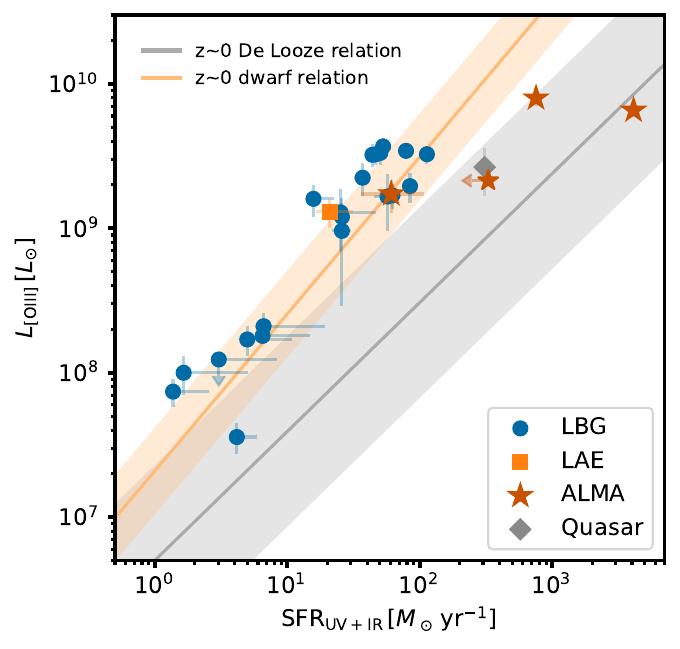}
    \end{subfigure}%
    \begin{subfigure}{.5\textwidth}
    \includegraphics[width = 1.0\textwidth]{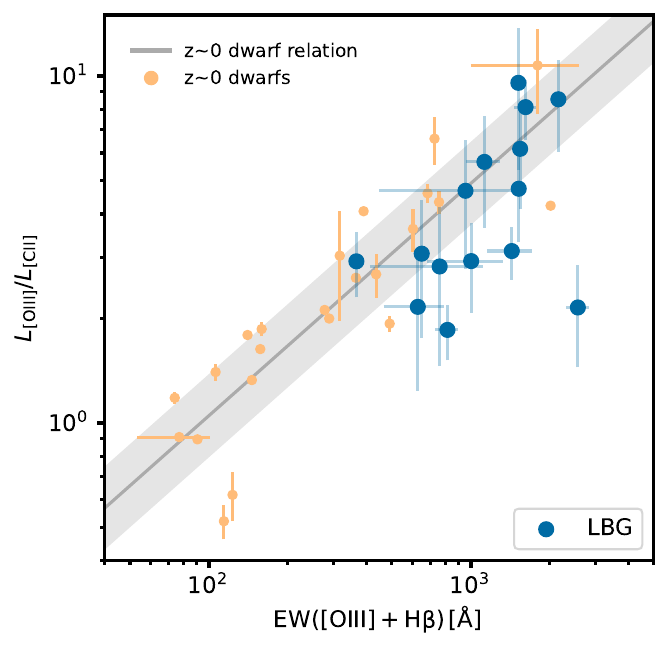}
    \end{subfigure}
    \caption{left: the literature compilation of $z>6.5$ sources with respect to the $z\sim0$ scaling relations and scatter measured by \citet{DeLooze14} for metal-poor dwarfs (orange line and shaded region) and the full local galaxy sample (grey line and shaded region). right: the FIR $L_{\rm \oiii}/L_{\rm \cii}$ ratio as function of the optical \oiii+H$\beta$ equivalent width, a common star-formation burstiness indicator, for $z\sim0$ dwarf galaxies (yellow points), including the scaling relation and scatter measured by Algera et al. (2025; grey line and shaded region) and the $z>6.5$ compilation. 
    }
    \label{fig:OIII-SFR}
\end{figure}

In the local Universe, there are clear differences in the $L_{\rm \oiii}$-SFR relation for different galaxy selections, with metal-poor dwarf galaxies following the relation 
\begin{equation}
      \log{{\rm SFR}/M_\odot\, \rm yr^{-1}} = (-6.71 \pm 0.33) + (0.92 \pm 0.05) \times \log{L_{\rm \oiii,dwarfs} / L_\odot}.  
\end{equation}
with 0.30 dex scatter, whereas a broader sample of dwarf galaxies, AGN, FIR-bright galaxies follow this relation 
\begin{equation}
      \log{{\rm SFR}/M_\odot\, \rm yr^{-1}} = (-6.71 \pm 0.33) + (0.92 \pm 0.05) \times \log{L_{\rm \oiii,all} / L_\odot}.  
\end{equation}
with a broad 0.66 dex scatter. 
\textbf{Figure \ref{fig:OIII-SFR}} shows a good agreement of $z>6.5$ LBGs and LAEs with the local dwarf galaxy relation, as well as serendipitous ALMA selected galaxies next to LBGs \citep{Fudamoto21}, while IR-bright sources and Quasars are broadly consistent with the general relation. 

\subsubsection{[OIII]$\lambda 88\mu$m as a metallicity indicator}
\label{sec:OIII-SFR}
In both metal-poor dwarf galaxies at $z\sim0$ and high-redshift LAEs and LBGs the high \oiiilam/SFR ratios have proven challenging to model with photoionisation codes (e.g. Cloudy). 
For a simple dust screen configuration \citet{Harikane20}, \citet{Witstok22} and \citet{Killi23} find that solar metallicity, in combination with relatively low gas density ($\log n/\rm cm^{-3} = 0.5-2.0$) in the star-forming regions, is preferred. 
Although these high metallicities were already difficult to reconcile with the early epoch in which these sources are found (i.e., in the pre-JWST era), JWST Oxygen abundance measurements are showing clear inconsistencies \citep[e.g.][]{Scholtz25, Harikane25a, Usui25}.

In $z\sim0$ dwarf galaxies, which have both high \oiiilam/SFR and low-oxygen abundance ($z\lesssim0.2Z_\odot$), \citet{Cormier15} propose a high filling factor of diffuse highly ionised gas (responsible for \oiiilam/SFR) due to a low covering factor of the PDR, such that the star-forming regions start leaking ionising photons that escape out to great distances. 
This is similar to the 2-phase ionised gas models proposed by \citet{Harikane25a, Usui25} to explain the observed \oiiilam/\oiii~5007{\AA} and \oiii~4363{\AA}/\oiii~5007{\AA} line ratios and discrepancies between optical and infrared derived electron densities. Such a model would also suggest that Oxygen abundance measurements from JWST alone could be significantly underestimated \citep[up to 0.8 dex;][]{Harikane25a}.

Despite the challenges in interpreting \oiiilam, it might be possible to use \oiiilam/SFR as a rough metallicity tracer, using the empirical relation by \citet{Jones20}:
\begin{equation}
12+\log({\rm O^{++}/H^+})=7.735+\log{\frac{L_{\rm \oiii}}{10^7\,L_\odot}} - \log{\frac{\rm SFR}{M_\odot\,\rm yr^{-1}}}
\end{equation}
using a SFR derived from the Balmer lines and $T_e=1.5\times10^4\,\rm K$ and $n_e=250 \,\rm cm^{-3}$. Assuming $12+\log({\rm O/H})\sim0.17 + 12+\log({\rm O^{++}/H^+})$ (i.e. an ionisation correction factor of 0.17 dex) \citet{Jones20} find good agreement for local metal-poor dwarfs  between \oiiilam/SFR and optical direct $T_e$ oxygen abundance measurements (0.26-dex scatter). 

Using \oiiilam/SFR as a metallicity indicator, ALMA results have shown that galaxies above $z>11$ are already enriched up to $\sim10-30\%$ of the solar oxygen abundance \citep{Zavala24, Schouws25, Witstok26}

\subsubsection{[OIII]$\lambda 88\mu$m/[CII]$\lambda 158\mu$m ratio}
\label{sec:OIII/CII}
The line ratio of the two brightest lines accessible with ALMA at $z>6.5$, \oiiilam/\ciilam, provides a rough indicator of the ISM conditions of a galaxy. Just as \oiiilam/SFR ratios are systematically higher than for typical star-forming galaxies at $z\sim0$ (Section \ref{sec:OIII-SFR}),  \oiii/\cii\ line ratios are some of the highest observed in the local Universe \citep[e.g.][]{Inoue16,Carniani20}. While \oiii\ emission requires an ionising radiation field above 35.1 eV, PDR-dominated \cii\ emission relies on FUV radiation between 6-13.6 eV. A natural assumption is therefore that the burstiness of the recent star-formation history dominates the line ratio at a fixed C/O abundance. Indeed, $z\sim0$ metal-poor dwarfs show a strong correlation with the equivalent width of optical emission lines, EW(\oiii+H$\beta$), often used as a proxy for star-formation burstiness \citep[][ Kumari et al. in prep]{Witstok22}. \textbf{Figure \ref{fig:OIII-SFR}} shows a compilation of $z>6.5$ sources with both \oiii/\cii\ and EW(\oiii+H$\beta$) measurements with respect to the local relation, fitted by~\citet{Algera25} to the DGS sample:
\begin{equation}
\log\!\left(\frac{L_{\mathrm{[O\,III]}}}{L_{\mathrm{[C\,II]}}}\right)
= (0.69 \pm 0.07) + (0.67 \pm 0.11) 
\times \left( \log\!\left(\frac{\mathrm{EW}_{\mathrm{[O\,III]}+ {\rm H}\beta}}{\text{\AA}}\right) - 3 \right),
\end{equation}
with 0.12 dex scatter. The agreement of our high redshift compilation with the local metal-poor dwarf sample suggests similar physics drives the \oiii/\cii\ ratio across cosmic time.

Possibly linked to the burstiness of star-formation, \citet{Harikane20} find from Cloudy modeling of  \oiii/SFR as a function of \cii/SFR that systematically higher ionisation parameters are needed to explain the high \oiii/\cii\ ratios, possibly combined with lower PDR covering factors. 
Algera et al. (2025) perform the first study that uses JWST spectroscopy alongside \oiii/\cii\ to disentangle the influence of metallicity (12+$\log(\rm O/H)$), ionisation parameter (\oiii~5007\AA/[O{\sc ii}]~3727,29\AA; O32) and burstiness (EW(\oiii+H$\beta$)) on high-redshift galaxies and $z\sim0$ dwarf galaxies simultaneously. While no significant correlation of \oiii/\cii\ with metallicity is found at high or low redshift, both ionisation parameter and burstiness have statistically significant correlations with \oiii/\cii\ in the low redshift dwarf sample, primarily driven by low \cii/SFR at high EW(\oiii+H$\beta$) and O32. However, comparing the high-redshift sample with the local relations, Algera et al. (2025) find high-redshift galaxies are significantly offset to higher \oiii/\cii\ at fixed O32. This suggests that ionisation parameter alone is not sufficient to explain the high  \oiii/\cii\ ratios in high redshift galaxies, and increased burstiness and/or lower covering factors of the PDR are also important drivers to explain the observed properties.

\subsection{Other Lines}
While \cii$\lambda 158\mu$m and \oiii$\lambda 88\mu$m have been the workhorse lines of ALMA high-redshift observations due to their inherent brightness, a range of fainter lines originating in the ionised, neutral and molecular phases of the ISM can be observed or constrained with ALMA bands 1-10 as galaxies at increasing high redshift are discovered with JWST (see also the \textbf{Supplemental Text}).  

\subsubsection{[OIII]$\lambda 52\mu$m}
\label{sec:OIII52}
The fine-structure line of double ionised Oxygen at $52\mu$m line has been heralded as the ideal line to measure oxygen abundance and constrain electron density and temperature when combined with the \oiiilam\ and \oiii~5007\AA\ lines \citep{Jones20}. However, only a handful of detections and upper limits  have been reported to date \citep{Killi23, Harikane25a} at redshift $z=6.2-7.2$. This lack of data is mainly due to the high frequency of the line requiring Band 9 observations in a relatively compact configuration to ensure the source remains unresolved (putting tight constraints on the observing schedule). Perhaps surprisingly, \citet{Killi23} and \citet{Harikane25a} report relatively faint \oiii$\lambda 52\mu$m lines with \oiii$\lambda 52\mu$m/\oiii$\lambda 88\mu$m$<1$, suggesting modest electron density of the ionised gas ($n_e< 100 \,\rm cm^{-3}$) in all but one galaxy. This is in contrast to the electron densities measured in the rest-frame optical with JWST spectroscopy, which suggests $n_e\sim 200-1700 \,\rm cm^{-3}$ \citep{Harikane25a}.
This discrepancy provides increasing evidence for the presence of diffuse, warm, yet highly ionised gas  surrounding  hot, dense H{\sc ii} regions (see Section \ref{sec:OIII-SFR}).  

\subsubsection{[NII]$\lambda 205\mu$m and [NII]$\lambda 122\mu$m}
\label{sec:NIIlines}
The single ionised Nitrogen ion has two fine-structure transitions that - if sufficiently bright - are easily accessible with ALMA in bands 6 and 7 at $z\sim7$.  Each transition provides different different constraints on  the ISM. 
\nii$\lambda 205\mu$m is most commonly used to understand the fraction of \cii$\lambda 158\mu$m emitted by ionised and neutral (PDR) gas in the ISM.
Nitrogen and Carbon have similar second ionisation potentials (29.6 and 24.4 eV respectively), while the first ionisation potential of Nitrogen (14.5 eV) is slightly above that of Hydrogen and traces exclusively ionised gas (in contrast to Carbon; see Section \ref{sec:CII}).
Moreover, \nii$\lambda 205\mu$m and \cii$\lambda 158\mu$m have similar excitation energies and critical densities ($\sim 45 \,\rm cm^{-3}$) when electrons are the collisional partners. As a result the \nii$\lambda 205\mu$m line emission scales directly with the ionised fraction of the \cii$\lambda 158\mu$m emission ($L_{\rm [NII]}\sim L_{\rm [CII], ion}$) with their ratio only dependent on the gas phase C/N abundance. The \cii$\lambda 158\mu$m/\nii$\lambda 205\mu$m ratio is therefore used to disentangle the \cii$\lambda 158\mu$m arising from different phases (ionised versus neutral gas) of the ISM. 
\citet{Witstok22} reported the first nondetection of \nii$\lambda 205\mu$m in an LBG at $z=6.7$, suggesting the \cii$\lambda 158\mu$m is arising predominantly from the neutral gas, despite the source having high \oiii$\lambda 88\mu$m/\cii$\lambda 158\mu$m that possibly indicates a high volume filling factor of (highly) ionised gas \citep{Cormier15}.
Deeper observations were taken on a sample of four \cii-bright luminous LBGs $z=6-7$ by \citet{Fudamoto26}, confirming $\sim$74-96\% of \cii$\lambda 158\mu$m emission is PDR dominated. Deep upper limits on the \nii$\lambda 205\mu$m line in a Quasar host galaxy at $z=7.5$ similarly suggests $>75\%$ of the \cii\ emission is PDR dominated \citep{Novak19}.

\nii$\lambda 122\mu$m has a high critical density and therefore the \nii$\lambda 205\mu$m/\nii$\lambda 122\mu$m  line ratio constrains the electron density of the ionised gas for $n_e$ between 10 and 3000 $\rm cm^{-3}$ \citep[e.g.][]{HerreraCamus16}. Furthermore the \nii$\lambda 122\mu$m/\oiii$\lambda 88\mu$m line flux ratio constrains the ionisation parameter and/or hardness of the radiation field for a given gas phase N/O abundance, with little dependence on electron density due to similar critical densities (310 and 510 $\rm cm^{-3}$, respectively). 

Several sources at $z=6-7$ have yielded non-detections \citep{Harikane20,Sugahara21} or faint emission \citep{Novak19,Killi23,Litke23} to date in contrast to bright \oiii$\lambda 88\mu$m detections. The limits on the \nii/\oiii\ line ratio suggests $\log{U_{\rm ion}}\gtrsim-3$, fully consistent with measurements of the ionisation parameter in high-redshift galaxies with JWST \citep[e.g.][]{Reddy23}. \citet{Novak19} detect \nii$\lambda 122\mu$m alongside constraints on \nii$\lambda 205\mu$m in a Quasar host galaxy at $z=7.5$. The line ratio of the two \nii\ lines (\nii$\lambda 122\mu$m/\nii$\lambda 205\mu$m $>$ 4.4) gives an electron density constraint of $n_e > 180\, \rm cm^{-3}$, consistent with the high electron densities measured by JWST at these redshifts \citep{Isobe23,Reddy23}.

\subsubsection{[OI]$\lambda 146\mu$m and [OI]$\lambda 63\mu$m} \label{sec:OI}
The neutral Oxygen atom has two fine structure transitions that are accessible to ALMA at $z>6.5$. \oi$\lambda 63\mu$m is a bright line in local starburst galaxies ($L_{\rm [CII]}/L_{\rm [OI]}\sim 1$), though more difficult to detect at high-redshift due to its high frequency typically requiring ALMA band 9 or 10 observations. \oi$\lambda 146\mu$m is a $\sim 10\times$ fainter line, but conveniently observable in ALMA band 6 at $z\sim7$.  

Due to the first ionisation potential of Oxygen being identical to Hydrogen (13.6eV) the \oi\ fine structure lines trace the neutral ISM without being affected by ionised gas. Both \oi\ lines have a higher critical density than \cii$\lambda 158\mu$m ($\sim$10$^5$ cm$^{-3}$ versus $\sim$10$^3$ cm$^{-3}$ in neutral gas) and a higher
excitation temperature needed to populate the upper level (228-329
K versus 91 K), therefore \oi\ traces warmer, denser
gas than \cii\ and their line ratio gives important insight into the ISM conditions of the star-forming gas within galaxies.

No robust detections of \oi$\lambda 63\mu$m have been obtained at $z>6.5$ to date \citep{Rybak23}, though \citet{Ishii25} detect the line at $z=6.04$ in a Quasar host galaxy with $L_{\rm [CII]}/L_{\rm [OI]}=1.3\pm0.5$, consistent with local starburst galaxies. 
New detections of bright very high-redshift galaxies with JWST might enable observations in ALMA band 7-8  for similar line strengths in the near future. Promisingly, several detections of \oi$\lambda 146\mu$m in LBGs \citep{Fudamoto24}, a SMG \citep{Litke23} and a Quasar \citep{Novak19} are now obtained at $z>6.5$. In all these galaxies a remarkable high \oi/\cii\ ratio has been established compared to the local galaxy population (including starbursts, metal-poor dwarfs and AGN), suggesting the ISM in high redshift galaxies is systematically denser ($\sim10-1000\times$) than MS galaxies at $z\sim0$. 

The \oiii$\lambda 88\mu$m/\oi$\lambda 63\mu$m or \oiii$\lambda 88\mu$m/\oi$\lambda 146\mu$m ratio, which is insensitive to variations of elemental abundances, may be used as an indicator of the filling
factor of ionized gas, traced by \oiii, relative to the filling factor of PDRs, traced by \oi\ \citep{Cormier15}. The bright detections of \oi$\lambda 146\mu$m to date compared to \oiii$\lambda 88\mu$m, suggest similar PDR filling factors as local (U)LIRG galaxies, but significantly higher PDR filling factors than local metal-poor dwarf galaxies \citep[][]{Fudamoto24}. 

\subsubsection{[NIII]$\lambda 57\mu$m}
The double ionised nitrogen ion has a fine-structure transition \niii\,57$\mu$m that falls in ALMA band 9 at $z\sim7$. The most promising use of \niii\,57$\mu$m is in combination with \oiiilam\ or \oiii\,52$\mu$m to derive N/O abundance.  Oxygen is an alpha-capture primary element produced by massive stars, whereas nitrogen is both a primary and a secondary nucleosynthetic element. Furthermore, Oxygen and Nitrogen have similar ionisation potentials, making their line ratios only weakly dependent on ionisation parameter and radiation hardness. 
Therefore the \niii/\oiii\ ratio is a good proxy for N/O relative abundance, which provides constraints on the star-formation history of the galaxy, and can be calibrated as a metallicity indicator \citep[e.g.][]{Peng21}.  

Both \citet{Nagao11} and \citet{PereiraSantaella17} present \oiii/\niii\,57$\mu$m based metallicity calibrations using Cloudy modeling. These models suggest [{N \sc iii}] is reasonably bright ($L_{\rm \oiii88} \sim L_{\rm \niii57}$) at solar metallicity, but for lower metallicity systems becomes too faint for the lower sensitivity bands of ALMA. However, recent JWST observations report the existence of `N-emitters'; sources with (super-)solar N/O despite sub-solar O/H abundance \citep[e.g.][]{Cameron23,Castellano24}. These sources could be bright in \niii\,57$\mu$m and particularly interesting at $z>10$ where the line moves into ALMA band 8.  

\subsubsection{CO and [CI]}
At $z\sim7$ rotational CO transitions $J=3-2, 6-7$ and higher are accessible with ALMA due to the recent installment of band 1 receivers. Moreover, the neutral gas tracer [{C\sc i}] has two fine-structure lines close to CO(4-3) and CO(7-6) in frequency. However, due to their inherent faintness studies remain limited to IR bright systems, such as SMGs or Quasar host galaxies. \citet{Jarugula21} detect CO $J=6-5,7-6, 10-9$ and  [{C\sc i}]\,370$\mu$m in two SMG selected galaxies at $z = 6.9$. They derive a CO ladder that peaks at $J=6-5$ and steadily declines towards higher $J$ transitions. From the CO-based gas mass measurements there is no evidence for evolution in depletion time, consistent with \cii\ based results in LBGs at the same redshift \citep{Aravena24}.
\citet{Venemans17a}, \citet{Decarli22} and \citet{Li22} detect high-$J$ CO lines and [{C\sc i}]\,370$\mu$m in Quasar host galaxies at $z>6.5$, while \citet{Kaasinen24} detect CO(2-1)  with the JVLA in three of them. The \cii/[{C\sc i}] ratio is particularly useful for differentiation between PDRs and XDRs, with all Quasar host galaxies preferring the PDR ISM conditions, suggesting the central black holes of these galaxies do not significantly impact the cold ISM. The CO ladder rises strongly from CO(2-1) to CO(6-5) and CO(7-6), suggesting a significant fraction of gas is in a high excitation state. 

In less luminous systems \citet{Hashimoto23} target a LBG at z=7.15 and find non-detections for CO(6-5), CO(7-6), and [{C\sc i}]\,370$\mu$m. Their limits on $L_{\rm \cii}/L_{\rm CO}$ are higher than found in Quasar host galaxies and suggest lower density in the PDR gas. \citet{Jones24} use the JVLA to constrain CO(2-1) in a LBG at $z=8.31$. The most distant successful detection of CO in a LBG is currently at $z=7.31$~\citep{Cescon26}.

\section{DUST CONTINUUM}\label{sect:dust}
There remain fundamental open questions in the study of dust in the early universe.
When did significant dust accumulate in the Universe?  How does the dust differ from that found in local galaxies?  How important is dust obscured star-formation in the census of galaxies in the early Universe? 
Here we describe the key observational constraints on the dust continuum emission to-date at $z > 6.5$, and discuss the implications of these measurements on the obscured component of cosmic star-formation and dust formation mechanisms. 

\begin{figure}
    \centering
    \includegraphics[width=0.8\linewidth, trim = 0 0.2cm 0.5cm 1cm,  clip]{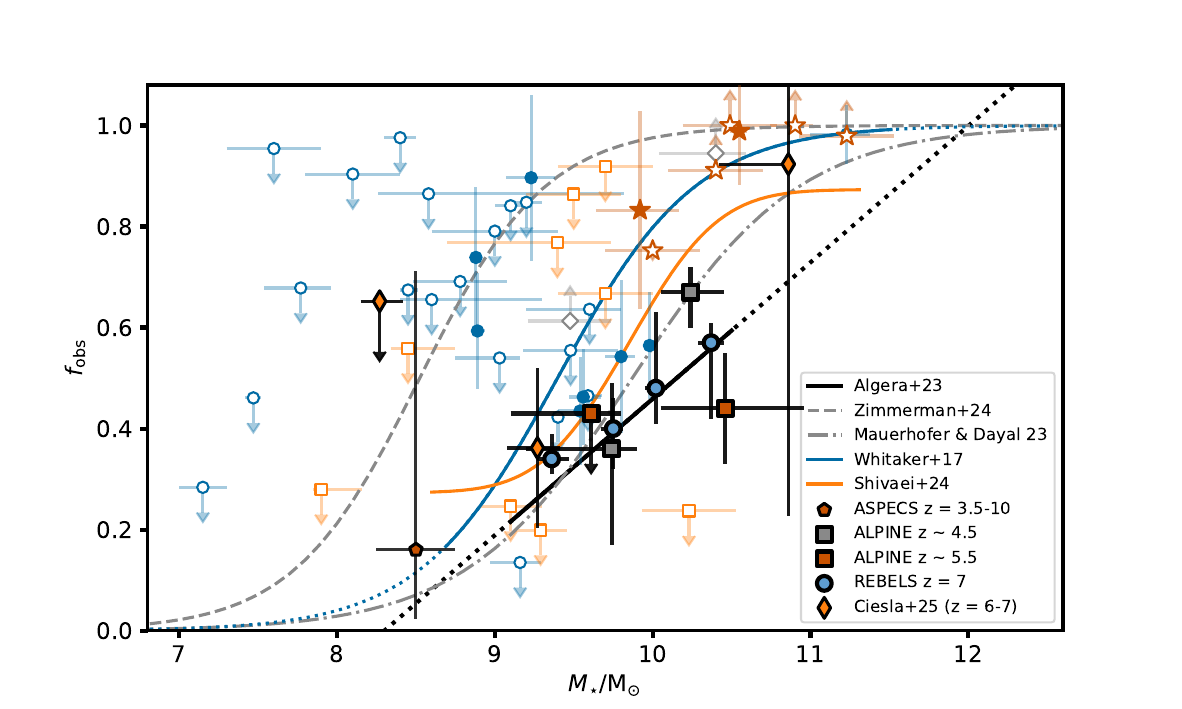}\\
    \includegraphics[width=0.8\linewidth, trim = 0 0.2cm 0.5cm 1cm,  clip]{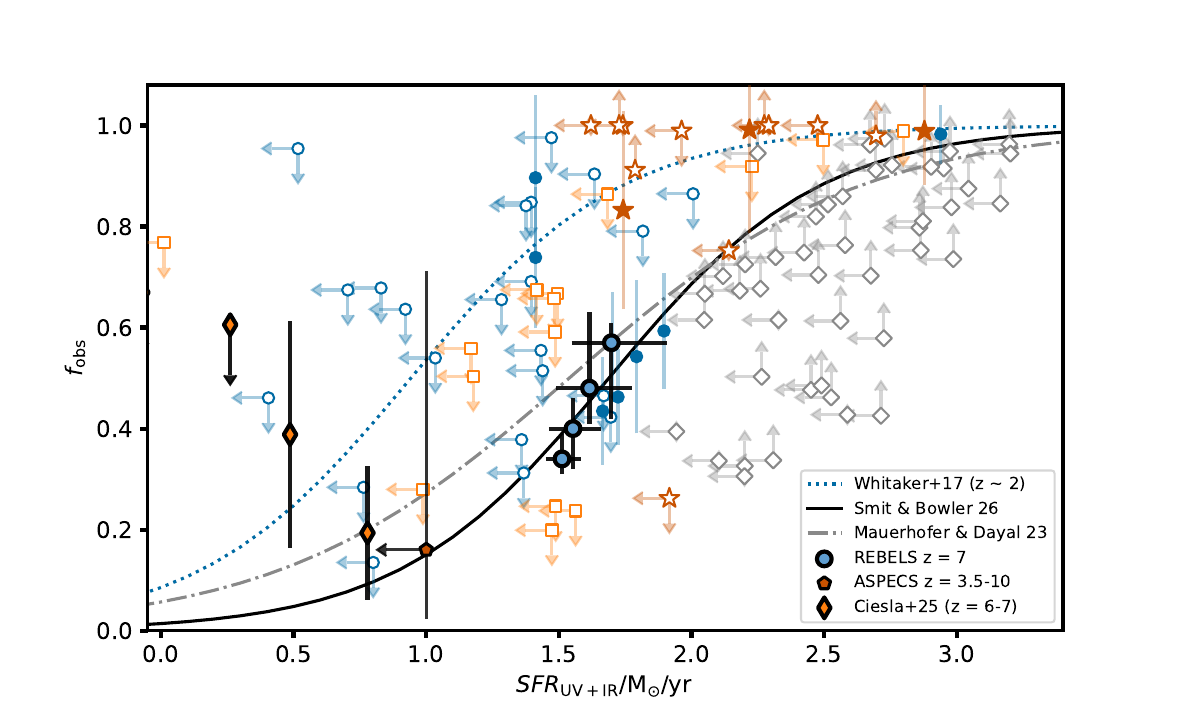}
    \caption{The obscured fraction of star formation as a function of stellar mass (total SFR) in the upper (lower) plot.  In both figures the open symbols correspond to objects where there is a limit set as $3\sigma$, placed on either the UV or FIR SFR (i.e. a non-detection). {\bf Top:} We show a subset of sources that have $M_{\star}$\ measurements.  We also show the stacked results from the REBELS survey~\citep{Algera23}, the ALPINE survey~\citep{Fudamoto20}, the ASPECS survey~\citep{Bouwens20} and A$^3$COSMOS~\citep{Ciesla25}.  {\bf Bottom:} The \fobs~as a function of total SFR (with our best fit relation following equation~\ref{equ:SFR}). } 
    \label{fig:fobs}
\end{figure}

\subsection{The fraction of obscured star formation}\label{sect:fobs}
\begin{marginnote}
\entry{Obscured fraction}{$f_{\rm obs}=\frac{{\rm SFR}_{\rm IR}}{({\rm SFR}_{\rm UV}+{\rm SFR}_{\rm IR})}$}
\end{marginnote}
\noindent
The importance of dust obscured star formation in galaxies can be quantified by measuring \fobs, the fraction of obscured star formation.
At $z \lesssim 2$, a stellar mass and SFR dependence on \fobs~has been observed, do such relations exist at $z > 6.5$? 

\subsubsection{\fobs -- $M_{\star}$}
The majority of sources in our compilation have upper limits on their dust continuum emission from ALMA or NOEMA as shown in \textbf{Figure~\ref{fig:fobs}}.
To mitigate this, several stacking analyses have been performed to determine an average \fobs\ in high-redshift galaxy samples.
For the REBELS sample we show the stacking results from~\citet{Algera23} at $z \simeq 7$, and compare this to the results at $z =4$--$6$ from the ALPINE survey presented in~\citet{Fudamoto20}.
To complement these targeted results on relatively high-mass galaxies, deep field programs provide stacked constraints on the \fobs\ of lower mass galaxies.
The~\citet{Bouwens20} results, derived in a broad redshift bin ($z = 3.5$--$10$), appears to indicate that galaxies at \lmstar$\simeq 8$--$9$ show a relatively low obscured fraction of $\sim 20$ percent consistent with the lack of dust continuum detections found within the ASPECS survey.
Similar results are found from the stacking results of~\citet{Ciesla25} who included spectroscopically confirmed sources in the A$^3$COSMOS data.\footnote{Note that these studies assume different FIR SEDs to the default in this review, with the REBELS analysis assuming the constant conversion factor used by~\citet{Inami22} and ALPINE assuming the template of~\citet{Bethermin20}.  In addition, \citet{Algera23} excluded starbursts. \citet{Bouwens20} assumed \betad $= 1.6$ and an evolving dust temperature.  To show the \fobs~derived from the ASPECS program by~\citet{Bouwens20} at \lmstar$ < 9$ we converted the provided IRX value.  We refit the~\citet{Ciesla25} fluxes with the same FIR SED as used in this review.
}

Taking together the individual sources and the stacking results, there is evidence for a trend in \fobs\ with stellar mass at $z > 6.5$.
We compare the $z > 6.5$ results to the studies of~\citet{Whitaker17} and~\citet{Shivaei24} at $z = 0$--$2.5$.
\citet{Whitaker17} performed a stacking analysis of a sample of mass complete SDSS galaxies, with the FIR emission being derived from the~\emph{Spitzer}/MIPS 24$\mu{\rm m}$ band and assuming a conversion from this band to the full \lir.
\citet{Shivaei24} instead derived the obscured fraction of star formation from SED fitting of MIRI selected sources at $z = 0.7$--$2$ with coverage of the PAH features (which are used to estimate the \lir) at around $\lambda_{\rm rest} = 6$--$11\mu{\rm m}$.
Both of these studies see a trend with increasing \fobs\ with stellar mass, with a large scatter around the relation.
\citet{Whitaker17} find no evolution with redshift (up to $z = 2$) in the \fobs-$M_\star$ relation, which is supported by the~\citet{Shivaei24} results.
The results presented here suggest that there is an evolution of the relation to lower \fobs\ at a given stellar mass at $z > 6.5$.
\citet{Algera23} and~\citet{Fudamoto20} showed that while the~\citet{Whitaker17} study would predict \fobs $> 0.8$ for the ALPINE and REBELS galaxies at \lmstar $\simeq 10$, instead the fraction is lower at around \fobs$\sim 0.45$. 

It is clear in \textbf{Figure~\ref{fig:fobs}} that there is a large scatter for individual objects around the relations derived from for example, the REBELS results~\citep{Algera23}.
One significant part of the scatter is likely due to sample selection effects, as can be seen by comparing the \fobs~ for LAEs and ALMA-selected sources.
LAEs show exclusively upper limits, with a group of three sources at \lmstar$\sim 10$ with very low obscured fractions of star formation ($f_{\rm obs} \lesssim 0.25$) as expected given the expected low dust content in these sources (\citealp{Ouchi20}).
Conversely, the ALMA selected serendipitous sources from~\citet{Fudamoto21}, which are spectroscopically confirmed via \cii, show extremely high $f_{\rm obs}$ as expected from their lack of rest-frame UV emission.\footnote{We note that the stellar masses for both LAEs and the serendipitous sources can be uncertain due to a lack of detections in the rest-frame UV and optical.}
We note that if stellar masses are underestimated as some resolved studies and studies using non-parametric SFHs are suggesting at $z > 6$ (e.g.~\citealp{GimenezArteaga23}) then the offset to the \fobs~predicted by the $z \simeq 2$ studies would increase.
Similarly, if the dust temperature is lower on average than the value we assume, then the \fobs~offset would also increase compared to $z \simeq 2$.

\subsubsection{\fobs -- ${\rm SFR}$}
While the trends in \fobs\ have been discussed mainly with respect to stellar mass, further information can be gathered from the relation as a function of total SFR (\sfrtot $=$ \sfruv $+$ \sfrir).
As a greater proportion of the ALMA compilation has UV+IR measurements (including the quasars), this diagram is populated by an increased number of sources.
Unsurprisingly, the LAEs show low \fobs~values, which populate the lower right-hand side of \textbf{Figure~\ref{fig:fobs}}, while far-IR detected sources appear to be highly star forming and obscured objects.
There is also clear selection effects on the population of the diagram, with the REBELS sample of LBGs (which makes up the majority of the central points), are UV bright and hence have \lsfr$> 1.5$ by prior selection, with upper limits on the dust obscured component in most cases.
We also show the the stellar mass binned \fobs\ values from~\citet{Algera23} converted into \sfrtot\ bins using the average \muv\ of the REBELS sample (to give the UV SFR).
The stacking results of~\citet{Ciesla25} were converted to total SFR assuming the same conversions as this review.

For the quasar host galaxies we show these sources on the $f_{\rm obs}$--SFR diagram only, as it is challenging to measure the stellar mass from the quasar-dominated rest-frame UV and optical emission.
If we assume that the quasar host accounts for less than 10 percent of the UV emission we can put a limit on $f_{\rm obs}$.
With this assumption these values support quasar hosts being extremely dust obscured (\fobs\ $> 0.5$) with high total SFRs of the order of $\simeq 100$--$1000\,M_{\odot}/{\rm yr}$, as identified in previous works (e.g.~\citealp{Venemans20}).

Not withstanding the multiple limits and sample-selection effects that go into \textbf{Figure~\ref{fig:fobs}}, there is evidence for a trend in the \fobs~with total SFR.
As shown in \textbf{Figure~\ref{fig:fobs}}, the \fobs\ from the stacked results shows a deficit with respect to the $z \simeq 2$ trend from~\citet{Whitaker17}.
\citet{Whitaker17} sees a trend to decreasing \fobs\ at a given SFR from $z = 0$--$2$.
The results at $z > 6.5$ show a continuation of this trend, such that at a given \sfrtot\ a smaller proportion of the total SFR is obscured at higher redshift.
Following these lower redshift studies, we find that a functional form of:
\begin{equation}\label{equ:SFR}
  f_{\rm obs} = \frac{1}{(1+a\,e^{b\,SFR)}}
\end{equation}
with $a = 70$ (compared to $a = 10.7$ at $z = 2$) and $b = -2.5$ (fixed to the $z = 2$ value) can well describe the data at $z > 6.5$, although there is significant scatter.
This form closely follows that found by the model predictions as illustrated in \textbf{Figure~\ref{fig:fobs}} for the \fobs-$M_{\star}$ relation (and infrared-excess - see Section~\ref{sect:irx}).
Using a cosmological hydrodynamic simulation with on-the-fly dust formation (including dust evolution and destruction), \citet{Zimmerman24} over-predicts the \fobs~at high-redshift comparison to the observations.
In contrast, the~\citet{Mauerhofer23} semi-analytic model shows good agreement, however this model assumes perfect mixing of gas and dust, and a uniform spherical dust geometry.
\citet{Zimmerman24} attribute their predicted increase in \fobs~at a given stellar mass to the compact sizes and higher gas fractions in galaxies at higher redshift, pointing out that observations may be biased low due to selection effects.

\begin{figure}
\begin{subfigure}{.5\textwidth}
    \includegraphics[width = 1.0\textwidth]{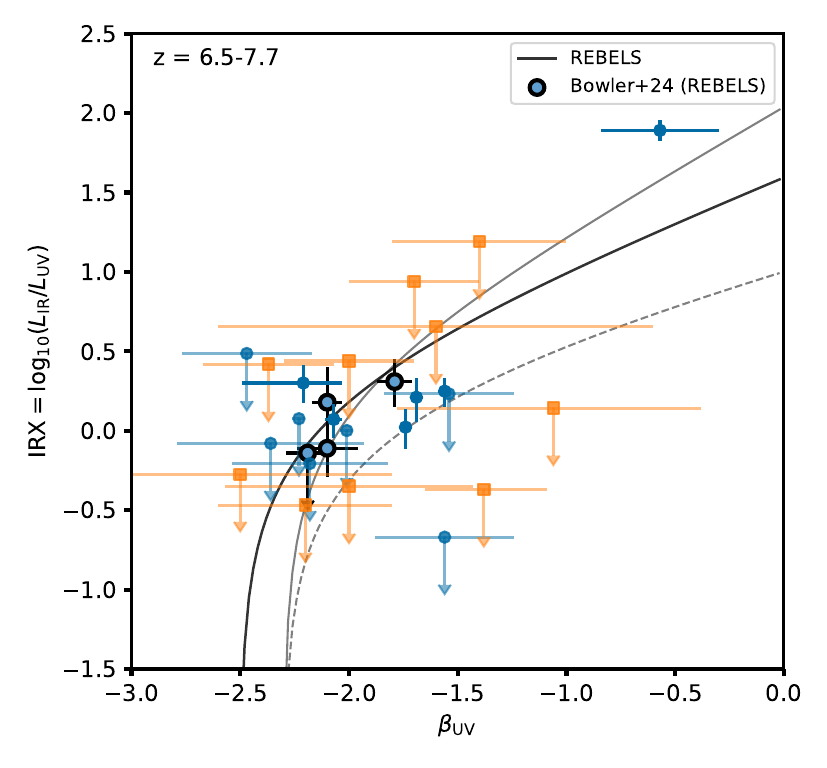}
    \end{subfigure}%
    \begin{subfigure}{.5\textwidth}
      \includegraphics[width = 1.0\textwidth]{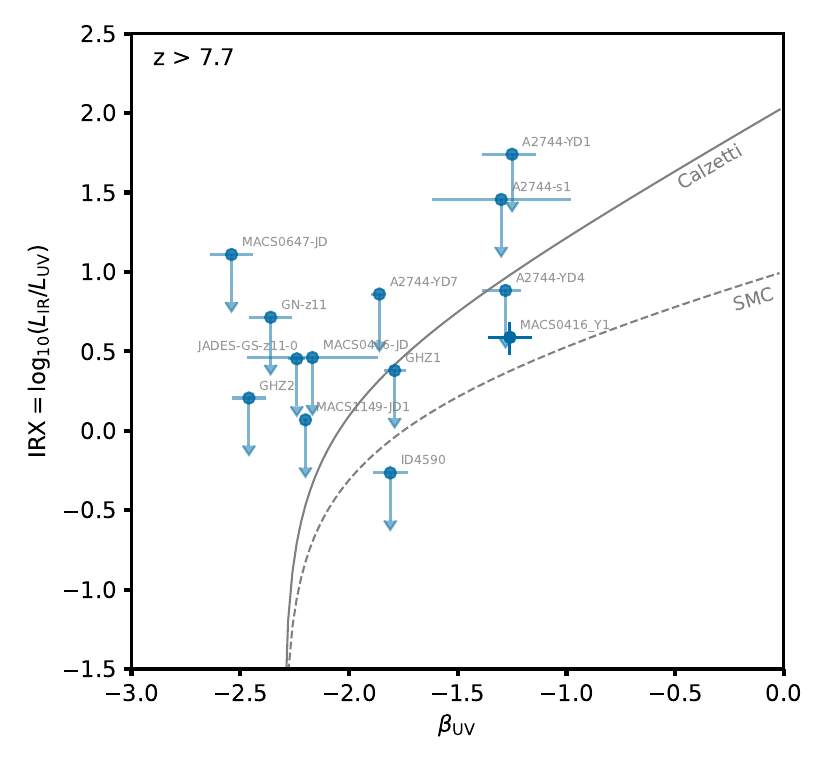}

    \end{subfigure}%
          
    \caption{The IRX-$\beta$ relation with rest-frame UV slope taken from the compilation of ALMA observed sources at $z = 6.5$--$7.7$ (left) and $z > 7.7$ (right).
    We compare to the local prediction from the~\citet{Calzetti94} attenuation law and the SMC extinction law, assuming a screen of dust.
    For clarity the majority of REBELS galaxies have been removed from the plot, and the stacked results from~\citet{Bowler24} shown instead (open circles).  REBELS galaxies that have improved \betauv~values from~\emph{JWST} are shown separately~\citep{Fisher25}.
    }
    \label{fig:irx}
\end{figure}

\subsection{Infrared Excess Scaling Relation}\label{sect:irx}
The importance of dust obscuration has been further quantified in the literature using the \irxb~and \irxm~relations.
IRX is defined as the logarithm of the ratio of the total IR to monochromatic UV luminosity.
\begin{marginnote}
\entry{Infrared-excess}{$IRX = L_{\rm IR}/L_{\rm UV}$}
\end{marginnote}
It combines the rest-frame UV and FIR parts of the galaxy SED, providing additional information on the dust properties (e.g. the rest-frame UV can give attenuation curves; see~\citealp{Salim20}, but cannot discern geometrical offsets between stars and dust).
We present the \irxm~relation {\bf Supplemental Figure 5}, since it can be directly related to the \fobs~relations discussed in Section~\ref{sect:fobs}.

The IRX has been shown to form a tight correlation with galaxy colour in the rest-frame UV (\betauv) in starburst galaxies in the local Universe~\citep{Meurer99}, which is as expected if the energy lost in the rest-frame UV is re-radiated in the FIR~\citep{Calzetti00}.
Deviations above the relation have been shown in various theoretical and observational works to be caused by geometrical offsets between the unobscured regions (leading to blue \betauv $\simeq -2$ but a large IRX that is dominated by fully obscured regions).
Conversely deviations below the relation can be a result of older stellar ages or a steeper attenuation curve (see e.g.~\citealp{Popping17a, Vijayan22}).
Initial observations with ALMA and NOEMA at $z \sim 5$ by~\citet{Capak15} revealed red galaxies with a low IRX, indicating low dust obscured SFR and suggestive of an SMC-like (hinting at low-metallicity) extinction curve.
ASPECS results similarly suggested a deficit in IRX at a given \betauv\ from stacking results (in the range $z = 3.5$--$10$;~\citealp{Bouwens20}).
The ALMA FFS produced only upper limits in stacking results from~\citet{Carvajal20}, with the highest redshift sources at $z = 5$.
Studies following-up brighter $z \simeq 7$ galaxies recovered a range of values supporting a deficit~\citep{Smit18, Hashimoto19} or finding some consistency with the local starburst relation~\citep{Bowler18, Schouws22}.

We show the ALMA compilation on the \irxb~relation in \textbf{Figure~\ref{fig:irx}}, where we split by redshift for clarity (at $z = 7.7$).
It is clear that the majority of the galaxies do not have a dust continuum detection.
The majority of sources are also relatively blue, with \betauv$\simeq -2$, potentially due to the LBG selection process (which favors blue sources over redder objects that are more likely to be confused with contaminant populations).
Interestingly, there are several $z > 7.7$ sources that show redder slopes \betauv$ \,\simeq -1.2$, including the highest redshift dust detection (MACS0416-Y1;~\citealp{Ma24, Harshan24}).
This could be a result of reduced blue-bias in the sample selection for certain spectroscopic follow-up campaigns with~\emph{JWST}.
Quasar hosts cannot be accurately represented as the rest-frame UV emission is dominated by AGN emission.
With the assumed FIR SED of this work, we see that there are sources both above and below the low-redshift Calzetti-like relation.
\citet{Bowler24} performed a stacking analysis of the REBELS and ALPINE sources, assuming a \Td$= 46\,{\rm K}$ and \betad$=2.0$ from $z = 4$--$8$.
They found consistency with the local~\citet{Calzetti00} starburst relation on average for galaxies at these redshifts.
This suggests little evolution in the dust properties of galaxies from the local Universe, however we note that the attenuation properties are only weakly related to the underlying extinction curve once a complex geometry is taken into account~\citep{Narayanan18}.
There remain a handful of sources below the SMC-like extinction curve prediction in the \irxb~diagram at $z = 6.5$--$7.7$.
This could be interpreted as evidence for an SMC-like attenuation curve. 

At $z > 7.7$ the errors on \betauv~are generally reduced, due to the availability of~\emph{JWST} spectra.
A handful of these high-redshift sources also show a deficit in comparison to the REBELS/Calzetti-like prediction (ID4590, MACS0416-Y1), suggesting if the FIR SED assumptions are realistic, that dust properties may be different in the very early universe.
However, the key uncertainties in this measurement are a) the assumed FIR SED, as temperature changes can cause great changes in the derived \lir~and hence IRX and b) errors on the slope \betauv, which while mitigated by~\emph{JWST} spectroscopy, are still dependent on calibrations and the method used to derive the slope.
Finally, there is the uncertain effect of star-dust geometry and the unknown intrinsic UV-slope of these galaxies.
Further deep multi-band ALMA observations are required to define this relation at $z \gtrsim 8$.

Several simulations and models predict an excess in the \irxb~relation at $z > 6.5$ in comparison to the local relation~\citep{Pallottini22, Ferrara22}, while others predict little evolution to higher redshifts~\citet{Narayanan18, Liang21, Vijayan22, Mauerhofer23} or a deficit as expected for the SMC-like dust~\citet{Ma19}.
\citet{Narayanan18} for example predict an attenuation curve close to the~\citet{Calzetti00} form, due to sources having a complex star-dust geometry (which tends to flatten the law) as well as a smaller range in the intrinsic ages of galaxies at high redshift.

 \begin{figure}
    \centering
    \includegraphics[width=0.95\linewidth]{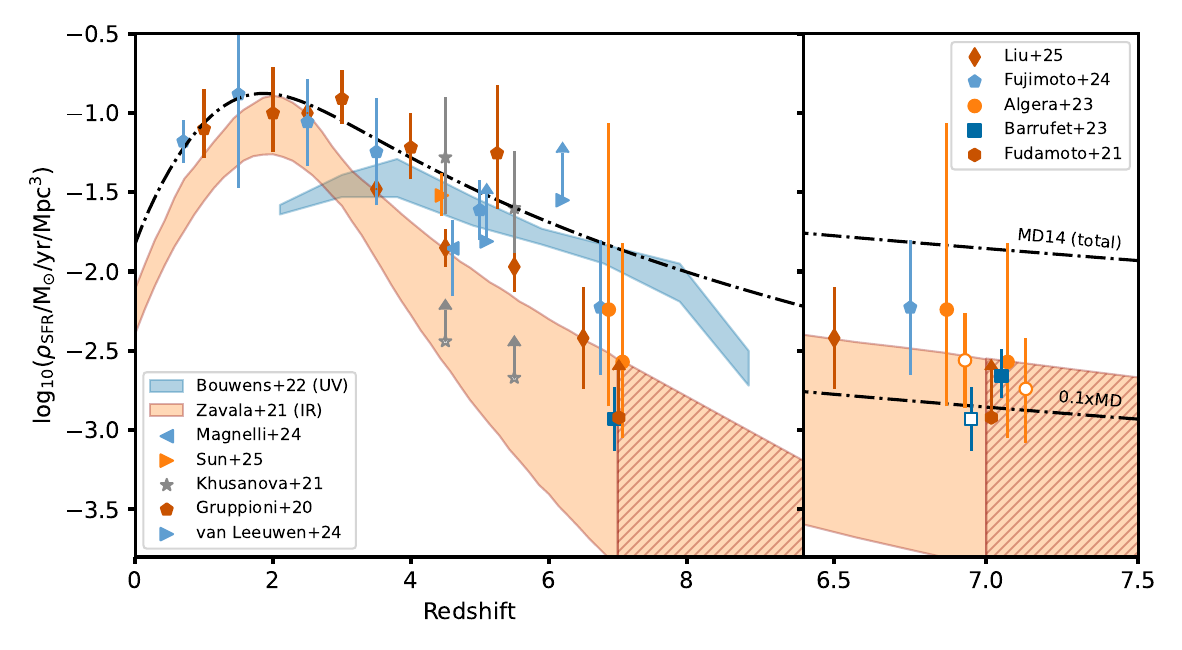}
    \caption{The cosmic SFR density as a function of redshift.  We show estimates of the obscured CSFRD determined for galaxies at $z > 6.5$ from~\citet{Algera23, Barrufet23, Fudamoto21, Fujimoto24, Liu26}.
    These are shown in comparison to the total expected value from~\citet{Madau14} (dot-dashed line), the UV (unobscured) measurement from~\citet{Bouwens22} and the IR (obscured) model from~\citet{Zavala21}.
    A zoom-in of the high-redshift results is shown on the right.
    The open points represent measurements from~\citet{Algera23} and \citet{Khusanova21} without extrapolation to lower stellar masses than the REBELS (ALPINE) sample, and hence represents a lower limit.
    The open points from~\citet{Barrufet23} are when the serendipitous sources from~\citet{Fudamoto20} are removed.
    }
    \label{fig:csfrd}
\end{figure}

\subsection{Cosmic Star Formation Rate Density}\label{sect:sfrd}
As discussed so far, the direct detection of the dust continuum emission provides an estimate of the obscured SFR in galaxies. 
This can be used to compute the obscured cosmic star-formation rate density (CSFRD), typically by using an estimation of the IR luminosity function (LF), which can then be integrated down to a particular limiting \lir\ and transformed into a CSFRD via a \lir\ to $SFR_{\rm IR}$ conversion factor.
Due to the dependence on stellar mass of the \fobs, coupled with the small field-of-view of ALMA/NOEMA, it is challenging to determine the IR LF from blind surveys that have small numbers of detections (if any at high redshift).
Instead, the IR LF is typically inferred from combining the rest-frame UV LF or the galaxy stellar mass function (GSMF) as a basis for the number density of sources, with a relation between the \lir~and \muv\ or \lmstar.
As this is highly survey dependent we do not compute this for the sources in our compiled catalogue, but instead look to the literature for estimates.

\citet{Algera23} used the REBELS sample to determine the CSFR density at $z \simeq 7$ using the GSMF coupled with a stacking analysis, to determine a relation between stellar mass and IR luminosity.
The slope of this relation was sensitive to how binning was performed, and hence we present two points in \textbf{Figure~\ref{fig:csfrd}} corresponding to the Monte-Carlo and simple binning approaches in this study.
The integral to compute the CSFRD was computed to the point at which~\citet{Algera23} determine the \fobs\ is consistent with zero (\lmstar $\sim 8.2$).
We also present the results when only the stellar mass range of the REBELS galaxies themselves is considered.
An independent estimate of the CSFRD from the REBELS sample was derived by~\citet{Barrufet23}, who determined the IR LF by instead determining the number density of sources from the UV LF.
The IR LF is still weakly constrained at $z \simeq 7$, however~\citet{Barrufet23} was able to measure three LF points across the expected knee (\llir $= 11.4$--$12.12$, with the knee at around \llir $=11.6$).  
The faint-end slope was fixed to $\alpha = -1.3$ and they integrated down to \lir $ = 10^{10.5}$\lsun.
From these REBELS analyses the obscured CSFR density is measured to be at least 10 percent of UV based estimates (see \textbf{Figure~\ref{fig:csfrd}} right panel).

The ACLS has been used to estimate the IR LF up to $z \simeq 7$~\citep{Fujimoto24}.
This survey found a single galaxy at $z > 6$, the `Cosmic Grapes' source at $z = 6.07$.
With this galaxy, and estimates from other sources in the literature (namely~\citealp{Watson15, Fudamoto21, Barrufet23}) they compute the IR LF at $z = 6$--$7.5$ and subsequently the obscured CSFRD at this epoch.
The faint-end slope was fixed to $\alpha = -0.94$, and they integrated down to \lir $ = 10^{10}\,$\lsun.
We show their obscured estimates in \textbf{Figure~\ref{fig:csfrd}}.
When they add these values to the unobscured component as determined by~\citet{Bouwens22}, the result is around 60 percent in excess of~\citet{Madau14}, however multiple uncertainties remain in the computation of the IR LF.
These include the conversion of \lir\ to $\rm SFR_{\rm IR}$, the computation of \lir\ from a single data point in the FIR and the small number statistics at $z > 4$ (where only 13 galaxies are detected in an area of 100 sq. arcmin).
\citet{Liu26} performed a stacking analysis of the \athree\ data in the PRIMER COSMOS field to determine the obscured CSFRD from $z = 3$--$7$, estimating that $\sim 20$ percent of the CSFRD can be obscured even at $z \simeq 7$.
We show their estimate from all massive galaxies in the COSMOS field, using the average relation between sSFR and \lir\ to include sources in the estimate even if they are not covered by ALMA data.
A lower limit on the obscured CSFRD from serendipitous sources was estimated from early REBELS data by~\citet{Fudamoto21}, who discovered two \cii\ detected but \emph{HST}-dark galaxies in the LBG-targeted ALMA pointings. 
They computed the CSFRD by taking the volume of the REBELS ALMA data available at that point, and the SFR estimates from the UV and IR luminosities (the results are consistent if \cii\ SFRs are used instead).
The CSFRD was then corrected by a factor of four to account for the clustering of the sources.

We compare to several estimates of the obscured CSFRD at $z = 4$--$6$, where several studies have suggested the obscured contribution could be dominant, although there is large scatter.
From ALPINE we show the results from~\citet{Khusanova21} who used stacking to obtain a \lir\ to \muv\ or stellar mass relation for the main sample, and~\citet{Gruppioni20} who searched for serendipitous sources.
The dependence on extrapolating beyond the range of the data can be seen for example in the two sets of~\citet{Khusanova21} points presented in \textbf{Figure~\ref{fig:csfrd}}.
We show the results from~\citet{vanLeeuwen24} who looked for serendipitous sources in the REBELS and ALPINE data (including the results of~\citealt{Loiacono21}).
From \athree\ we present the results from~\citet{Magnelli24} and~\citet{Liu26}, and from  ASPIRE we show the results of~\citet{Sun25} determined from 8 serendipitous sources at $z = 4$--$6$ around the targeted $z > 6.5$ quasars.
\citet{Zavala21} extrapolated the IR luminosity function and evolution determined from the MORA survey to predict $20$--$25$ percent of the CSFRD at $z = 6$--$7$ should be obscured.

While there remain uncertainties in the computation of the CSFRD at $z > 6.5$, it is clear that even with conservative assumptions (e.g.~\citealp{Algera23}) there is a significant component ($> 10$ percent, and up to 40 percent) that is obscured.
As we have shown in earlier sections, there is a stellar mass and SFR dependence of the \fobs, and hence this obscured component is mainly contributed by galaxies at \lmstar $ > 9$ and \sfrtot $> 10\,$\sfrunits.
The obscured component of the CSFRD at $z > 7$ remains poorly constrained, with the lack of sufficiently deep data on samples of galaxies at these redshifts.
However, considering the extrapolation of the~\citet{Zavala21} and~\citet{Madau14} relations, we could expect of the order of 10 percent to still be obscured at $z \simeq 8$.

\subsection{Dust Temperature}\label{sect:dustt}

\begin{figure}
    \centering
    \includegraphics[width=0.99\linewidth]{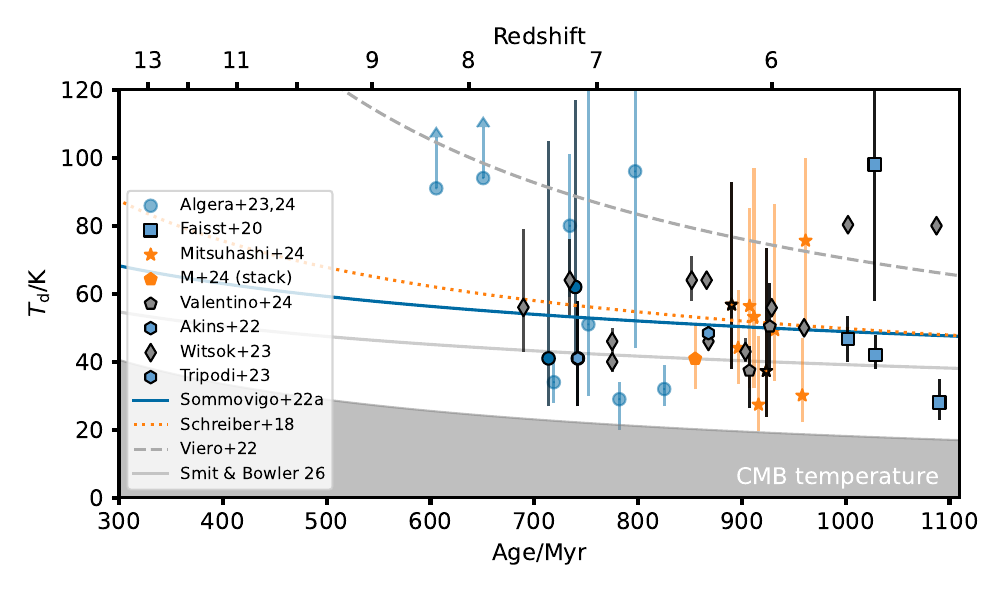}
    \caption{The measured SED dust temperature from galaxies with multi-band ALMA observations.
    The $z > 6.5$ constraints are from~\citet{Algera23}, taking the values with \betad$=2.0$.
    Sources with detections in $>2$ bands in the FIR are outlined in black.
    \citet{Mitsuhashi24} provided measurements from the SERENADE sample at $z \simeq 6$, where they allowed \betad\ to be free in the fitting.
    We also show the Cosmic Grapes source studied in~\citet{Valentino24}.
    The values for the four sources studied by~\citet{Faisst20a} (who compiled Band 6, 7 and 8 data) are taken from~\citet{Algera23} assuming \betad$= 2.0$. 
    }
    \label{fig:tdust}
\end{figure}

With the advent of multi-band ALMA measurements it has become possible to place constraints on the FIR SED shape, in particular the dust temperature (and in some cases emissivity) at very high redshifts.
To do this requires at a minimum of two bands (typically band 6 and band 8 at $z \simeq 7$, as these bands can also be used to target the redshifted \cii~and \oiii~lines respectively).
\textbf{Figure~\ref{fig:tdust}} shows a compilation of \Td\ measurements at $z \gtrsim 5$ in comparison to several empirical and simulation predictions for the dust temperature evolution.
At $z > 6.5$ there are eleven galaxies, two quasars and two SMGs (in the same system - SPT-E and SPT-W) at the time of writing that have published \Td~measurements from multi-band observations.
An early compilation was made by~\citet{Witstok22}, who fitted a sample of five LBGs that had constraints in Band 6 and 8.  
COS-3018, REBELS-P9, REBELS-P7 have detections and showed a wide range of dust temperatures (\Td$ = 30$--$100\,{\rm K}$).
COS-2987 and Z-007 were not detected in any ALMA band and hence \Td~was not constrained.
\citet{Algera24} performed fitting of three REBELS sources (REBELS-12, 25, 38) which have Band 6 and Band 8 observations, plus they presented a refitting of other sources using a consistent fixed \betad $= 2.0$.
They instead found low dust temperatures of $\sim 30\,{\rm K}$, which importantly increases the derived dust masses (see Section~\ref{sect:dustmass}).

Several sources have observations in a large number of bands and hence have more precise \Td~(and sometimes \betad) measurements.
In~\citet{Algera24a}, data for the IR luminous galaxy REBELS-25 was fitted from Bands 3,4,5,6,8,9.
The resulting temperature of $T_{\rm d} = 32^{+9}_{-6}$ (and \betad$=2.5 \pm 0.5$) supports a lower dust temperature.
The B14-65666/Big Three Dragons source has a fitted temperature of \Td$=62_{-23}^{+55}$, where the lack of a turn-over in the FIR bands to lower wavelength results in greater errors than for REBELS-25.
A \Td$=41^{+17}_{-14}{\rm K}$ and \betad$=1.7^{+1.1}_{-0.7}$ was found for the A1689--zD1~\citep{Akins22}, consistent with the fitting for the same source in~\citet{Bakx21}.
It has observations in Bands 3,6,7,8,9, with a good constraint on the peak of the SED.
Surprisingly, two $z \sim 8$ sources show evidence for extremely high dust temperatures of $> 90\,{\rm K}$.
MACS-0416-Y1 is a lensed galaxy that has been well studied across multiple bands with ALMA and is currently the highest redshift dust detection ($z = 8.3113$;~\citealp{Bakx20}).
The measurements for galaxy A2744-YD4 ($z = 7.88 $;~\citealp{Laporte17, Hashimoto23b}) are also consistent with a high-dust temperature, however it only has $3$--$4\sigma$ detections in Band 6 and 7~\citep{Algera24}.
The compact galaxy/quasar hybrid source found at $z = 7.19$ presented in~\citet{Fujimoto22} has a measured high dust temperature (\Td$\sim 80\,{\rm K}$), as do the other quasar host galaxies presented in e.g.~\citet{Novak19} and~\citet{Venemans20}.
Note that the dust temperature when the peak is not constrained is degenerate with the emissivity index, such that a lower \betad\ results in a higher temperature (see~\citealp{Sugahara22} for an example).
Furthermore, typical fits assume optically thin dust, however if the wavelength at which the dust becomes optically thick ($\lambda_0$) approaches the expected peak of the FIR emission this can also bias the derived temperature (typically to higher values).

At slightly lower redshifts we show the stacked value from the $ z = 6$ SERENADE sample of \Td$=40.9^{+10.0}_{-9.0}$ (assuming a prior on the emissivity centered on \betad$=1.8$).
At $z \simeq 5$ we show the four sources with multi-band observations from~\citet{Faisst20a, Villanueva24}.
We show the \Td~derived for the $z = 6.327$ quasar presented in~\citet{Tripodi23} from multi-band ALMA observations.
\citet{Witstok23} fitted to a compilation of 17 FIR detected sources at $z = 4$--$7$ that have a minimum of 4 bands in the FIR. 
They derived an average \betad $=1.8 \pm 0.3$, suggesting little evolution in the effective dust properties from lower redshift (see also~\citealp{Ward24}), and a mild evolution in the dust temperature.\footnote{Note that due to requirement of being detected in multiple FIR observations, this sample is biased towards FIR bright (\lir $> 10^{12}$\lsun) sources, which potentially show higher dust temperatures than those found in lower SFR galaxies.}

There remain great uncertainties in the typical dust temperature of galaxies at $z > 6.5$, and from the data available to date we cannot discern a strong trend.
There are several extrapolations of observed dust temperature to higher redshifts that predict a rise of different slopes (e.g.~\citealp{Schreiber18, Viero22}).
The CMB temperature can become significant in the first billion years, and this leads to a heating of the dust as well as a reduced contrast between the observed continuum emission and the background~\citep{daCunha13}.
It is clear from \textbf{Figure~\ref{fig:tdust}} that the CMB temperature becomes significant at $z > 8$, where it will exceed 30{\rm K}.
Models are able to reproduce the evolution with redshift, highlighting the importance of an increasing sSFR or SFR surface density in this rise~\citep{Ma19, Liang19, Sommovigo21, Vijayan22}.
They also highlight the effects of a distribution of dust temperatures and multi-phase emission (e.g. cold vs. warm, mass or luminosity weighted; e.g.~\citealp{Lower24, Sommovigo25}).
In this work we assume an offset relation from~\citet{Sommovigo22} that intersects the measured dust temperature of the SERENADE stack from~\citet{Mitsuhashi24}.
 
\begin{figure}
    \centering
    \includegraphics[width=0.85\linewidth, trim = 0 0cm 0 2cm]{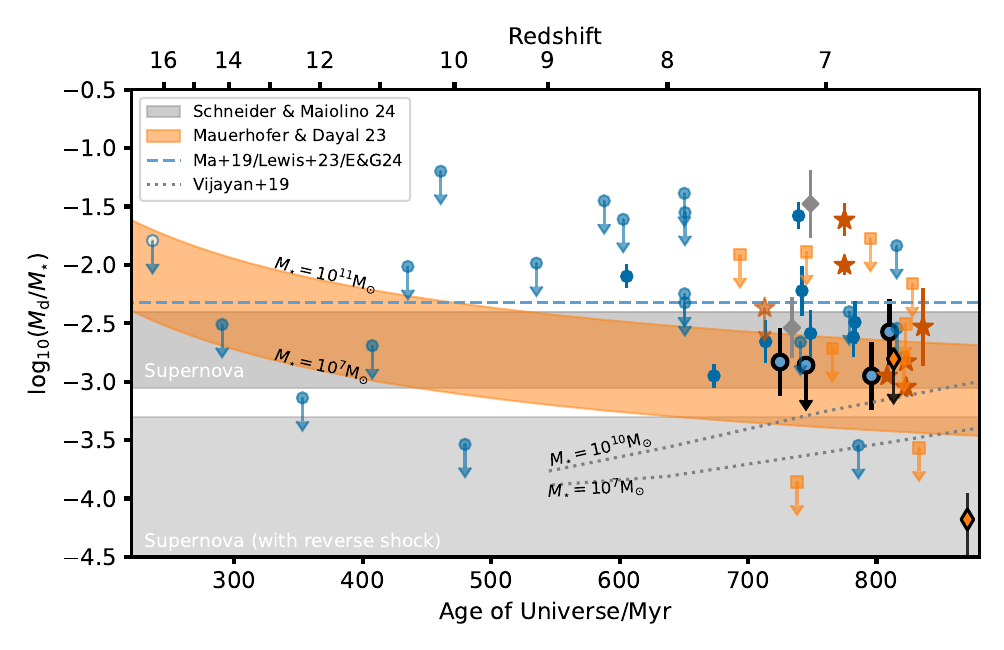}\\
    \includegraphics[width=0.85\linewidth, trim = 0 0.2cm 0 1cm]{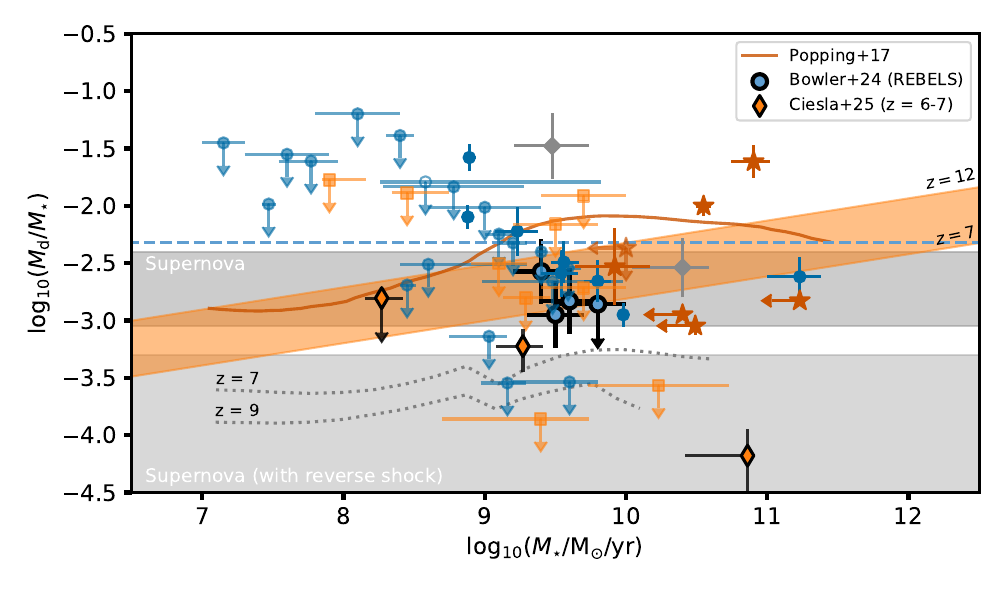}
    \caption{The ratio of the dust mass to stellar mass for the sample of $z > 6.5$ sources, with the upper plot showing this as a function of age and the lower plot as a function of stellar mass.
    Predictions of various models and simulations are shown as the different lines.
    For clarity, the majority of the REBELS galaxies (except those with new \emph{JWST} stellar masses from~\citealt{Fisher25}) have been replaced by the stacked results from~\citet{Bowler24}.
    Open symbols represent sources without spectroscopic redshifts.
    }
    \label{fig:dustmass}
\end{figure}

\subsection{Dust Mass and Dust Formation Mechanisms}\label{sect:dustmass}
The dust mass in galaxies is an important constraint for models of dust production.
The relatively short timescales available for dust formation at $ z \gtrsim 7$, as well as the different physical conditions in these galaxies (e.g. lower metallicity, compact sizes, higher sSFR), provides a key test and challenge. 
Several observational constraints have suggested that dust masses at high redshift may be considerable (e.g.~\citealp{Mancini16, Lesniewska19, Witstok22}), and potentially in excess of model predictions.
The formation of dust at these redshifts is likely to be dominated by core-collapse supernova (CCSN), which can rapidly produce dust several Myrs after the onset of star formation.
The impact of Asymptotic Giant Branch (AGB) stars is thought to be sub-dominant, due to the long main sequence lifetimes of these stars, which results in the delayed production of dust to beyond 150-400 Myrs.
Other stellar processes, such as red supergiant stars, produce yields that are too low. See~\citealp{Schneider24} for a detailed discussion of the production mechanisms.
Once created, dust grains are then reprocessed within the ISM.
In particular, the role of dust destruction from the CCSN reverse shock is thought to be significant, but is highly uncertain.
Other processes, such as astration (where dust is trapped within new stars), ejection and grain growth are also included within models.
The relative importance of these processes depends on multiple properties of the galaxy and ISM at high redshift, in particular the IMF (which controls the number of high mass stars), the stellar and gas phase metallicity (which controls stellar evolution), heating from the CMB, and the SFH.
The grain composition and size distribution, as well as the geometry, could also differ (e.g.~\citealp{Narayanan25}).

In \textbf{Figure~\ref{fig:dustmass}} we compare the ratio of the dust mass to stellar mass from our compilation to the predictions from models and simulations.
We stress here that the dust masses derived at high-redshift are subject to several strong systematic uncertainties, namely the dust absorption co-efficient, and the dust temperature (see Section~\ref{sect:dust}).
Furthermore, the stellar masses are uncertain (although to a lesser degree), and the sample selection biases (to UV-bright sources) must be considered.
Despite this, we see that the \mdsm\ for individual sources and stacks are of the same order of magnitude as simulation predictions.
In the case of a lower dust temperature (e.g. 25K for the cold component as advocated in~\citealp{Liang19}), the dust masses would be significantly higher.
Typically the semi-analytic models and simulations produce a constant ratio of \mdsm, as the dust mass is related to the CCSN rate, which is correlated with the SFR (and hence the stellar mass for a simple SFH).
For example,~\citet{Ma19}, \citet{Lewis23} and~\citet{Esmerian24} all predict an approximately constant ratio of $5 \times 10^{-3}$, with  ratios in the range $10^{-3} $--$ 10^{-2}$ common (e.g.~\citealp{Triani20, DiCesare23}).
Some models predict a slight stellar mass dependence, such as~\citet{Popping17}.
Similarly~\citet{Mauerhofer23} find a weak dependence on stellar mass, related to the higher redshift halos retaining more gas and dust at a given stellar mass, while~\citet{Vijayan19} see the reverse trend, although the scatter is significantly larger than the difference (shown in \textbf{Figure~\ref{fig:dustmass}}).
Some works argue that high stellar yields without significant grain growth can reproduce observations (e.g.~\citealt{Dayal18, Triani20, Mauerhofer23}), while others advocate for the importance of grain growth (e.g.~\citealp{Vijayan19, Esmerian24}).

Predictions for the expected \mdsm\ can be produced via a relatively simple estimate that takes into account the typical dust yield per stellar event (e.g. per CCSN; $y_{\rm SN}$) and integrates over a given SFH and IMF.
In \textbf{Figure~\ref{fig:dustmass}} we show the ranges of expected \mdsm\ from this method as presented in~\citet{Schneider24} (see also~\citealt{Lesniewska19, DiCesare23}).
The AGB predictions from this method are below the vertical range of the plot.
The upper range of these predictions, for CCSN without a reverse shock, are in good agreement with the data and the models.
However it is clear that with dust destruction, there is tension with the observations.
In particular, the SMG SPT0311-58W is in excess of the predictions by almost an order of magnitude.
We conclude that on average the observations are in reasonable agreement with the maximal predictions from models and empirical arguments, especially given the assumptions on both the derived FIR SED and dust formation mechanisms.

\begin{figure}

    \includegraphics[width=0.9\textwidth]{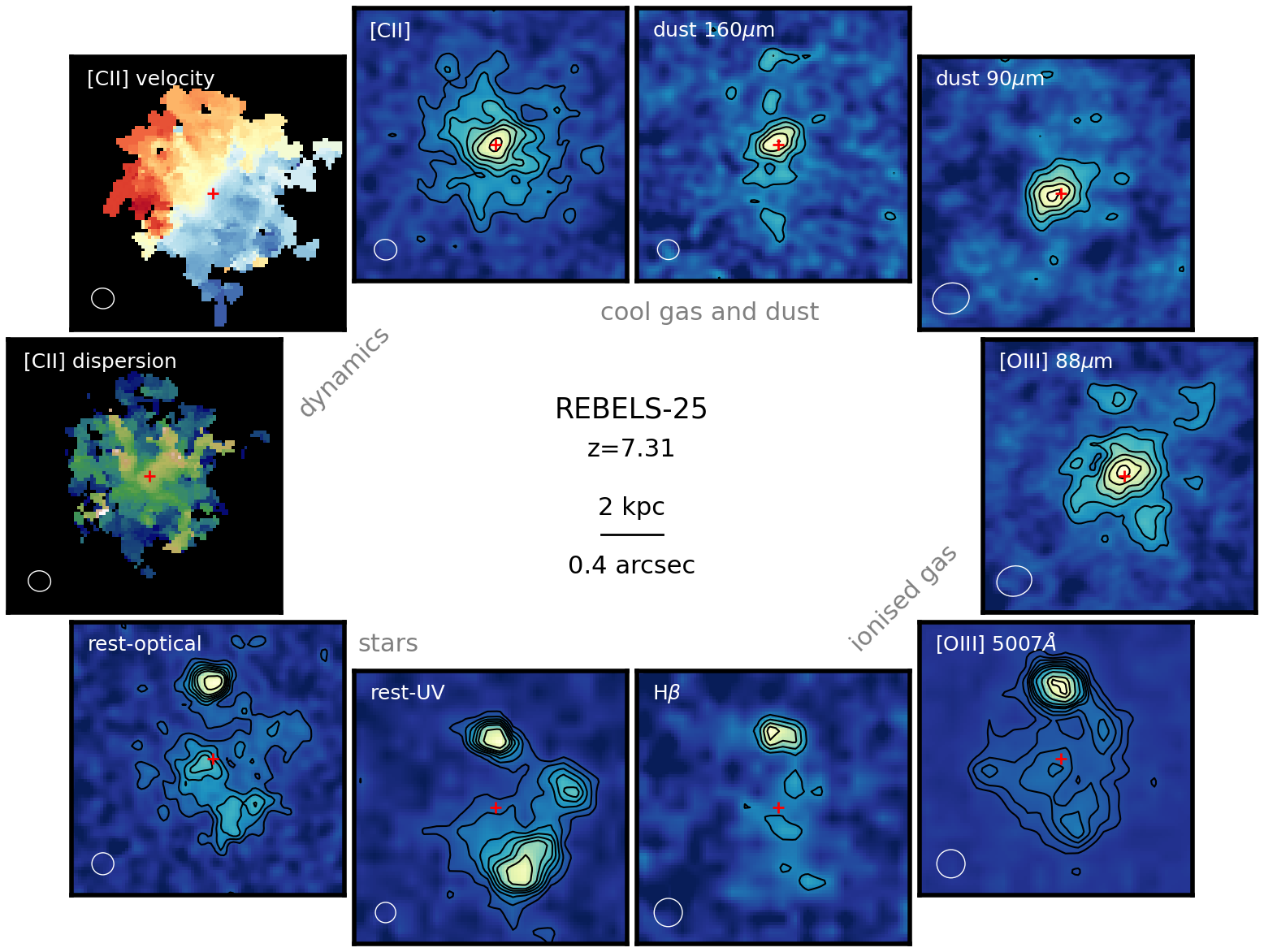}
    \caption{Compilation figure of different galaxy components in the LBG REBELS-25 ($M_\ast\sim10^{10}M_\odot$, $z=7.3$), inspired by a similar figure of a massive galaxy at $z\sim2$ in \citet{F"orsterSchreiber20}. The top six panels are obtained with ALMA observations of the cool gas (\cii), including its velocity (rotation) and dispersion (turbulence) of the gas \citep{Rowland24}, dust (continuum at 90$\mu$m and 160$\mu$m) and hot gas (\oiiilam). The bottom four panels are obtained with \emph{JWST} observations using NIRCam imaging (rest-optical stellar light) and the NIRSpec/PRISM IFU (rest-UV stellar light and hot gas seen in H$\beta$ and \oiii~5007\AA). For the panels of \cii\
dispersion, \cii\ velocity, \cii\, and dust 160$\mu$m, data are taken from \citet{Rowland24} (CC BY 4.0).
    }
    \label{fig:R25}
\end{figure}

\section{RESOLVED STUDIES}
\label{sec:resolved}
While the vast majority of ALMA observations at $z>6.5$ have been obtained in relatively compact, low angular resolution configurations, they have provided tentative new insights into the sizes, ISM complexity and kinematics of galaxies. 
Only recently have higher resolution ALMA observations been studied and one of the major challenges for ALMA in the next decades will be to increase the statistical samples of galaxies where dust and cold gas can be measured on the scales that are now available for stars and hot gas with~\emph{JWST}.

\subsection{Morphology}
Typically ALMA and NOEMA observations have been at significantly poorer spatial resolution that the available rest-frame UV and optical images. Despite this, measurements of the offset of the rest-frame FIR emission from the stellar emission and the spatial extent have been made.

\subsubsection{Offsets}
Some of the earliest ALMA observations of $z>6.5$ galaxies already suggested the possibility of offsets between the rest-frame UV and \cii\ emission \citep[e.g.][]{Maiolino15, Pentericci16} and a significant fraction of $z\sim5-6$ galaxies showed unambiguous offsets of both the continuum and \cii\ emission \citep{Capak15,Willott15}. 
In contrast, \oiii\ detections have consistently been reported with a good alignment between the UV and line emission \citep[e.g.][]{Inoue16,Witstok22}.

Given the importance of dust in the excitation of \cii\ in PDRs, some alignment of the two tracers is expected. 
UV-bright components with weak dust continuum could either be metal-poor star-forming regions with little dust build-up or regions where stellar feedback has removed the dust and gas reservoir. 
Alternatively, great differences in dust temperature within galaxies could make some star-forming regions appear to have weak dust emission in the 160$\mu$m continuum (the majority of detections), while being bright at $\lesssim90\mu$m.    

\citet{Bowler22} and \citet{Inami22} systematically studied the offsets of dust 160$\mu$m continuum in bright $z\simeq7$ galaxies.
\citet{Bowler22} identified strong UV-continuum colour gradients for galaxies with UV-dust offsets - with redder regions aligned with the dust continuum - confirming the robustness of UV-dust offsets and showing that the UV-bright clumps alone provide a biased view of the full galaxy morphology and colours. 
\citet{Inami22} found $\sim0.5-1.5"$ ($2.5-8$ kpc) offsets between the rest-UV and far-IR emission peaks in galaxies from the REBELS survey and suggest this could imply spatially decoupled phases of obscured and unobscured star formation within a single galaxy or else merging systems. 
Similar results were found for individual sources (e.g.~\citealp{Hashimoto19, Matthee19, Schouws22, Tamura23}).

With the launch of~\emph{JWST} and in particular the NIRSpec/IFU mode, more detailed morphological~\emph{JWST}-ALMA comparisons have become available \citep[e.g.][]{Hashimoto23b, Arribas24,Jones24,Scholtz25}.
As can be seen in \textbf{Figure~\ref{fig:R25}}, dust obscuration can introduce significant offsets ($\sim0.3-0.4"$) between the ALMA and \emph{JWST} line and continuum tracers - even for emission from the the same ion (\oiiilam\ and \oiii~5007\AA). In particular the rest-frame UV emission is concentrated in a number of unobscured clumps, while the rest-optical emission lines are highly biased towards a bursty UV-bright region. The gas traced by \ciilam\ and \oiiilam\, in contrast, shows no evidence for clumpiness related to the rest-UV or optical clumps.

\subsubsection{Sizes}

Possibly linked to the observed offsets of \cii\ and dust continuum is the difference in size measurements between \cii\ and UV stellar continuum. \citet{Fudamoto22} stacked low-resolution observations of star-forming galaxies at $z\sim4-7$ and found a typical $\gtrsim2\times$ larger half-light radius of \cii\ compared to the stacked UV stellar light. They suggest that main-sequence star-forming galaxies at $z\sim4-7$ are morphologically dominated by gas rather than stars. 
Alternatively, if \cii\ traces many different phases of the ISM, both obscured and unobscured star-forming regions could be \cii\ emitting, naturally increasing the size of the \cii\ emitting regions in case of observed offsets between unobscured UV clumps and dust-continuum regions.   

\citet{Mitsuhashi24} use spatially resolved observations of LAEs and LBGs at $z=4-6$ to show that typical main-sequence star-forming galaxies have dust reservoirs $\sim2\times$ more extended than their UV stellar counterparts. 
This is in strong contrast to SMGs and DSFGs at the same redshift, which typically have very compact, central dust clumps, with UV emission emerging at greater radii \citep[][and references therein]{Hodge20}. 
\citet{Ikeda25} report that the \cii\ sizes of the same galaxies are more extended than the dust in roughly half the sample. 
They conclude that most of the \cii\ extent can be explained by the extended emission from dusty regions, however, emission from extended neutral gas such as seen in local galaxies could furthermore increase the \cii\ sizes. 

\subsubsection{Halos}
\label{sec:halos}
The spatial extent of \cii\ with respect to the UV stellar light might be conflated by a gas `halo' around the stellar disk and for the purposes of this review we only use the term `halo' when gas is found on scales of the CGM, e.g. $\gtrsim 10\,\rm kpc$, which can be separated from the ISM-scale emission of \cii. 
Using stacking of low-resolution observations~\citet{Fujimoto19} and~\citet{Ginolfi20} found indications of \cii\ halos in $z=5-7$ and high-SFR $z=4-6$ galaxies respectively. 
While such a component could be seen as evidence for cold outflows, higher resolution data is required to rule out multiple close components due to minor and major mergers. 
\citet{Ikeda25} studied both individual and stacked high-resolution \cii\ emission of $z=4-6$ galaxies and found no convincing evidence for \cii\ halos in either analysis. 

In contrast, \citet{Akins22} studied high-resolution \cii\ and \oiii\ emission in a $z=7.13$ lensed galaxy and found clear evidence for a 2-component system in \cii\, with a $\sim12\,\rm kpc$ component that is not seen in \oiii.  They argue outflows might be explain the origin of the observed halo. \citet{Novak20} stacked high-redshift quasar host galaxies at $z\gtrsim6$ and found a 2-component profile with an extended $\sim10\,\rm kpc$ component in both the dust and \cii\ and argue that rather than outflows, the extended \cii\ component is still tracing the ISM of these extended systems. 
More high-resolution, high S/N observations will be required to settle this discussion and obtain critical insight into the typical CGM conditions of high-redshift galaxies.

\subsection{Kinematics}\label{sect:kinematics}
The spectral resolution provided by ALMA and NOEMA is typically higher than that obtained with rest-frame UV and optical spectroscopy with e.g. JWST.  
This allows a sensitive probe of the velocity structure of the gas by measuring the peak and shape of FIR emission lines.

\subsubsection{Disks and Mergers}
\label{sec:disks}
Velocity gradients were first detected in low angular resolution observations, where disks and mergers are difficult to distinguish. \citet{Smit18} argued that smooth velocity gradients, combined with compact UV stellar morphologies, in two LBGs at $z\sim7$ could be evidence of rotation. In contrast, \citet{Hashimoto19} found a velocity gradient in a $z \sim 7$ LBG that aligned with the two clump morphology in the UV light and argued a major merger. 

\citet{Posses23} performed the first high-resolution \cii\ analysis of an LBG at $z>6.5$ and found evidence for a disk (see also~\citealp{Parlanti23}), accompanied by a minor merger. 
Furthermore, \citet{Rowland24} present a kinematically cold gas disk (v/$\sigma\sim10$) seen in \cii\ at $z=7.3$ in the most FIR-bright LBG from the REBELS survey (see \textbf{Figure \ref{fig:R25}}). The disk dispersion is $\sim3\times$ ($\sigma\sim30\,\rm km/s$) higher than observed at $z\sim0$ in H{\sc i}/H$_2$ gas surveys. Such cold gas kinematics are unexpected when extrapolating dynamical measurements at $z=0-3$ using hot gas tracers, where turbulence increases with redshift and the resulting rotational support decreases \citep{"Ubler19}. However, this discrepancy might be explained by the different gas phases traced in these studies, where \cii\ is more likely to trace the thin, cold gas disks of galaxies, while hot gas is turbulent due to the impact of stellar feedback \citep[e.g.][]{Rizzo21}. 
This is tentatively seen in the chaotic motions traced by \oiiilam\ emission in a lensed LBG at $z=8.3$ \citep{Tamura23}, in contrast to the smooth velocity gradient seen in \ciilam\ emission \citep{Bakx20}. 
In contrast to the examples of possible disk galaxies at $z>6.5$, \citet{Spilker22} present $\sim100$--$350$ pc scale \cii\ gas kinematics of a lensed SMG at $z=6.9$ and find chaotic motions  with highly turbulent clumps ($\sigma\sim70$--$120\,\rm km/s$), suggesting both mergers and disk fragmentation could play a role in the bursty nature of this extreme system.    

\citet{Wang24a} present a survey of resolved \cii\ emission in 31 $z\sim7$ Quasar host galaxies and find a diversity of kinematics with $\sim40\rm \%$ having smooth velocity gradients, consistent with rotation, while the rest showed dispersion dominated, compact or disturbed kinematics. This is a higher number of rotators than found in low-angular resolution observations of \cii\ in  LAEs and LBGs at $z\sim4-6$, where $\sim40\rm \%$ is classified as mergers and only $\sim13\rm \%$ as possible rotating disks (Le F\`{e}vre et al. 2020; Jones et al. 2021). However, \citet{Lee25} find a $50\pm9\%$ disk fraction using the high resolution \cii\ observations in a subset of these $z\sim4-6$ galaxies from the CRISTAL survey. These differences suggest that resolved observations are key to the identification of disks, while only modest variations in disk fraction arise due to the different selection criteria of the samples. 

\subsubsection{Outflows}
Similar to the search for \cii\ halos, various studies have tried to measure the presence of outflows from the broad, high-velocity components in the integrated and spatially resolved spectra of high-redshift galaxies. \citet{Ginolfi20} stacked integrated spectra of \cii\ in galaxies at $z=4-6$ and found evidence for a broad component in galaxies with SFR$>25\,M_\odot\,\rm yr^{-1}$. In contrast,  \citet{Novak20} stack high-redshift quasar host galaxies at $z\gtrsim6$ and find no evidence for a broad component

In individual galaxies \citet{Schouws23}, \citet{Hygate23} and \citet{Fudamoto24} report broad components in low angular resolution observations at $z=6.7, z=7.3$ and $z=6.2$ respectively, and \citet{HerreraCamus21} find evidence for an outflow in resolved observations of a $z=5.5$ galaxy even if the signature is lost in the integrated spectrum. \citet{Hygate23} and \citet{HerreraCamus21} find mass-loading factors $\gtrsim0.5-1$, consistent with local relations and predictions from star-formation-driven outflows.

\citet{Akins22} report a broad \oiii\ component found in a brightly lensed LBG at $z=7.1$. No such broad component is seen in the \cii\ gas and there is no indication of an extended \oiii\ `halo' (as seen in \cii, see Section \ref{sec:halos}), demonstrating the complexity of hot and cold gas outflows in ALMA observations.

\section{DISCUSSION AND FUTURE DIRECTIONS}\label{sec:discussion}
In Section \ref{sec:background} we outlined the underlying changes within the cosmic web in the expanding Universe from $z\sim0$ to $z\sim7$, and defined several fundamental questions that we can ask about the ISM of early galaxies. 
Observationally, ALMA is finding some trends that are broadly consistent with this expected physical picture. 
For example, line ratios at $z>6.5$ indicate higher density ($\sim10-1000\times$) neutral/PDR gas compared to typical $z\sim0$ galaxies (Section \ref{sec:OI}).   
Moreover, even in the coldest disks found to date at $z>6.5$, turbulence is $\sim3\times$ higher than observed in the local Universe (Section \ref{sec:disks}). 
Despite these changes, the relation between the gas traced by \ciilam\ and the SFR stays roughly constant over 13 billion years of cosmic time, suggesting the basic scaling relations fundamental to the star-formation process could already be in place in the EoR (Section \ref{sec:CII}).   

Arguably some of the most notable surprises are the evidence for very early metal build up out to $z\sim14$ (Section \ref{sec:scanning} and \ref{sec:OIII-SFR}) and the rapid dust build-up given the modest contribution of AGB stars at early times (Section \ref{sect:dustmass}).
Intriguingly, scaling relations suggest that dust properties appear to evolve little from $z \sim 7$ to today, although the degree of dust obscured star-formation is lower at high redshift (Section~\ref{sect:fobs} and~\ref{sect:sfrd}). 

ALMA is only starting to scratch the surface when it comes to constraining the ISM properties that will allow ionising photons to escape into the IGM. However, interesting results have emerged that suggest that even though most galaxies studied with ALMA to date have modest obscuration fraction (Section \ref{sect:fobs}), ALMA traces diffuse highly ionised gas outside of the dense star-forming regions (Section \ref{sec:OIII-SFR}), possibly suggesting a lower covering factor of the PDRs that allows ionising photons to escape through low density channels (Section \ref{sec:OIII/CII}).

In the next decade ALMA has mapped out high-priority upgrades to the telescope in the ALMA2030 Development Roadmap. ALMA's frequency coverage will be completed with the installation of ALMA band 2 (possibly starting observations October 2026), which increases the number of CO transitions that can be accessed at high redshift (see \textbf{Figure \ref{fig:ALMAbands}}). Top priority for ALMA2030 is the Wideband Sensitivity Upgrade (WSU), that will already be implemented in the band 2 receivers and will cover $4\times$ larger instantaneous bandwidth than is currently available in band 1 and 3. Under development are upgrades to band 6 and 8, which in particular will increase the continuum imaging speed by at least a factor of 3, and $2\times$ larger instantaneous bandwidth for line-scanning purposes. These upgrades guarantee ALMA will continue  cutting edge discoveries in high-redshift galaxy formation beyond 2030.

\begin{summary}[SUMMARY POINTS]
\begin{enumerate}
    \item Over the last decade, since ALMA has come online, FIR dust continuum has been detected out to $z = 8.3$ and cold gas through \ciilam\ out to $z=9.1$ and hot gas through \oiiilam\ out to $z = 14.2$. ALMA studies above $z = 6.5$ are predominantly targeting luminous sources ($L>L^\ast$) and therefore massive systems or small samples of lensed galaxies, due to sensitivity requirements.
    \item \ciilam\ correlates well with various cold gas tracers. Meanwhile the $z= 0$ SFR-\cii\ relation is already broadly in place at $z\simeq7$,
     suggesting the integrated Kennicutt-Schmidt relation might already be in place and star-formation efficiency remains roughly constant from the first billion years to the present day.
    \item The high \oiii/\cii\ values compared to the local general galaxy population are linked to burstiness of the SFH. Elevated \oiii/SFR values suggest relatively high metal enrichment in bright galaxies out to $z>11$, possibly in combination with a low covering factor of the PDR, such that star-forming regions leak ionising photons out to great distances. 
    \item Dust emission is more common in massive galaxies, as revealed by the trend in $f_{\rm obs}$ with stellar mass from fitting to the rest-optical.  Modeling the distribution functions in the FIR lead to estimates of the cosmic star-formation rate density, which shows that $> 10$ percent of the SFR density at $z \simeq 7$ is obscured.  
    \item Scaling relations such as \irxb\ appear consistent with local relations.  Dust temperature shows a large scatter, but the average temperature appears to be moderately elevated as compared to the local universe (\Td$\sim 45\,{\rm K}$).
    \item Dust masses compared to the stellar mass suggest that rapid dust enrichment occurs, likely from supernova with little dust destruction.
    \item In individual sources cold disks are already in existence at $z = 7$, with their gas disks being more extended and smooth than their rest-frame UV
    and optical counterparts.  
\end{enumerate}
\end{summary}\label{box:summary}

\begin{issues}[FUTURE ISSUES]
\begin{enumerate}
\item ALMA will continue to push the redshift frontier in terms of spectroscopic line confirmations, given that bright rest-frame optical lines move out of the JWST/NIRSpec wavelength coverage above $z>12$, while bright lines such as \oiiilam, \ciilam\ and \oi\,63$\mu$m either stay visible in high-transmission bands of ALMA or move into them at these redshifts. 
\item With an increase in bright galaxies confirmed at the highest redshifts, new lines become accessible that will constrain ISM properties, such as \oiii~52, \oi~63 and [N{\sc iii}]~57, probing ionized and neutral gas density and abundance patterns.
\item ALMA upgrades will dramatically improve line-scanning and enable dust detections or upper limits at increasingly high redshift, constraining the timescale available for dust grain production.
\item The FIR SED will continue to be refined with multi-band ALMA observations, constraining the dust temperature and emissivity and hence the IR luminosity and dust mass.  In parallel, observations with~\emph{JWST} will determine the dust properties via measurements of the attenuation curve, dust bump and Balmer ratios.
\item With new telescopes like~\emph{Euclid} and~\emph{Roman}, the samples of massive galaxies known at $z>6.5$ will increase by factors of $>1000$, with the low mass end probed by rare strongly lensed sources.  This will provide ideal targets for resolved studies and kinematics, for statistical samples of sources within the EoR.
\end{enumerate}
\end{issues}

\section*{DISCLOSURE STATEMENT}
The authors are not aware of any affiliations, memberships, funding, or financial holdings that
might be perceived as affecting the objectivity of this review. 

\section*{ACKNOWLEDGMENTS}
 RB and RS acknowledge support from STFC Ernest Rutherford Fellowships [grant numbers ST/T003596/1 and ST/S004831/1]. 
 We thank Sander Schouws, Nimisha Kumari, Hiddo Algera, Lucy Rowland, Katherine Ormerod, Matus Rybak, Massimo Stiavelli, Takahiro Morishita, Maru\v{s}a Brada\v{c}, Cosimo Marconcini, Miguel Pereira Santaella, Rychard Bouwens, Mauro Stefanon, and Robert Crain for providing data and/or feedback on the manuscript.

\bibliographystyle{ar-style2_overleaf}
\bibliography{bibtex_parsedARAA}



\section*{Supplementary material}

## Contents



# 1. Observational methods

## 1.1. Concepts of (Sub)Millimetre Line Observations

**1.1.1. Excitation mechanisms and the photo-electric effect.** Due to the variety of gas phases that ALMA accesses at high-redshift, multiple excitation mechanisms are important. In ionised gas collisionally excited lines (CELs; such as [OIII]) and recombination lines (such as H$\alpha$) are both key coolants of the gas. ALMA can only observe collisionally excited fine structure lines that emerge from the splitting of the lowest energy level due to spin-orbital interaction, where hot electrons in the ionised gas are the main collisional partner. A key characteristic of fine structure lines is the small energy difference between the excited level and ground level, such that their excitation potential ($< 500\,$K; see Table 1 in the main text) is far below the typical gas temperatures of HII regions ($\sim 8000 - 20,000\,$K). This results in a nearly constant line emissivity with gas temperature. In comparison, ionised lines observed in the rest-frame UV or optical have excitation temperatures above that of typical star-forming regions ($> 20,000\,$K) and therefore show strong temperature dependence in their emissivity.

For PDR and neutral gas, collisionally excited fine structure lines are still the main coolant, however, the main collisional partner becomes HI (or even $H_2$). The key process that heats the gas to have enough kinetic energy for efficient collisional excitation is the photo-electric effect; the ejection of high-energy electrons from small dust grains and PAHs due to absorption of FUV photons penetrating the PDR. Efficient emission of [CII] is therefore dependent on the presence of dust and the dust composition, as well as the radiation from young OB stars.

The most important lines from the molecular gas phase that ALMA can access at high-redshift are rotational excited states of CO. Lower excited states (e.g. $J = 3 - 2$), tracing relatively cold molecular gas (excitation potential 33K), have recently become accessible in Band 1 at $z \sim 7$. However, higher excited lines ($J = 6 - 5$, $J = 7 - 6$), which have higher excitation potential and critical density (see section 1.1.2), can be brighter in warm and/or very dense gas and can be targeted with Band 3.

**1.1.2. Critical density.** Critical density, $n_{\rm crit}$, is the density at which the rate of collisional de-excitations is equal to the rate of spontaneous radiative de-excitations. For $n << n_{\rm crit}$, the cooling rate per volume and the line emissivity scale as $n^2$, while for $n >> n_{\rm crit}$ the cooling rate is linear with $n$.

For low-density cold molecular gas, the CMB is an important source of excitation and

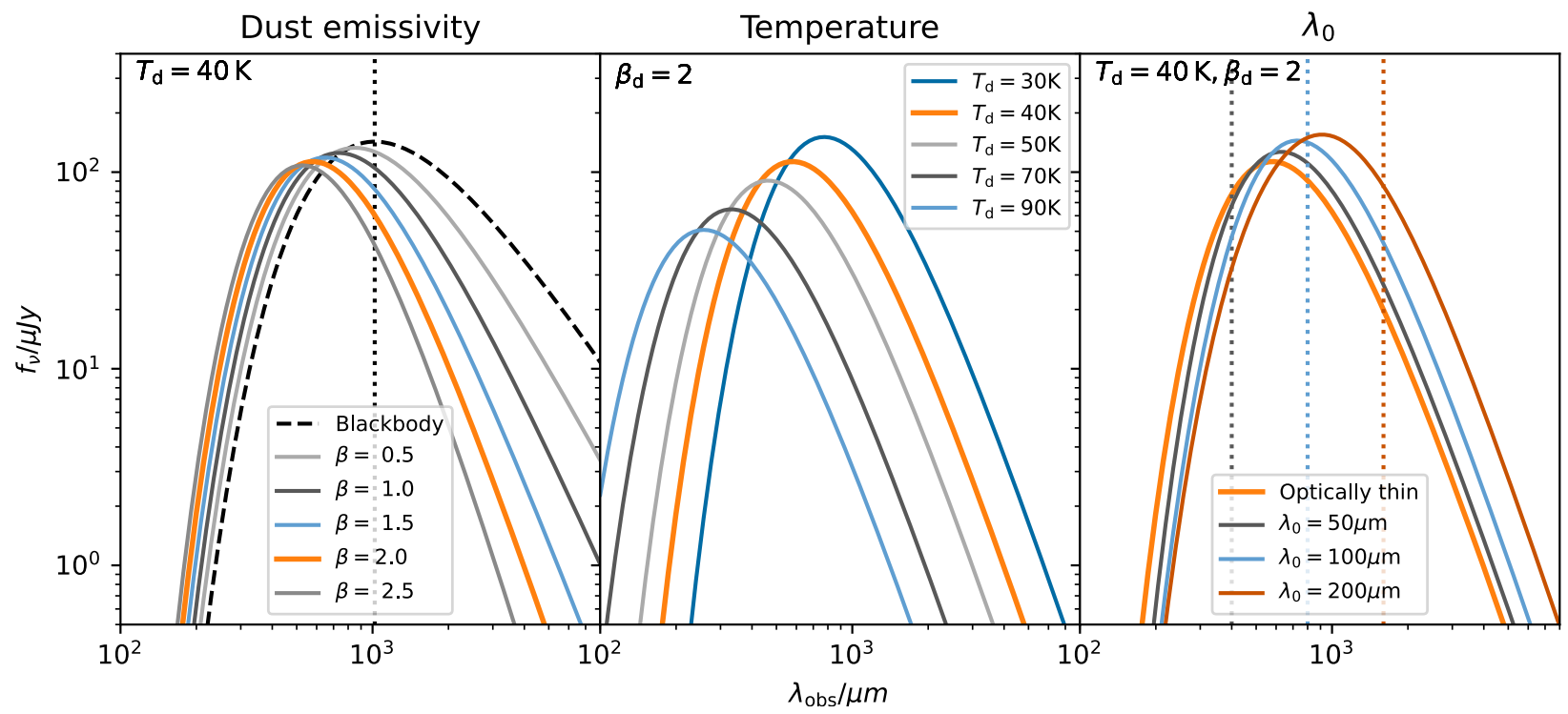


Supplemental Figure 1: Diagram illustrating the modified blackbody functions used to derive an IR luminosity for a galaxy at $z = 7$. There are three parameters that can be varied, and in each case we fix the other parameters to those shown in the figure. Each curve corresponds to a $L_{\mathrm{IR}} = 10^{11}\,\mathrm{L}_\odot$. **Left:** The dust emissivity ($\beta_\mathrm{d}$ changes the slope of the function at long wavelengths. **Middle:** The dust temperature ($T_\mathrm{d}$) changes the peak of the function. **Right:** The wavelength at which the optical depth is unity ($\lambda_0$) changes the peak.

stimulated emission (particularly import at high-redshift where the CMB temperature rises; da Cunha et al. 2013). Gas below $n << n_{\mathrm{crit}}$ corresponds to the regime where collisional excitation is subdominant to radiative excitation and a molecular line becomes "invisible" against the CMB background. As such, $n_{\mathrm{crit}}$ is often used in the sub-millimetre literature as a minimum density for line visibility, and conversely, the detection of a line is used as a rough density diagnostic. When $n >> n_{\mathrm{crit}}$, and collisional excitation dominates, a line is said to be "thermalized" (Carilli & Walter 2013).

In HII regions, some of the brightest lines (forbidden lines such as [OIII] $\lambda$5007Å) have critical densities far above the density of the gas that emits the line ($n << n_{\mathrm{crit}}$). However, a few lines (e.g. [O III] $\lambda 88\mu$m) have lower critical densities and in the density regime between the critical densities of two lines their line ratios are used as a density diagnostic. Furthermore, given that the line luminosity normalised by the ionising photon production or SFR declines with larger $n$ above the critical density due to collisional de-excitations, a line is sometimes called "quenched" when $n > n_{\mathrm{crit}}$ in the optical/near-infrared literature. While both uses of $n_{\mathrm{crit}}$ have their reasons, their contrasting nomenclature can be a source of confusion.

**1.1.3. Ionization Parameter and Habing field.** For an HII nebula within a star-forming region, the ionization parameter, $q$ or $U$ ($U = q/c$) - a measure of the local ionizing photon flux ($h\nu > 13.6\,\mathrm{eV}$) to the local Hydrogen density ($n_H \sim n_e$) - is essential to understand the ionization structure of the nebula (Kewley, Nicholls & Sutherland 2019). Outside of the Strömgren sphere where the PDR begins, the radiative properties of the gas are similarly determined by the FUV ($6\,\mathrm{eV} < h\nu < 13.6\,\mathrm{eV}$) radiation field and the gas density; $G_0/n_H$ (Wolfire, Vallini & Chevance 2022). The Habing field $G_0$ is traditionally used to indicate the FUV photon flux, where 1 unit Habing is $10^8$ photons $\mathrm{cm}^{-2}\,\mathrm{s}^{-1}$ or $1.6 \cdot 10^{-3}\,\mathrm{erg\,s}^{-1}\,\mathrm{cm}^{-2}$.

The radiation field in the ionized and PDR gas are related to each other for a given

initial mass function (IMF) and star-formation history. Abel et al. (2009) estimate $\log U \sim \log G_0 - 6$ for a model that assumes pressure equilibrium, a 4Myr continuous star-formation history and a Salpeter IMF. However, ionizing radiation is only produced by the most massive, most short-lived stars, while FUV emission also emerges from longer lived stars. Any burstiness of star-formation can therefore introduce large differences between the ionised and PDR gas properties. When AGN or X-ray binaries are nearby, X-rays control the chemical and thermal structure of the neutral/molecular gas (X-ray dissociation region; XDR), which introduces subtle differences in the predicted emission line ratios (Wolfire, Vallini & Chevance 2022, and references therein).

## 1.2. Grey body models

In Supplemental Figure 1 we illustrate the changes in the FIR SED when each of the three main parameters are altered. These are the dust temperature, $T_{\rm d}$, the emissivity index $\beta_{\rm d}$, and the wavelength below which we are optically thick, $\lambda_0$.

## 1.3. Interferometry

The ALMA large array consisting of 12 meter diameter antennas (the '12m array') typically carries out observations with 40-50 antennas (the exact number varies between cycles and observing blocks) and as every antenna is correlated with every other antenna in the array, the effective collecting area of ALMA is divided over ~1000 baselines ($N \times (N-1)/2$ for $N$ antennas). For each individual baseline an object on the sky can only be detected when it is smaller than the resolution determined by the baseline length ($B$); $\theta \sim \lambda/B$ (with $\theta$ in radians and assuming a source close to zenith). Indeed, an interferometer measures structural information by providing a range of baselines, some of which 'resolve out' the structure ($\theta_{\rm source} > \lambda/B$) and provide only non-detections, while simultaneously more compact baselines ($\theta_{\rm source} < \lambda/B$) measure the source flux. Supplemental Figure 3 gives an overview of the distribution of baselines for each ALMA configuration.

**1.3.1. UV-plane.** Interferometry signals are recorded in the so-called 'UV-plane' or Fourier space; with the complex visibilities ($V = {\rm real} + i \cdot {\rm imag}$) consisting of an amplitude ($A$; providing flux information) and a phase (providing positional information); $V = A \cdot e^{(i*{\rm phase})}$ and $A = \sqrt{{\rm real}^2 + {\rm imag}^2}$. The UV plane distance ($\sqrt{u^2+v^2}$ or simplified $u$) is linked to the projected baseline length ($B_\perp = B \cdot \sin\phi$, with $\phi$ being the angle between the baseline vector and the source on the sky) as $u = B_\perp/\lambda$. As a result, a point source provides a constant signal for all baselines, whereas a uniform signal on the sky is fully resolved out.

Approximating a typical high-redshift galaxy as a 2D Gaussian signal, which is then recorded in the UV plane as a 2D Gaussian, with standard deviation

$$\sigma_{\rm uv} = \frac{1}{2\pi \cdot \sigma_{\rm im}}, \quad 1.$$

where $\sigma_{\rm uv}$ is the UV-plane distance in units of $\lambda$ and $\sigma_{\rm im}$ is in radians. The conversion of the image-plane FWHM to the UV plane FWHM is then given by

$$\mathrm{FWHM}_{\rm uv} = \frac{8\ln 2}{2\pi \cdot \mathrm{FWHM}_{\rm im}} \frac{3600 \cdot 180}{\pi} \frac{1}{1000} = \frac{2592 \ln 2}{\pi^2 \cdot \mathrm{FWHM}_{\rm im}}, \quad 2.$$

where $\mathrm{FWHM}_{\rm uv}$ is UV-plane distance in units of k$\lambda$ and $\mathrm{FWHM}_{\rm im}$ in arcseconds.

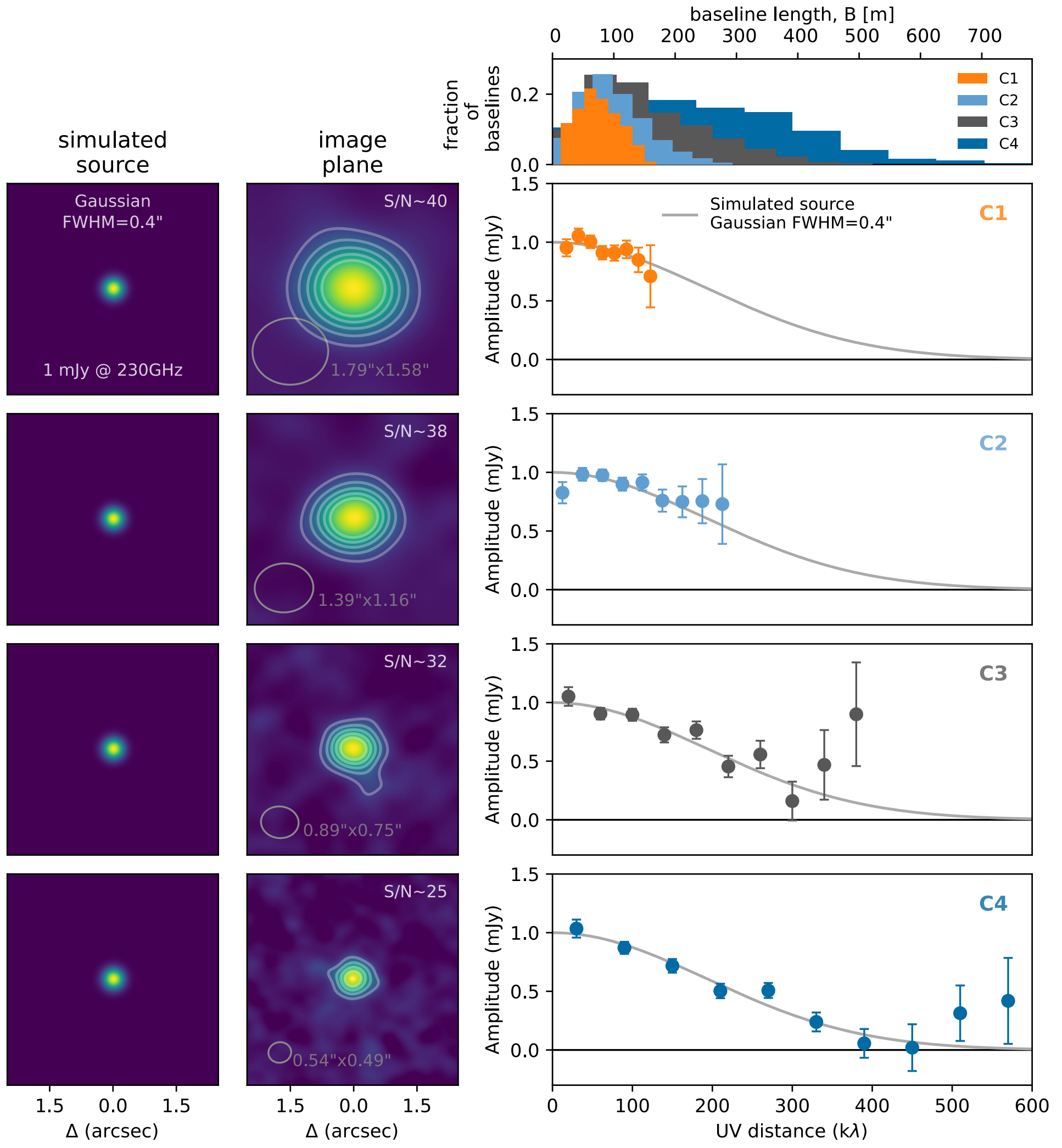


Supplemental Figure 2: Simulated Gaussian source with $R_e \sim$ 1kpc at $z \sim 7$ (FWHM=0.4") at 230GHz (left columns). We use the four most compact configurations (C1-C4; tabulated Synthesized Beams of 1.47″-0.40″ at 230GHz, but imaged with a larger Beam using natural weighting) to simulate the observations in the image plane, at fixed rms/Beam (middle panels). The peak S/N starts at 40 for C1 and goes down by $\sim$ 5%, 20% and 40% for C2, C3 and C4 respectively. White contours are drawn at 5, 10, 15, 20 and 25$\sigma$ and the beam size is indicated in the bottom left corner of each panel. The changing S/N can most easily be understood by comparing the baseline distance distribution of each configuration with the size of the source in the UV plane (right column). Grey lines show the intrinsic signal and filled points the binned mock observations, while the top panel shows the fraction of baselines at each baseline distance.

Arguably a better description of high-redshift star-forming galaxies is that of an exponential disk, or Sérsic profile with $n = 1$. The fitting function in the UV-plane is given by

$$V(u) = \left( \frac{S}{1 + 4\pi^2 \cdot R_d^2 \cdot u^2} \right)^{3/2}, \qquad 3.$$

where $u$ is the UV-plane distance in $\lambda$, S is the total flux in the image plane and $R_d = R_{1/2}/1.6785$ is the image plane size in radians.

**1.3.2. ALMA configurations.** How ALMA 'sees' a source in the UV plane depends on the chosen configuration, given that UV distance equals $\sim B/\lambda$. The ALMA array configuration changes throughout the year (currently from configuration C1 to C10) and the baselines cover a range from 15 meter to 16.2 kilometers, corresponding to 0.042-3.19" at 100 GHz (Band 3) and 0.0048-0.69" at 460 GHz (Band 8). In Supplemental Figure 2 we simulate a simple circularly symmetric Gaussian source with FWHM=0.4" that could reflect a luminous $z \sim 7$ LBG with $R_e \sim$ 1kpc. We use the ALMA simulator to observe this source in four configurations, C1-C4, with the same added noise level. The signal in the UV-plane is not affected by a PSF (see right most panels of Supplemental Figure 2), however, with the Fourier transform to the image plane, the PSF or 'Beam' emerges from the finite sampling of the UV plane by the baseline distribution (see middle panels of Supplemental Figure 2).

High-redshift targets are typically faint and using even a small subset of ALMAs collecting area to obtain structural information comes at a significant observational cost. In Supplemental Figure 2, the effect of the changing configuration from C1 to C4 changes the Beam or PSF from ~1.7" to ~0.5", while the S/N goes down from S/N=40 for C1 by ~ 5%, 20% and 40% for C2, C3 and C4 respectively. In the UV plane configuration C4, which has $\mathrm{FWHM_{source}} \sim \mathrm{FWHM_{beam}}$, provides a reasonable size measurement given sufficient S/N, however, from the perspective of a detection experiment the size of the galaxy acts as an effective 'transmission curve' for the incoming signal, with a significant fraction of baselines only recording <50% of the maximum amplitude. For the most compact configuration, C1, all structural information is lost, however, all baselines detect a signal close to the maximum amplitude. Additionally, the choice of configuration is a consideration when matching multi-wavelength observations. According to the ALMA Technical handbook the nominal rms absolute positional accuracy, $\sigma_{\mathrm{posacc}}$, is approximated by

$$\sigma_{\mathrm{posacc}} = \mathrm{FWHM_{beam}}/(0.9 \cdot \mathrm{S/N_{peak}}). \qquad 4.$$

Therefore, the simulated C4 configuration in Supplemental Figure 2 improves by a factor of $\sim 2\times$ in positional accuracy over C1.

**1.3.3. Beam and PSF.** The Beam that emerges from the transform of the UV plane signal to the image plane is significantly different from PSFs used in typical optical and near-infrared observations, subsequently affecting the data-processing. The aperture diffraction pattern of a single dish can never have negative minima, allowing for a simple estimate of the encircled energy at a given radius and correction for the missed light from the wings of the PSF for a total flux estimate. In contrast, the synthesized Beam that emerges after correlating antennas in an interferometer, the so-called "dirty Beam", includes both positive and negative sidelobes that significantly distort the appearance in the "dirty image" (the real image convolved with the dirty Beam) of a source in the image plane. To correct

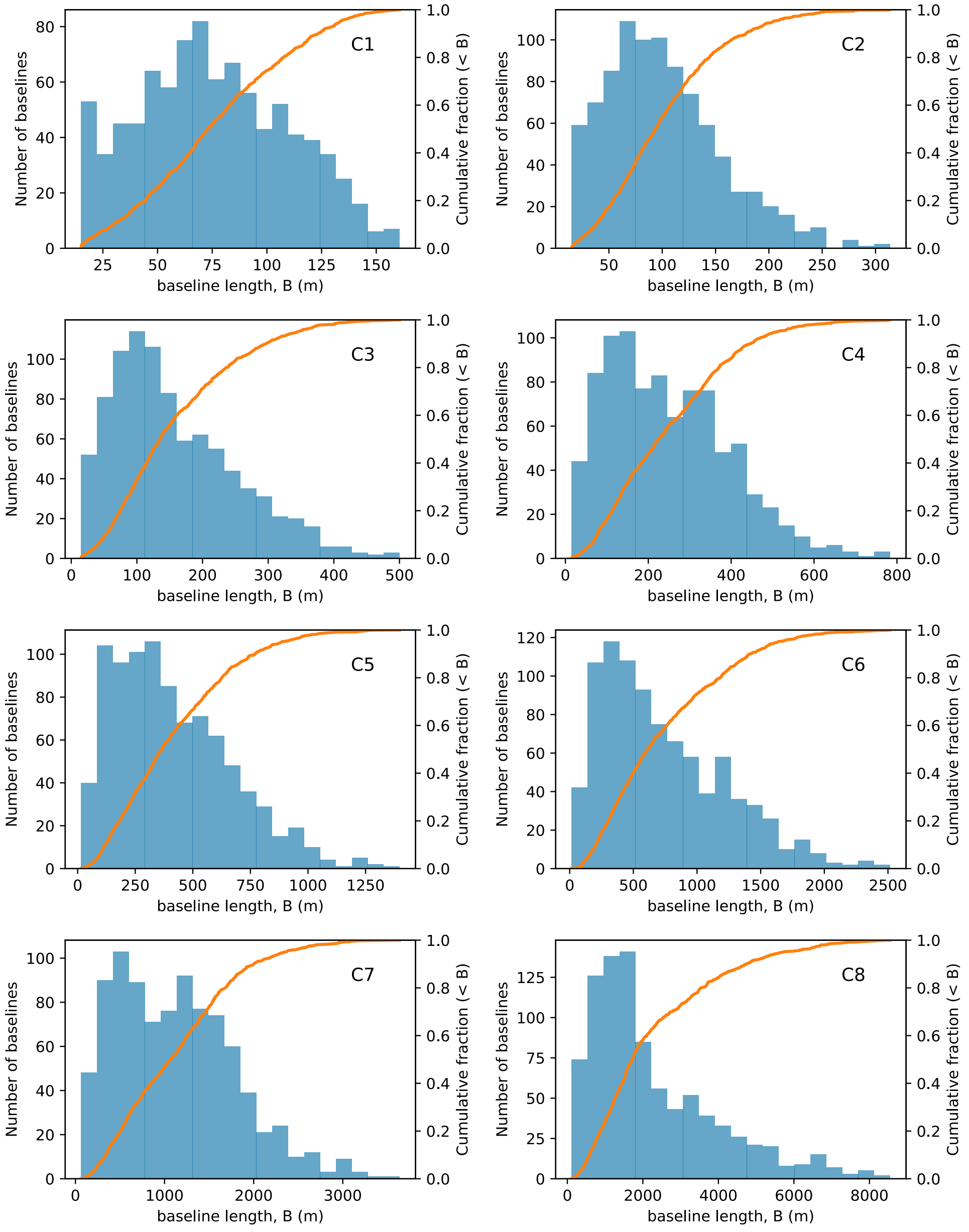


Supplemental Figure 3: Baseline distributions of ALMA configurations C1-C8 in Cycle 11. The cumulative fraction of baselines (right axes) provides a quick estimate of the collecting area of ALMA that will observe a source close to zenith of given size $\theta_{\rm source}$ (in radians). Any single baseline will resolve out a source $\theta_{\rm source} > \lambda/B$ (see text).

for these sidelobes, a process of 'cleaning' is required to obtain a modeled image of the source as it would appear with a perfect 2D Gaussian PSF - the so-called "clean Beam". Furthermore, brightness units in interferometry are typically given in Jy/Beam. The *dirty* Beam is normalised to unity at the peak and therefore an unresolved source in the dirty image has a peak flux in Jy/dirty Beam equal to the integrated flux of the source in Jy. The *cleaned* Beam can be compared to a normal PSF, therefore in the cleaned image the Jy/clean Beam units can be converted to Jy/pixel by dividing by

$$(\pi \cdot \mathrm{BMAJ} \cdot \mathrm{BMIN})/(4 \ln 2 \cdot d_{\mathrm{pix}}^2), \qquad 5.$$

where BMAJ and BMIN are the FWHM of the beam along the major and minor axis, while $d_{\mathrm{pix}}$ is the pixel size in the same angular units.

**1.3.4. Imaging process.** In the imaging process different weightings can furthermore be given to shorter or longer baselines, such that the image plane Beam or PSF changes and brings out diffuse or compact components respectively. So-called 'natural' weighting optimizes S/N by giving equal weight to all baselines and is the default for high-redshift observations. So-called 'Briggs' weighting emphasizes longer baselines, scaling with the Robust parameter to increase the image-plane angular resolution (ranging from 2 to -2, with Robust=2 being equal to natural weighting). On the other hand, adding a 'taper' to natural weighting acts as a smoothing in the UV-plane by reducing weight on the long baselines. Briggs Robust=0.5 weighting is the default assumed in ALMA setup and pipeline data-processing (Desired Angular Resolution, noise rms requested, delivered data products), however, it increases the noise by ∼10-15% in exchange for a ∼20-30% smaller PSF compared to natural weighting.

**1.3.5. Spectral resolution.** Unlike optical to near-infrared spectrographs ALMA does not have a line spread function (LSF) and the spectral resolution is typically limited by the number of channels that are averaged in order to obtain the required S/N. For typical high-redshift observations where 8-16× spectral averaging is used for the spectral window setup, the spectral resolution is roughly equal to the channel width of the reduced data. Without an LSF, the width of the line (FWHM) is an important parameter in the required exposure time for a given line flux (see below).

## 1.4. ALMA observations

The following parameters and definitions are useful for designing observations of faint galaxies at $z \gtrsim 7$ and setting up the ALMA observing tool (OT):

- **4-bit spectral mode**: This mode increases sensitivity for spectral line observations, at a cost of continuum sensitivity and spectral range (e.g. for line scanning). A typical spectral window (spw) setup for a spectral line detection experiment is 4-bit 937.5 MHz bandwidth, with 15.625 MHz spectral resolution (16× spectral averaging). A typical spw setup for a continuum detection experiment is 2-bit 1875 MHz bandwidth, with 31.250 MHz spectral resolution, while a line scanning setup would use 2-bit 1875 MHz bandwidth, with 15.625 MHz spectral resolution.
- **Largest Angular Structure**: The diameter of a source covering most of the source flux, e.g. the Full Width at 10% Maximum or the diameter of a Kron factor $k$=2.5 aperture. The OT will check that this angular scale is smaller than $\theta_{\mathrm{MRS}}$ (see below).

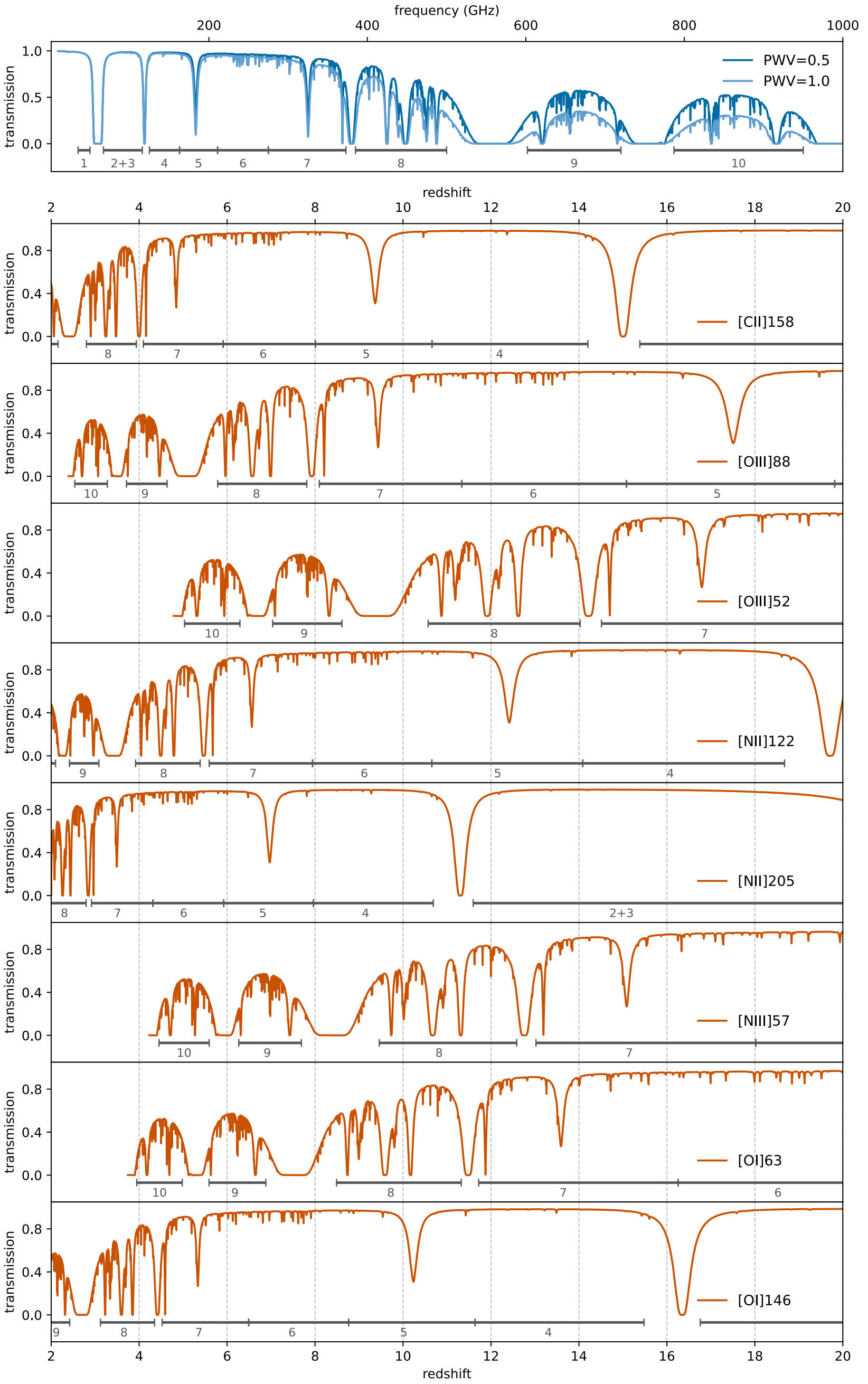


Supplemental Figure 4: *Top panel:* transmission of the atmosphere as a function of frequency for precipitable water vapour (PWV) of 0.5 mm, corresponding to the top 25% best weather conditions, and more typical weather conditions with PWV = 1.0 mm. *Bottom panels:* transmission of the atmosphere for PWV = 0.5 mm as a functon of redshift for 8 different emission lines that could be measured with ALMA. Grey bars show the ranges of the different ALMA bands.

- **Maximum Recoverable Scale -** $\theta_{\rm MRS}$: The angular scale resolved out by ∼95% of ALMA baselines for a given configuration. This parameter is meaningful for very bright sources (S/N>>10) but rarely applies to high-redshift observations, where much smaller angular scales are resolved out at a significant S/N level (Supplemental Figure 2).
- **(Synthesized) Beam**: Full Width at Half Max (FWHM) of a Gaussian fit to the primary peak of the "dirty Beam" in the image plane.
- **Desired Angular Resolution**: Requested synthesized beam, assuming a Briggs Robust=0.5 weighting scheme; i.e. a 20-30% smaller beam than would be accomplished with natural weighting. For a *detection experiment* (maximizing S/N without gaining structural information) $\rm FWHM_{beam} \gtrsim 2{\times}FWHM_{source}$ is optimal. Using $\rm FWHM_{beam} \sim FWHM_{source}$ will give some modest structural information at the cost of ∼40% lower peak S/N.
- **Primary Beam**: Sensitivity of the observation drops with distance from the target coordinates. The field of view of ALMA for a given wavelength, $\theta \sim \lambda/D_{\rm antenna}$ ($D$=12 m for ALMA large array), corresponds to the FWHM of the primary beam.
- **Bandwidth used for Sensitivity**: The spectral width over which the root-mean-square noise is estimated - typically user supplied in km/s for line observations (e.g. $\rm FWHM_{line}/3$) and AggregateBandWidth for continuum observations.
- **Minimum line S/N in the OT**: For a simple ALMA observational setup to obtain a line flux measurement of an unresolved source - for a given line width, $\rm FWHM_{line}$, in km/s - the peak line flux density, $S_{\rm line}\,{\rm dv/FWHM_{line}}$, is required to be detected with S/N=3 over a spectral bandwidth equal to $\rm FWHM_{line}/3$.

**1.4.1. ALMA bands and atmosphere transmission.** An important consideration for observing with ALMA is the atmosphere transmission in different bands and the weather conditions. Supplemental Figure 4 shows the transmission of the atmosphere for different frequencies and redshift for a given emission line. While lower ALMA bands can be observed in a wide range of weather conditions, band 7-8 typically requires precipitable water vapour (PWV) below 0.5 mm - corresponding to the 25% best weather conditions on the Chajnantor Plateau. For ALMA band 9-10 even better conditions of PWV < 0.25 mm is required, corresponding to only 10% of the total ALMA observing time. As a comparison; on Cerro Paranal and many other observing sites, 10% best weather conditions corresponds to < 0.4" V-band seeing conditions.

## 2. Compilation of a catalogue from the literature

In Table 1 we show the highest redshift objects in the literature, that have been observed with ALMA to-date. In the catalogue produced for this review, we extracted various properties from the literature, and computed several ourselves in a uniform way.

From the literature we extract key information — IDs, coordinates, redshifts, magnification factors — together with basic physical properties (UV luminosity, stellar mass, rest-frame UV slope $\beta_{\rm UV}$), emission line properties (luminosities, equivalent widths, line widths), and dust temperatures ($T_{\rm d}$ or $T_{\rm peak}$).

While continuum fluxes and frequencies are extracted from the literature, $L_{\rm IR}$ is uniformly computed as discussed in the main text, using the 160$\mu$m flux or upper limit if multiple bands are available. Using the derived $L_{\rm IR}$ and continuum fluxes, we furthermore

compute $f_{\rm obs}$, IRX and dust mass. For our literature compilations we complement our catalog with the REBELS stacked properties from Bowler et al. (2024). We compute the following dust masses from the Bowler et al. (2024) stacks: $\log_{10} M_{\rm d}/{\rm M}_\odot = 6.65 \pm 0.09$ and $6.83 \pm 0.08$ in the $6.5 < z < 6.9$ bin ($M_{\rm UV} = -22.34$ and $-21.71$ mag respectively) and $\log_{10} M_{\rm d}/{\rm M}_\odot < 6.94$ and $\log_{10} M_{\rm d}/{\rm M}_\odot = 6.65 \pm 0.09$ in the $6.9 < z < 7.7$ bin ($M_{\rm UV} = -22.75$ and $-21.97$ mag respectively)

The UV and IR SFRs were derived using ${\rm SFR_{UV}} = \kappa_{\rm UV} \cdot L_{\rm UV}$ and ${\rm SFR_{IR}} = \kappa_{\rm IR} \cdot L_{\rm IR}$, where $\kappa_{\rm UV} = 0.72 \times 10^{-28}\,{\rm M_\odot\,yr^{-1}\,erg^{-1}\,s\,Hz}$ and $\kappa_{\rm IR} = 2.83 \times 10^{-44}\,{\rm M_\odot\,yr^{-1}\,erg^{-1}\,s}$ following Madau & Dickinson (2014), but converting to a Chabrier (2003) IMF.

Source selection and category are divided in four labels: LBG, LAE, Quasar and ALMA. ALMA refers to sources selected in the rest-frame FIR through serendipitous detections with ALMA and NOEMA or from (sub)millimetre single-dish surveys. From the sample of 79 LBG-selected sources 48 are from REBELS and its pilot programs (Smit et al. 2015; Laporte et al. 2017b; Smit et al. 2018; Bouwens et al. 2022; Inami et al. 2022; Schouws et al. 2022; Witstok et al. 2022; Schouws et al. 2023; Algera et al. 2023, 2024b,a; Harikane et al. 2025; Scholtz et al. 2025, Schouws et al. in prep). For the remaining LBG-selected galaxies there are 14 lensed sources: A1689-zD1 (Bradley et al. 2008; Watson et al. 2015; Knudsen et al. 2017; Bakx et al. 2021; Wong et al. 2022; Witstok et al. 2023; Killi et al. 2023), A1703-zD1 (Bradley et al. 2012; Smit et al. 2014; Schaerer et al. 2015; Molyneux et al. 2022), A2744-YD1, A2744-YD4, A2744-YD7, A2744-s1 (Laporte et al. 2017a; Morishita et al. 2023; Hashimoto et al. 2023), ID4590 (Heintz et al. 2023; Fujimoto et al. 2024; Giménez-Arteaga et al. 2023), MACS0416-JD , MACS0416_Y1 (Laporte et al. 2015; Infante et al. 2015; Kawamata et al. 2016; Tamura et al. 2019; Bakx et al. 2020; Laporte et al. 2021; Algera et al. 2024b; Ma et al. 2024), MACS0647-JD (Coe et al. 2013; González-López et al. 2014; Hsiao et al. 2023), MACS1149-JD1 (Wilkins et al. 2016; Hashimoto et al. 2018; Laporte et al. 2019; Carniani et al. 2020; Stiavelli et al. 2023), MS0451-H (Knudsen et al. 2016; Carniani et al. 2020; Jauzac et al. 2021), RXJ1347 (Smit et al. 2014; Huang et al. 2016; Bradač et al. 2017), and RXJ1347-018 (Glazer et al. 2024); and 17 unlensed sources: B14-65666 (Furusawa et al. 2016; Hashimoto et al. 2019; Bowler et al. 2022; Algera et al. 2024b; Sugahara et al. 2025), BDF-3299, BDF-521, SDF-46975 (Vanzella et al. 2011; Maiolino et al. 2015; Carniani et al. 2017, 2020), COSMOS13679, NTTDF6345, UDS16291, UDS4812 (Pentericci et al. 2016, 2018; Carniani et al. 2018), EGS-1952445714, EGS-5711424617 (Smit et al. 2015; Molyneux et al. 2022), GHZ1 , GHZ2 (Naidu et al. 2022; Castellano et al. 2022; Yoon et al. 2023; Bakx et al. 2023; Popping 2023; Castellano et al. 2024; Zavala et al. 2024; Napolitano et al. 2025), GN-z11 (Oesch et al. 2016; Bunker et al. 2023; Fudamoto et al. 2024), GS-z14 (Carniani et al. 2024, 2025; Schouws et al. 2025b,a), JADES-GS-z11-0 (Curtis-Lake et al. 2023; Witstok et al. 2026), S5-z17-1 (Harikane et al. 2023; Fujimoto et al. 2023; van Marrewijk et al. 2025), and z8-GND-5296 (Finkelstein et al. 2013; Schaerer et al. 2015).

There are 10 LAE-selected galaxies: CR7, VR7, MASOSA (Sobral et al. 2015; Matthee et al. 2017, 2019, 2020b,a), HCM6A (Hu et al. 2002; Kanekar et al. 2013; Schaerer et al. 2015), Himiko (Ouchi et al. 2013; Schaerer et al. 2015; Bowler et al. 2017; Kiyota et al. 2025), IOK-1 (Ono et al. 2012; González-López et al. 2014; Ota et al. 2014; Schaerer et al. 2015), MUSE852 (Uzgil et al. 2021), REBELS-07/SXDF-NB1006-2 (Ouchi et al. 2013; Inoue et al. 2016; Ren et al. 2023; Harikane et al. 2025), SDFJ132408.3+271543, SDFJ132415.7+273058 (Taniguchi et al. 2005; González-López et al. 2014).

Supplemental Table 1: A summary of the ultra-high redshift objects ($z > 7.7$) galaxies that have been observed with ALMA. The complete table from $z = 6.5$ is presented in the supplementary information.

| ID | Redshift | $M_{\rm UV}$ /mag | $L_{\rm IR}$ /$10^{10}$L$_\odot$ | $L_{\rm CII}$ /$10^8$L$_\odot$ | $L_{\rm OIII}$ /$10^8$L$_\odot$ |
|---|---|---|---|---|---|
| S5-z17-1 | $16.41^{+0.66}_{-0.55}$ | $-21.60 \pm 0.30$ | $< 80.9$ | – | – |
| GS-z14 | 14.179 | $-20.81 \pm 0.16$ | $< 11.6$ | $< 0.2$ | $2.1 \pm 0.5$ |
| GHZ2 | 12.333 | $-20.50 \pm 0.01$ | $< 5.4$ | – | $1.7 \pm 0.4$ |
| JADES-GS-z11-0 | 11.122 | $-19.30 \pm 0.05$ | $< 3.2$ | – | $1.0 \pm 0.3$ |
| GN-z11 | 10.603* | $-21.60 \pm 0.10$ | $< 47.9$ | $< 0.5$ | – |
| MACS0647-JD | 10.170* | $-20.30 \pm 0.20$ | $< 35.9$ | $< 0.5$ | – |
| GHZ1 | 9.875* | $-19.96 \pm 0.03$ | $< 4.9$ | – | $< 0.4$ |
| MACS1149-JD1 | 9.110 | $-19.10 \pm 0.06$ | $< 1.1$ | $0.1 \pm 0.0$ | $0.7 \pm 0.2$ |
| ID4590 | 8.496* | $-20.30 \pm 0.12$ | $< 1.5$ | $0.2 \pm 0.0$ | $0.4 \pm 0.1$ |
| REBELS-P2 | $8.47^{+0.73}_{-0.72}$ | $-21.74 \pm 0.36$ | $< 25.5$ | – | – |
| REBELS-04 | $8.45^{+0.10}_{-0.09}$ | $-22.33 \pm 0.03$ | $11.6 \pm 3.4$ | – | – |
| MACS0416-JD | 8.340* | $-19.60 \pm 0.05$ | $< 4.2$ | – | – |
| REBELS-P4 | $8.32^{+0.66}_{-0.92}$ | $-21.92 \pm 0.43$ | $< 28.7$ | – | – |
| MACS0416_Y1 | 8.311 | $-20.83 \pm 0.05$ | $17.6 \pm 3.3$ | $1.4 \pm 0.2$ | $12.0 \pm 3.0$ |
| REBELS-24 | $8.13^{+0.66}_{-0.51}$ | $-22.04 \pm 0.24$ | $< 22.8$ | – | – |
| REBELS-11 | $8.09^{+0.65}_{-0.37}$ | $-22.77 \pm 0.22$ | $< 14.9$ | – | – |
| A2744-YD1 | 7.879* | $-18.40 \pm 0.10$ | $< 26.7$ | – | – |
| A2744-YD7 | 7.877* | $-20.20 \pm 0.20$ | $< 18.5$ | – | – |
| A2744-YD4 | 7.875* | $-20.20 \pm 0.10$ | $< 19.5$ | – | – |
| A2744-s1 | 7.874* | $-16.70 \pm 0.50$ | $< 2.9$ | – | – |
| REBELS-13 | $7.79^{+0.84}_{-0.50}$ | $-22.90 \pm 0.23$ | $< 13.3$ | – | – |
| REBELS-37 | 7.642 | $-22.25 \pm 0.03$ | $16.0 \pm 2.6$ | – | – |

[a]Upper limits are $3\sigma$.; [b]* indicates redshift is from non-ALMA source (e.g. Lyman-$\alpha$).

There are 47 sources selected as Quasars or other type of AGN: COS-87259 (Endsley et al. 2022, 2023; Bouwens et al. 2025), GNz7q (Fujimoto et al. 2022), J0109-3047, J0305-3150, J1048-0109, J1120+0641, J1205-0000, J1243+0100, J1342+0928, J2348-3054, P006+39, P036+03, P167-13, P231-20, P323+12, and P338+29 (Venemans et al. 2012; Bañados et al. 2015; Venemans et al. 2016; Mazzucchelli et al. 2017; Decarli et al. 2018; Novak et al. 2019, 2020; Venemans et al. 2020; Izumi et al. 2021); and 31 quasars from Wang et al. (2024a).

The ALMA category has two very luminous single-dish-selected sources: SPT0311-58E and SPT0311-58W (Strandet et al. 2017; Marrone et al. 2018; Litke et al. 2023; Álvarez-Márquez et al. 2023; Arribas et al. 2024); three sources selected close to LBGs: REBELS-12-2, REBELS-29-2, and REBELS-39-2 (Fudamoto et al. 2021; van Leeuwen et al. 2024; Algera et al. 2025); and 13 sources selected close to quasars; SMG-A, SMG-B, SMG-C (Wang et al. 2024b), J0305-3150C1, J0305-3150C2, J0305-3150C3, J1342+0928C1, P167-13C1, P183+05C2, P231-20C1, P231-20C2, P231-20C3, and P231-20C4 (Willott, Bergeron & Omont 2017; Decarli et al. 2017; Mazzucchelli et al. 2019; Neeleman et al. 2019; Venemans et al. 2020; Bakx et al. 2024; van Leeuwen et al. 2024).

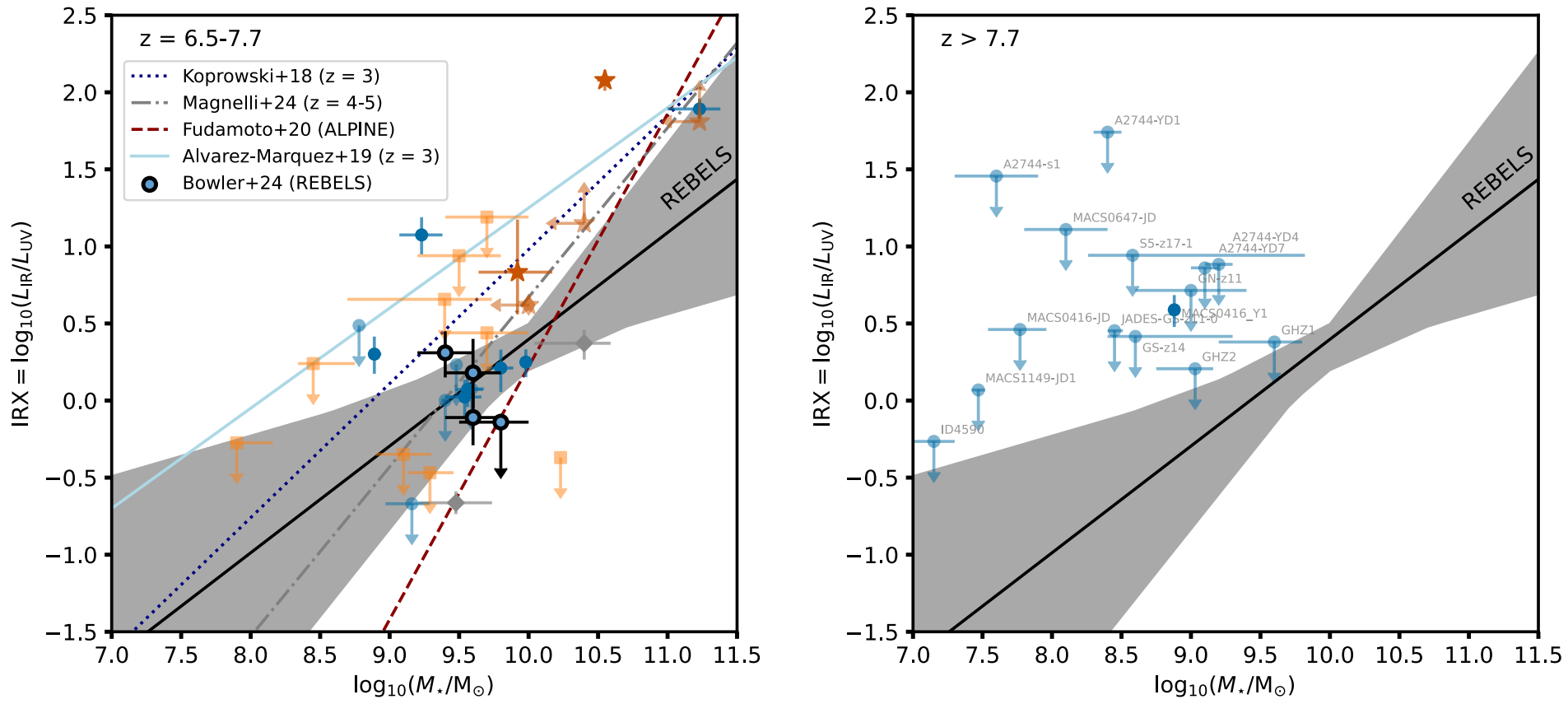


Supplemental Figure 5: The IRX–$M_\star$ scaling relation with rest-frame UV slope and stellar mass from the compilation of ALMA observed sources at $z = 6.5$–$7.7$ (left) and $z > 7.7$ (right). In the $z < 7.7$ plot we show the derived relations at $z \simeq 3$ (Koprowski et al. 2018; Álvarez-Márquez et al. 2019) and the ALPINE/A$^3$COSMOS samples at $z \simeq 4$ (Fudamoto et al. 2020; Magnelli et al. 2024). For clarity the majority of REBELS galaxies have been removed from the plot, and the stacked results from Bowler et al. (2024) shown instead (open circles). REBELS galaxies that have improved $\beta_{\rm UV}$ and $M^\star$ values from *JWST* are shown separately (Fisher et al. 2025).

## 3. Mid-infrared Fine Structure Lines at the highest redshifts

For galaxies at $z > 8$ bands 9 and 10 start tracing rest-frame mid-infrared (MIR; 5-40 $\mu$m) lines. First detected in a handful of $z = 0$ galaxies with the Long Wavelength Spectrometer (LWS) on board the Infrared Space Observatory (ISO) and later with Spitzer Infrared Spectrograph (IRS). Some of these lines are too far in the infrared spectrum to be studied by JWST/MIRI at $z = 0$ and therefore limited reference samples are available.

[OIV] $\lambda 26\mu$m is perhaps the most targeted line, which due to its high ionisation potential of 54.94 eV is an ideal AGN diagnostic. AGN dominated sources typically show luminosities $L_{[\rm OIV]}/L_{[\rm OIII]} \gtrsim 1$, while metal-poor dwarf galaxies at $z \sim 0$ range over $L_{[\rm OIV]}/L_{[\rm OIII]} \sim 0.01 - 0.1$ (e.g. Spinoglio et al. 2017). Similarly, [Ne V] $\lambda 24\mu$m can be used as an AGN diagnostic, having an even higher ionisation potential of 97.12 eV, but is often $\sim 3 - 5\times$ fainter than [OIV].

[Si II] $\lambda 35\mu$m is a neutral/PDR gas tracer (ionisation potential of 8.15 eV) and becomes an important coolant of PDR gas at very high densities and high FUV radiation fields (Meijerink, Spaans & Israel 2007). Silicon is of interest because it is a major constituent of dust grains, with [Si II]/[OI] tracing the relative abundances of silicon and oxygen.

[S III] $\lambda 33\mu$m and [S III] $\lambda 18\mu$m are fine-structure lines tracing ionised gas in star-forming regions. The [S III] $\lambda 33\mu$m/[S III] $\lambda 18\mu$m line ratio is sensitive to electron density $n_e$ between 100 and 10,000 cm$^{-3}$. However, [S III] $\lambda 18\mu$m will likely require the next generation FIR space telescope (such as the proposed PRIMA mission) to be observed, given its only accessible to ALMA in band 10 beyond redshift 15. [S III] $\lambda 33\mu$m is frequently brighter than [O III] $\lambda 88\mu$m and accessible in band 9 and 10 over the redshift range $z \sim 8 - 14$.

## 4. IRX–$M_\star$

Previous works by Fudamoto et al. (2020) from the ALPINE survey have shown a deficit in the IRX at a give stellar mass in comparison to $z \simeq 3$ results. This effect, shown in Supplemental Figure 5 is confirmed in the stacking results of Bowler et al. (2024) at $z \simeq 7$. We see in general that LAEs lie further below the average relation, which is unsurprising due to their lower dust content as derived from rest-frame UV and optical analyses (Ouchi, Ono & Shibuya 2020). Conversely, the ALMA detected sources that populate the high mass end with $\log_{10}(M_\star/\mathrm{M}_\odot) \sim 10$ show high IRX values consistent or slightly above the average relation. These results indicate that the proportion of obscured star-formation is lower by a factor of $\geq 3$ at a given stellar mass. This is consistent with the $f_{\rm obs}$ trend with stellar mass, and suggest that although the dust properties in terms of attenuation (in the IRX–$\beta$ relation) appear close to that found at lower redshift, the degree of dust obscuration is lower. It is clear that at $z > 7.7$ the majority of sources are undetected in the dust continuum, but that this is expected given the low stellar masses inferred for these faint sources ($\log_{10}(M_\star/\mathrm{M}_\odot) \lesssim 9$). The highest redshift continuum detected source, MACS0416_Y1 at $z = 8.31$ (Bakx et al. 2020), appears in excess of the $z = 4$–8 stacked relation from Bowler et al. (2024). Larger samples are clearly needed to determine the normalisation and scatter in the IRX–$M_\star$ relation at $z \simeq 8$ and beyond.



\end{document}